\documentstyle[epsf,12pt]{article}

\newlength{\figsize}
\begin{document}

\begin{titlepage}

\begin{tabbing}
\` hep-lat/9804008 \\
\` Oxford preprint OUTP-98-29P \\
\end{tabbing}
 
\vspace*{.1in}
 
\begin{center}
{\large\bf SU(N) gauge theories in 2+1 dimensions\\}
\vspace*{.7in}
{Michael J. Teper\\
\vspace*{.3in}
Theoretical Physics, University of Oxford,\\
1 Keble Road, Oxford, OX1 3NP, U.K.\\
}
\end{center}

\vspace*{0.8in}

\begin{center}
{\bf Abstract}
\end{center}

We calculate the mass spectra and string tensions
of SU(2), SU(3), SU(4) and SU(5) gauge theories in 
2+1 dimensions. We do so by simulating the corresponding lattice
theories and then extrapolating dimensionless mass ratios to 
the continuum limit. We find that such mass ratios are, to a 
first approximation, independent of the number of
colours, $N_c$, and that the 
remaining dependence can be accurately reproduced by a simple 
$O(1/N^2_c)$ correction. This provides us with a prediction 
of these mass ratios for all SU($N_c$) theories in 2+1 
dimensions and demonstrates that these theories
are already `close' to $N_c=\infty$ for $N_c\geq 2$. 
We find that the theory retains a non-zero confining string
tension as $N_c \to \infty$ and that the dimensionful coupling 
$g^2$ is proportional to $1/N_c$ at large $N_c$, when expressed 
in units of the dynamical length scale of the theory.
During the course of these calculations we study in detail the 
effects of including over-relaxation in the Monte Carlo, of using
a mean-field improved coupling to extrapolate to the continuum
limit, and the use of space-time asymmetric lattice actions
to resolve heavy glueball correlators.


\end{titlepage}

\setcounter{page}{1}
\newpage
\pagestyle{plain}

\section{Introduction}
\label		{intro}

The non-perturbative physics of QCD continues to be largely impervious 
to analytic attack. Thus 't Hooft's proposal to consider SU($N_c$) gauge 
theories (with quarks) as perturbations in powers of $1/N_c$ 
around $N_c=\infty$ 
\cite{tHo}.
remains of great interest. In many ways the $N_c=\infty$
theory is much simpler than the physically interesting 
$N_c=3$ theory, and the fact that the phenomenology
of the SU($\infty$) quark-gluon theory appears to be strikingly 
similar to that of (the non-baryonic sector of) QCD
\cite{tHo,Wit}.
motivates the suggestion that the physically interesting SU(3) 
theory might be largely understood if we could solve the much simpler 
SU($\infty$) theory. Unfortunately an analytic solution 
of the latter still
eludes us, even if much progress has been made in understanding
aspects of its structure
\cite{Cole,ManN,EK,MalWit}.

This situation has motivated a number of computational explorations
\cite{Das}.
Almost all of these have used the fact that the lattice SU($\infty$) 
theory can be re-expressed as a single plaquette theory
\cite{EK}.
Although these calculations have produced interesting results, 
the approach suffers from a basic problem: it tells
us nothing about the corrections to the $N_c=\infty$ limit
and so cannot address the critical question of how close
SU(3) is, in fact, to SU($\infty$). The twisted Eguchi-Kawai
approach also suffers from the fact that the space-time volume 
described by the theory is finite and related to $N_c$.

In this paper we take a more direct approach to the problem.
We calculate the continuum properties of SU(2), SU(3), SU(4), ...
theories by simulating the corresponding lattice theories.
We then compare these properties and see how well they can
be described by simple corrections to a common $N_c=\infty$ 
limit. This approach has the advantage that it will tell
us just how close the physically interesting SU(3) theory
is to the simpler SU($\infty$) theory. It has, of course,
a potential disadvantage: if the theories we consider are far 
from the $N_c=\infty$ limit then we will have learned nothing
about the physics of that limit. Fortunately, as we shall see,
this turns out not to be the case. 
 
Ultimately we would like to consider SU($N_c$) gauge theories
coupled to light quarks (in the fundamental representation) in
3+1 dimensions. But this is beyond our current computational
resources. So our first restriction is to disregard the
quarks and focus instead on the pure gauge theory.
Since the non-perturbative physics of QCD is largely driven
by the self-coupling of the gauge fields this is certainly a
physically relevent problem. Moreover one expects the pure
gauge theory to have leading corrections that are $O(1/N_c^2)$
rather than the $O(1/N_c)$ that one expects with quarks. Thus 
the onset of large-$N_c$ physics should be easier to spot.

Our second restriction is to consider the pure gauge theory
in 2+1 rather than in 3+1 dimensions. Although it is less 
obvious that this leaves us with a physically relevant
problem, we shall argue in the next Section that gauge theories 
in 3 and 4 dimensions are sufficiently similar that this is 
likely to be so. Moreover it turns out that in $D=2+1$ we
can calculate the continuum properties of gauge theories with 
such accuracy that there is little ambiguity in our final
conclusions. This is not the case in D=3+1 where the
preliminary calculations of this kind
\cite{MTN}
are very inaccurate in comparison.

Motivating our calculations are several questions of particular
interest. What we know about the large-$N_c$ limit of gauge 
theories essentially comes from considering Feynman diagrams
to all orders. Such considerations indicate that SU($N_c$) gauge 
theories possess a smooth $N_c \to \infty$ limit if one varies
the coupling so as to keep $g^2 N_c$ fixed. Moreover the leading
correction to this limit should be $O(1/N_c^2)$. It is obviously
of interest to check these expectations in a fully non-perturbative
calculation. In addition, the phenomenology of large-$N_c$ theories
assumes that the theory remains confining in that limit. It is
important to check that this is really the case. Finally we
wish to see how large are the corrections at, say, $N_c=3$.
Is it really the case that SU(3) is close to SU($\infty$)? 

In addition we aim to calculate the detailed mass 
spectrum of the SU($\infty$) theory. We note that
models and theoretical approaches are usually simpler
in that limit. For example, the flux tube model 
of glueballs
\cite{patisg,tepmor,rjmt}
would naively appear to be identical for $N_c > 2$.
However, because the model does not incorporate the
effects of glueball decay, it should really be tested
against the $N_c\to\infty$ spectrum since it is only in that 
limit that there are no decays. A second example 
is provided by the recent progress in calculating the
large $N_c$ mass spectrum using light-front quantisation
techniques
\cite{Dalley}.
Thus our large-$N_c$ spectrum can serve as a useful
testing ground for models and attempts at analytic
solutions of the theory
\cite{Dalley,Ooguri}.
Of course in the process we shall
also calculate the spectra of the theories at finite $N_c$
and these too can be used as a testing ground for models and
analytic approaches. We remark that the most recent examples 
of the latter 
\cite{Hamer,Luo,Nair}
%
are intriguingly successful. For example, in 
\cite{Luo} 
the SU(2) and SU(3) $0^{++}$ glueball masses are
within $15\%$ of our values and the
$0^{--}$ is even closer. In  
\cite{Nair} 
the string tension is calculated for all $N_c$ and is
within $2\%$ of the values in this paper. 

We now briefly outline the contents of this paper.
In the next section we discuss some general properties of
SU($N_c$) theories in 2+1 dimensions. The aim is 
not only to set the framework for the subsequent calculations,
but also to convince the reader that these theories are
sufficiently similar to their $D=3+1$ counterparts that
what we learn in this paper about the former probably tells
us something about the latter. We then move on to discuss
the technical details of how we carry out our calculations.
This includes details of our operators and of the variational
principle that underlies our extraction of excited states from 
the matrix of correlation functions. We also discuss the problem 
of identifying the continuum spin of a particle on a lattice
with only cubic symmetry, and provide a simple criterion
for doing so.
In the following section we present our calculations of the
string tension. It is here that we aim to demonstrate that
the theory remains confining as $N_c\to\infty$ thus 
providing the crucial ingredient for extracting the
phenomenology of those theories. It turns out that the
string tension is the physical quantity that we calculate
the most accurately. So it is in this section that we
shall test most precisely the expectation that there
is a smooth limit reached by keeping $g^2N_c$ constant,
and that the leading corrections to that limit are
$O(1/N_c^2)$. The next section contains our calculations
of the glueball spectrum. (Since we have only gluons
in our theories, all the physical particles are colour
singlet composites of gluons i.e. glueballs.)
We shall briefly comment on the features of this spectrum;
in particular upon how it compares to what we know about the
$D=3+1$ spectrum. It would of course be nice to be able
to spot some striking regularities in the mass spectrum of the
SU($\infty$) theory. However one can only read significance
into the details of the spectrum if one has some 
framework that relates those details to the underlying
dynamics. For such an analysis, within the context of a model
in which glueballs are composed of closed loops of 
chromoelectric flux, we refer the reader to,
\cite{tepmor,rjmt}.
We shall find that the glueball spectra are in fact very 
similar to each other for all $N_c \geq 2$. To provide a 
contrast it is amusing to compare this to the spectrum that
one obtains in the $D=2+1$ U(1) theory -- the theory that is
furthest from U($\infty$) in the sequence of U($N_c$) theories. 
This we do in an Appendix E. We conclude with a brief discussion 
of our results.

We have relegated to the Appendices details of some 
self-contained aspects of
our calculations which should be of general interest.
Appendix A contains a detailed evaluation of the effects of 
over-relaxation in the Monte Carlo update. This is done
by directly comparing the statistical errors on the masses
that one obtains in (realistically large) calculations
that contain various ratios of over-relaxation to heatbath.
We are not aware of any previous comparison of this kind.
We find that over-relaxation brings a modest but worthwhile
improvement, particularly for the smaller values of the 
lattice spacing. 
Appendix B contains some analyses relevant to our
choice of operators. In particular we study how
sensitive is our operator basis to variations in
the parameter governing the `blocking'; whether
anything is gained by the inclusion of operators
incorporating `baryonic' vertices (for $N_c \geq 3$);
and we give more details about
our variational calculations of
the excited states. These are no doubt known to some experts,
but we hope they will be of use to others in the field.
In Appendix C we compare continuum extrapolations of the
string tension in the bare and mean-field improved couplings. 
We show that using the latter considerably improves the 
accuracy of the final results. This is of some interest
because we are not aware of any such previous comparison.
Appendix D contains details of the calculations
we have performed with asymmetric lattices which have
timelike lattice spacings that are much smaller than the
spacelike spacings. The primary purpose of this calculation is  
to check that our procedure for estimating the masses 
of the heavier states is in fact reliable. However our
methods for determining the ratio of the lattice spacings
are likely to be of more general interest and so we
develop this calculation in some detail. This 
enables us to calculate the SU(2) spectrum of the
theory close to its `Hamiltonian' limit, and to
compare it with the symmetric case as a test of universality.
Finally, in Appendix E we summarise some properties of the
U(1) mass spectrum. This is to provide a contrast to the
SU($N_c$) spectra that are the subject of this paper.

During the course of this work, we have published some 
preliminary summaries of some of the topics in this paper,
as well as on related topics not covered herein. Our early 
results on the SU(2) spectrum appeared in
\cite{MT2G}.
(See 
\cite{CaselleG}
for an interesting comparison with the spectrum of the 
gauge Ising model.)
The preliminary SU(2) string tension was discussed in  
\cite{MT2K}.
That paper also contained a study of the width and vibrational 
properties of the corresponding flux tube, which is not repeated 
here. Also not covered in this paper are the SU(2) and SU(3) 
deconfining temperatures
\cite{MT2T}.
(See 
\cite{CaselleT}
for a more extensive discussion.)
Some preliminary results on the $N_c$ dependence in 2+1 dimensions
have appeared in  
\cite{MTN}.
This paper also contains some {\it very} preliminary calculations for
gauge theories in 3+1 dimensions.

\section{Some features of D=2+1 gauge theories}

In the first part of this section we discuss some of
the fundamental dynamical properties of SU($N_c$) gauge theories 
in 2+1 dimensions. 
Since we are much more familiar with
the corresponding properties of the same theories in
3+1 dimensions, it will be illuminating to compare the two
theories as we go along. 

The second part of the section focuses on the consequences of
the fact that in D=2+1 parity and angular momentum do not commute.
We show that this leads to some parity doubling in the spectrum;
but that precisely what gets doubled is sensitive to the
ultraviolet and infrared cut-offs that we impose in our
calculations. 

\subsection{D=2+1 $\sim$ D=3+1?}

Gauge theories in 2+1 dimensions possess a dimensionful
coupling: $g^2$ has dimensions of mass and so provides
a scale even for the classical theory. By contrast in 
$D=3+1$ $g^2$ is dimensionless; the theory is classically 
scale-invariant. In addition the Coulomb interaction
in $D=2+1$ is $\propto g^2 \log r$; so the theory is
already confining at the `classical' level -- albeit
only logarithmically. Nonetheless, these 
apparently quite striking differences are misleading: 
the $D=2+1$ theory shares with its $D=3+1$ counterpart
its most important dynamical properties, as we now 
remind the reader.

\vskip 0.1in
{\noindent}$\bullet$ {\it Ultraviolet freedom.}
Both theories become free at short distances.
In 3 dimensions  the coupling, $g^2$, has dimensions
of mass so that the effective dimensionless expansion
parameter on a scale $l$ will be 
\begin{equation}
g_3^2(l) \equiv lg^2 \stackrel{l\to 0}{\longrightarrow} 0    
\label{A1}
\end{equation}
In 4 dimensions the coupling is dimensionless and runs
in a way we are all familar with:
\begin{equation}
g_4^2(l) \simeq {c\over{\ln(l\Lambda)}}
 \stackrel{l\to 0}{\longrightarrow} 0        
\label{A2}
\end{equation}
In both cases the interactions vanish as $l \to 0$,
although they do so much faster in the super-renormalisable
$D=2+1$ case than in the merely asymptotically free
$D=3+1$ case. 

\vskip 0.1in
   
{\noindent}$\bullet$ {\it Infrared slavery.}
The counterpart of the couplings becoming weak 
at short distances is that they become strong at large
distances -- ``infrared slavery''. This is immediate
if we let  $l\uparrow$ in eqns(\ref{A1},\ref{A2}). 
Thus in both 3 and 4 dimensions the interesting physics 
is nonperturbative.
   
\vskip 0.1in

{\noindent}$\bullet$ {\it The coupling and the mass scale.}
In 3 dimensions the coupling has dimensions of mass and
so explicitly sets the mass scale for the theory:
\begin{equation}
m_i = c_i g^2           
\label{A3}
\end{equation}
where $m_i$ is any dynamically generated mass in the theory.
(For example, a glueball mass.)
In 4 dimensions the coupling is dimensionless and so, naively,
things appear quite different. However in fact here too the 
coupling sets the overall mass scale. It does so through the 
phenomenon of dimensional transmutation: the classical scale 
invariance is anomalous, the coupling runs and this introduces 
a mass scale through the rate at which it runs:
\begin{equation}
m_i = c_i \Lambda       
\label{A4}
\end{equation}
where $\Lambda$ is as in eqn(\ref{A2}). So in both 3 and 4 
dimensions the value of the coupling determines the overall
mass scale.

\vskip 0.1in
   
{\noindent}$\bullet$ {\it Confinement.}
Both theories confine with a linear potential.
This is not something that we can prove of course.
However lattice simulations provide convincing evidence
that this is indeed the case. Note that although the
$D=2+1$ Coulomb potential is already confining, this
is a weak logarithmic confinement,
$V_C(r) \sim g^2 \ln(r)$, which has nothing to do with
the nonperturbative linear potential, $V(r) \simeq \sigma r$,
that one finds at large $r$.

\vskip 0.1in

In addition to these theoretical similarities, we shall
see that the calculated mass spectra also show some
striking similarities. For example: the lightest glueball
is the scalar $0^{++}$ with a similar mass, 
$m_{0^{++}} \sim 4\surd\sigma$, in both cases. 

While the above comparisons provide some support for
the argument that what we learn about $D=2+1$ gauge
theories might have something to teach us about 
the more interesting $D=3+1$ theories as well, 
it is important to emphasise that the 
theories do differ in important respects and
are certainly not the same. 
For example, there are no instantons in $D=2+1$ non-Abelian 
gauge theories. This would surely matter a great deal if we 
were to include quarks. Another difference is the fact 
that the rotation group in two space dimensions is Abelian.
This has some important consequences to which we now turn.

%
%
%
\subsection{Spin and parity doubling}

In two space dimensions rotations commute: the group is
Abelian. So states of spin $J$ do not come in multiplets
in the way that they do in 3+1 dimensions where the
rotation group is non-Abelian. We shall use $(x,y)$ for 
the spatial coordinates and $\theta$ for the angle of 
rotation. We can then define a parity transformation, $P$, 
by $(x,y) \to (x,-y)$. We note that the angular 
momentum operator, $x\partial_y - y\partial_x$, flips sign
under parity. That is to say, if some state $|\phi\rangle$
has angular momentum $j$ then the state $P|\phi\rangle$
will have angular momentum $-j$. 

This last fact has an important consequence for the spectrum.
Suppose $|j\rangle$ is some state of angular momentum $j$
and energy $E_j$. Consider the two linear combinations 
\begin{equation}
|j,\pm\rangle=|j\rangle \pm P|j\rangle .
\label{A5}
\end{equation}
If they are both non-null they will form a pair of states
that have opposite parity, since we easily see that
$P|j,\pm\rangle = \pm |j,\pm\rangle$. 
Moreover these two states will be degenerate,
since $P$ commutes with the Hamiltonian $H$, and
so we have the phenomenon of parity-doubling. Of course, so
far the argument could be equally applied to the
case of $D=3+1$. The crucial question is whether both
combinations are indeed non-null. Now as long
as $j\not=0$ the states  $|j\rangle$ and  $P|j\rangle$ are
orthogonal because they have eigenvalues $\pm j$ respectively 
with respect to $J$. In that case it immediately follows that 
neither of the linear combinations in eqn(\ref{A5}) can be 
null. This argument clearly fails for $j=0$ just as it
fails for any $j$ in 4 dimensions. Thus we conclude that
the $j\not=0$ states come in degenerate pairs of
opposite parity. While for the $j=0$ states there is no
reason to expect parity doubling.
 
The above argument assumes the continuum rotation group.
Our calculations, on the other hand, will be performed upon 
a square spatial lattice whose explicit symmetries are rotations
under $\pi/2$. Does this make any difference? Indeed it
does. States of angular momenta $\pm j$ are distinguished
by the phases $\exp\{\pm ij\theta\}$ that they acquire under
a rotation of $\theta$. We note that for $j=2$ in particular,
these phases are identical if we restrict ourselves to
rotations of $\theta=n\pi/2$. Thus on a square lattice
there is no reason to expect parity-doubling for $j=2$,
any more than for $j=0$. Of course as we reduce the
lattice spacing, $a$, we expect to increasingly recover 
continuum rotational invariance on physical length scales. 
The extent to which we do will be reflected in the
extent to which we recover $j=2$ parity doubling in
our mass spectrum.

The rotational invariance is not only broken by the
square lattice: it is also broken by the fact that
our space-time is a finite hypertorus. If the lattice
is symmetric in the two spatial directions (as it
usually will be), this once again leaves us with
rotations of $\pi/2$. As the volume becomes large
compared to the physical length scale of the theory,
we expect to recover full rotational invariance,
and re-obtain $j=2$ parity-doubling.

We thus expect to find parity doubling for $j\not=0$
states in the $D=2+1$ theory. However some of this 
parity doubling may be lost to the extent that either 
the lattice spacing or the periodic boundary conditions  
affect physical length scales. Thus the restoration
of parity doubling, in particular for $j=2$ states,
will provide us with direct evidence for the 
separation of the physical length scale from both the
ultraviolet and the infrared cut-offs.

\section{Methodology}

We work on a cubic lattice with periodic boundary conditions.
The lattice spacing is labelled $a$ and the length of the 
lattice in the $\mu$-direction is $L_{\mu}$ in lattice units.
The field variables are SU($N_c$) matrices. They reside on the links
of the lattice and are represented by $U_l$ or by $U_{\mu}(n)$,
using an obvious notation. The ordered product of the matrices
around a plaquette of the lattice is represented by $U_p$ or
by $U_{\mu\nu}(n)$. We use the standard plaquette action:
\begin{equation}
S = \beta \sum_{p}\{1-{1\over N_c}ReTr U_p\}
\label{B1}
\end{equation}
and this appears as a weighting factor $e^{-S}$ in the Euclidean
Path Integral. In the continuum limit this becomes the usual
Yang-Mills action with
\begin{equation}
\beta = {{2N_c} \over {ag^2}}.
\label{B2}
\end{equation}
Note the factor of $a$ that is there because $g^2$ has dimensions
of mass; the dimensionless bare coupling, being a coupling on
the scale $a$, is just $ag^2$.

We shall perform a few calculations on lattices with different
spatial and temporal lattice spacings: $a_s$ and $a_t$ respectively. 
In that case we use an action
\begin{equation}
S = \beta_s \sum_{p_s}\{1-{1\over N_c}ReTr U_{p_s}\}
+ \beta_t \sum_{p_t}\{1-{1\over N_c}ReTr U_{p_t}\}
\label{B3}
\end{equation}
where the spatial and temporal plaquette matrices, $U_{p_s}$
and $U_{p_t}$, are multiplied by different
couplings whose values are chosen to reproduce the
desired ratio of lattice spacings, $a_s/a_t$. How this
choice is made is described in detail in Appendix D.

The main technicalites involve the Monte Carlo and the calculation
of masses. We treat these in turn.

\subsection{Monte Carlo}

The Monte Carlo consists of a mixture of heat bath and over-relaxation
sweeps. We discuss these in turn.

\subsubsection{heat bath}

For SU(2) we use the standard Kennedy-Pendleton
\cite{KenPen}
heat bath algorithm. This is extended to higher groups using the 
Cabibbo-Marinari 
\cite{CabMar}
algorithm where effectively one updates some of the SU(2) subgroups 
of the SU(N) matrices. 

An important practical question here is how many of these
subgroups to update. Clearly the more subgroups one updates
the faster we will explore phase space. However one does not
want to carry this past the point of diminishing returns.
To determine an appropriate number of these subgroups
we have chosen a criterion which involves monitoring
how efficiently the action of the SU($N_c$) fields is 
reduced by cooling the fields 
\cite{cool,MTnewton},
when the cooling is applied through 
different numbers of SU(2) subgroups. We recall that to 
cool an SU(2) lattice link we simply replace the matrix 
that is on that link by the matrix which minimises the action. 
This matrix is easy to determine
\cite{cool,MTnewton}.
A link appears in 4 plaquettes and hence its contribution
to the action can be written as
\begin{equation}
\delta S_l = - {\beta \over 2} Tr\{U_l \Sigma\}
\label{B4}
\end{equation}
where the matrix $\Sigma$ is the sum of the `staples'
enclosing the link $l$.
Each staple is an SU(2) matrix and hence $\Sigma$ is
proportional to an SU(2) matrix. Then it is easy to
see that the matrix that minimises $\delta S_l$ is
given by
\begin{equation}
U_l =  {{\Sigma^{\dagger}} \over {|\Sigma|}}
\label{B5}
\end{equation}
We note that this is just the choice of matrix that
the heat bath algorithm makes if we set $\beta=\infty$.
Once we have applied this procedure to every link of the 
lattice we have performed a cooling sweep. And we can
systematically reduce the action by performing a sequence 
of such cooling sweeps. We can extend this to SU($N_c$) fields by 
using the Cabibbo-Marinari algorithm and cooling within the
chosen SU(2) subgroups. In this case the algorithm no longer
exactly minimises the action. Instead the rate at which it reduces
the action is a measure of how rapidly it moves through phase space.
So our procedure is to generate some (plausibly) thermalised SU($N_c$) 
fields, and then to cool these fields using various numbers of
SU(2) subgroups. An example of this, for the case of SU(5),
is shown in Table~\ref{table_cools}. We see that if we use very 
few subgroups the decrease in the action is very slow. 
(Compared, for example, to what happens 
in the SU(2) theory.) As we increase the number of subgroups
the action decreases more rapidly, indicating that the
algorithm explores  phase space more efficiently. If we were
to increase the number further then clearly at some point 
we would find that it led to little further
change in the rate of decrease in the action. At this point we 
would certainly be into diminishing returns. We thus try to choose
the smallest number of subgroups that will reduce the action 
reasonably fast. We then use these same subgroups in the Monte 
Carlo. In practice we have used 3, 4 and 8 subgroups in the case 
of SU(3), SU(4) and SU(5) respectively. (There is obviously some
ambiguity in the precise choice.)

\subsubsection{over-relaxation}

In addition to heat-bath sweeps one can also use over-relaxation
sweeps 
\cite{ORadler,ORcreutz,ORrevs}. 
In SU(2) this corresponds to replacing our old link matrix,
$U_{old}$, by a new link matrix, $U_{new}$, defined by 
\begin{equation}
U_{new} =  {{\Sigma^{\dagger}} \over {|\Sigma|}}
U_{old}^{\dagger} 
{{\Sigma^{\dagger}} \over {|\Sigma|}},
\label{B6}
\end{equation}
where the notation is as in eqn(\ref{B4}). It is easy to see
that this change does not alter the action. Moreover it can
be extended to SU($N_c$) using the Cabibbo-Marinari algorithm.

An over-relaxation step involves a large change in the
link matrix and so it is plausible that it will increase
the rate at which we traverse our phase space
\cite{ORadler,ORcreutz,ORrevs}. 
Indeed, in 4 dimensions, there is evidence that this is so
for large Wilson loops
\cite{ORtests}. 
However what we are interested in is the calculation
of the low-lieing mass spectrum and so what we want to know
is how over-relaxation affects such a calculation. In
Appendix A we present a rather detailed study of this
both in SU(2) and in SU(3). (This is, we believe, the only study 
of this kind for gauge theories in 3 or 4 dimensions.)
We find that for physical quantities, such as masses, 
a suitable mix of over-relaxation and heat bath sweeps
decorrelates field configurations significantly, although
not dramatically, faster than pure heat bath.
There is an additional gain
that arises from the fact that an over-relaxation step is faster
then a heat bath step, which in any case involves the
calculation of all the staples. (This gain is greater
in 3 than in 4 dimensions since there are fewer staples to 
calculate in the former case.) 

Thus in the calculations of this paper we shall typically 
choose to make 4 or 5 over-relaxation sweeps for each 
heat bath sweep.

\subsection{Calculating masses}

Our mass calculations are entirely conventional. 
The starting point is the observation that
\begin{eqnarray}
\langle \phi^{\dagger}(t) \phi(0) \rangle
& = & \sum_n |\langle vac |\phi|n \rangle|^2 \exp\{-E_n t\} 
\nonumber \\
& \stackrel{t\to \infty}{\longrightarrow} & 
|\langle vac |\phi|0 \rangle|^2 \exp\{-E_0 t\}  
\label{B7}
\end{eqnarray}
where $|0 \rangle$ is the lightest state that couples to
the operator $\phi$ and $E_0$ is its energy. (We use
operators that are localised within a single time-slice.)
So if we want the mass of the lightest colour singlet 
state with quantum numbers J,P,C we simply construct a
$\vec{p}=0$ operator with those quantum numbers, calculate
its correlation function and then obtain the mass(=$E_0$) 
using eqn(\ref{B7}). If the quantum numbers are trivial, the
lightest state might be the vacuum, in which case
we use vacuum subtracted operators. Of course it might
be that for some quantum numbers the lightest state
is a multi-glueball state. We shall come back to this
possibility later on, but shall, for
convenience, ignore it for now.

On the lattice $t=na$ and so what we obtain, not surprisingly,
is $aE_n$, the energy in lattice units. Note that if 
we are on a lattice with a finite periodic temporal extent,
then the expression in eqn(\ref{B7}) needs to have
an additional term for the propagation around the `back' of 
the torus. Such a term will always be included in
the numerical calculations of this paper,
although we shall, for simplicity, persist in writing all 
our expressions as though the temporal extent were infinite.
We also note that the temporal extent of our lattice, 
$T\equiv aL_t$, will always be chosen large enough for the 
partition function, $Z$, to be accurately given by its
vacuum contribution: $Z \simeq \exp\{-E_{vac}T\}$.
Thus the energies we calculate will always be with
respect to the energy of the vacuum.

In principle we can obtain from eqn(\ref{B7}) any number of excited
states as well. In practice, however, fitting sums of exponentials
to a function is a badly conditioned problem. So one needs to 
develop a more sophisticated strategy, as described later on
in this section. 

Again in principle, one can use in eqn(\ref{B7}) 
any operator with the desired
quantum numbers. However, a numerical calculation
has finite statistical errors and because the function 
$\langle \phi^{\dagger}(t) \phi(0) \rangle$
is decreasing roughly exponentially in $t$ it will, at 
large enough $t$, disappear into the statistical noise.
Thus in prectice we need to be able to extract $E_0$
from eqn(\ref{B7}) at small values of $t$. This 
requires the coefficient $|\langle vac |\phi|0 \rangle|^2$
to be large. That is to say, we need to use operators
that are close to the wave-functional of the state
in question. 

If we want to use good operators, we obviously need some
simple way to decide which operator is in fact better.
We shall use a variational criterion. However
before coming to that we say something more about
the operators we actually use. This splits naturally
into a discussion of glueball operators and those
from which we extract the string tension.

\subsubsection{operators for glueballs}

We are interested in colour singlet operators because, as we shall
see, our theories are confining. Now, 
the trace of an ordered product of link matrices around any 
closed path of the lattice is a colour singlet. So we 
can build our operators out of such loops.
Moreover, under charge conjugation the trace will go to
its complex conjugate: so the real part is $C=+$ and the
imaginary part is $C=-$. For $N_c \not = 2$ we can
also construct colour singlet operators containing
`baryonic' vertices. We shall not use such operators
in the calculations of this paper, but include a discussion
of their properties in Appendix B.

As a simple example, consider the set of spatial plaquettes
$U_{xy}(\vec{x},t)$ and form the operator
\begin{equation}
\phi(t) = \sum_{\vec{x}} ReTr U_{xy}(\vec{x},t).
\label{B8}
\end{equation}
It is a colour singlet. Moreover it is translation
invariant and so has $\vec{p}=0$. (To obtain a non-zero momentum
we would include a factor of $\exp\{i\vec{p}\vec{x}\}$.)
It is $C=+$ because we take the real part of the trace.
It is obviously invariant under parity and so is $P=+$.
Finally it is obviously invariant under the $n\pi/2$
rotational symmetry of our lattice: so it has $J=0$. This
operator will therefore project onto states that have $J^{PC}=0^{++}$ 
and $\vec{p}=0$. So from its correlation function we can, using 
eqn(\ref{B7}), extract the lightest $0^{++}$ glueball mass.

Suppose we now consider the ordered product of link matrices
around an arbitrary closed curve $C$ that starts and ends
at the point $(\vec{x},t)$. Call it $U_C(\vec{x},t)$. 
Then the linear combination
\begin{equation}
\phi(t) = \sum_{\vec{x}}
\sum_n e^{ij \theta_n} ReTr \{ U_{R(\theta_n)C} 
\pm U^{\dagger}_{PR(\theta_n)C} \}
\label{B9}
\end{equation}
will have $J=j$, $C=+$ and $P=\pm$. Here the angles being
summed over are $\theta_n=n\pi/2$. $R(\theta)$ is a
rotation operator, so that $R(\theta)C$ is the contour 
obtained when we rotate $C$ by an angle $\theta$. 
Similarly $PC$ is the parity transform of $C$. In the
second term $U$ is conjugated because the order around
the curve is reversed under parity. If we replace
$ReTr$ by $ImTr$ we get $C=-$. Symmetries of the
curve $C$ will be reflected in the operator in
eqn(\ref{B9}) being null for some values of $J^{PC}$.
%
%

There is obviously an ambiguity in our assignment of $J$.
We use the continuum notation because we are interested
in the continuum spectrum and we expect that the lattice
will recover continuum rotational invariance as $a\to0$,
at least on physical length scales $\xi/a\to\infty$.
If in constructing our operators we limit ourselves to
rotations of  $n\pi/2$ then any continuum spin $J$ that gives
the same value of  $\exp\{ijn{\pi\over 2}\}$ will couple
to this operator. That is to say, the state that we label
by J=0 actually contains states with J=0,4,8,12,.. and
similarly for the states we label by J=1 and J=2  
(which is all we have with rotations of $\pi/2$).
A similar ambiguity occurs in 4 dimensions. It is 
usually assumed that in this tower of states it is the 
state with the smallest value of $J$ that has the smallest 
mass. Thus, in our case, we shall claim to calculate
the lightest J=0,1 and 2 states. We shall
return to this point later.

It might be useful to indicate the operators $U_C$
that we actually use. Clearly we need only specify the
curves $C$. The first set consists of square and rectangular
curves. In particular, the $1\times 1$, $2\times 2$ 
$3\times 3$ squares and the $1\times 2$, $1\times 3$,
$2\times 3$ rectangles. These curves are obviously symmetric 
under parity reflection. Taking into account that parity also
conjugates the matrix, it is easy to see from eqn(\ref{B9})
that we can only get $J^{++}$ and $J^{--}$ states.
Moreover the square loops can give us only $J=0$
while linear combinations of the rectangular loops can 
give both $J=0$ and $J=2$. However all the loops are symmetric 
under rotations of $\pi$ and so cannot give $J=1$. To obtain
$J=1$ and $P=-C$ states we need other operators; in particular
we need curves that are not symmetric under P. To describe
such curves it is convenient to use an obvious shorthand 
notation in which the plaquette in the $x,y$ plane would
be written as $xyx^{\dagger}y^{\dagger}$. In this notation
the curves we use are a path ordered product of 2 plaquettes 
i.e. $xyx^{\dagger}y^{\dagger}x^{\dagger}y^{\dagger}xy$, and
the `twisted' version of this
$xyx^{\dagger}y^{\dagger}y^{\dagger}x^{\dagger}yx$; the path 
ordered product of the $1\times 2$ loop and a plaquette i.e.
$xyyx^{\dagger}y^{\dagger}y^{\dagger}x^{\dagger}y^{\dagger}xy$,
and the twisted version of this
$xyyx^{\dagger}y^{\dagger}y^{\dagger}y^{\dagger}x^{\dagger}yx$;
and finally  the path ordered product of two $1\times 2$ loops
i.e.$xyyx^{\dagger}y^{\dagger}y^{\dagger}
xxy^{\dagger}x^{\dagger}x^{\dagger}y$
and the twisted version of this,
$xyyx^{\dagger}y^{\dagger}y^{\dagger}
y^{\dagger}xxyx^{\dagger}x^{\dagger}$.
From suitable linear combinations of rotations, parity inversions 
and real or imaginary parts of these loops we can
construct operators with $J=0,1,2$, $P=\pm$ and $C=\pm$.

At this point we have described in some detail the
symmetry properties that the operators need to have.
However all the operators we have described 
so far are ultraviolet: they are based on loops of size 
$O(a)$. Such operators will have an approximately 
equal projection onto all states of the specified 
quantum numbers. The number of excited states increases
rapidly as $a\to 0$. Thus the normalised projection
onto the ground state decreases rapidly. This means that
as $a\to 0$ we have to go to much larger $t$ in eqn(\ref{B7})
to see the ground state dominating the correlation function.
But we cannot do so because of the statistical noise in
our Monte Carlo calculation. This means that we rapidly 
lose the ability to calculate ground states as we approach
the continuum limit.

An efficient remedy for this has been known for a long time
\cite{MTblock,APEsmear,CMT}.
What one needs are operators that extend over physical
length scales and are smooth on such scales. Only
such an operator has a chance of looking like
the ground state wave-functional if, as one expects,
the latter is smooth on physical length scales.
Guided by an intuition developed in the context of
$q\bar q$ wave-functions, one would expect the first 
excited state to have a node. This could be approximated
by a linear combination of large smooth operators.
Higher excited states would be characterised by more
nodes. Hence by more complicated linear combinations.
While this argument is by itself no more than plausible,
it turns out that this strategy works remarkably well.

We use the iterative `blocking' or `fuzzing' algorithm
that has been used extensively in $D=3+1$ spectrum
calculations
\cite{MTblock,CMT}.
We shall not repeat the details here; for a recent 
detailed account (in the context of SU(2) gauge fields
coupled to fundamental scalars in $D=2+1$)
see
\cite{fund}.
Briefly, at the first `blocking' level one has the usual link 
matrices: $U^1_{\mu}(\vec{x},t) \equiv U_{\mu}(\vec{x},t)$. 
At the second level we construct a 
`blocked' link matrix, e.g. $U^2_{x}(\vec{x},t)$,
by summing the paths $xx$, $yxxy^{\dagger}$ and
$y^{\dagger}xxy$ and projecting back to the `nearest'
SU($N_c$) matrix. All the paths start from the point $(\vec{x},t)$
and end at the point $2a$ away in the $x$ direction. But these
blocked links are not just longer; they are fatter (in the
spatial directions) as well. One iterates this procedure:
the blocked link matrices  $U^N_{\mu}$ are formed in exactly
this way from the $U^{N-1}_{\mu}$. (All this for spatial
$\mu$ only.) Thus these operators join sites that are $2^{N-1}a$ 
apart, and are correspondingly fat as well. We can form
path ordered products of these blocked links: for example 
around a super-plaquette, $C \equiv xyx^{\dagger}y^{\dagger}$,  
where now each step is of length $2^{N-1}a$. The trace of
this will be a colour singlet. (After taking expectation values;
there may be small non-gauge invariant pieces that depend on
how we projected from the sum of paths back into SU($N_c$).
See Appendix B for a brief discussion.)
Clearly the blocking algorithm is far from unique. In
Appendix B we compare a particular subset of such algorithms
in order to motivate the particular version we have used.

Thus we can form large smooth operators on any size-scale
we like. When we reduce $a$ by a factor of 2, we need
only iterate the blocking procedure one extra time. 
We form operators, using a sufficient range of blockings
(as determined by preliminary test calculations), on
all the paths described earlier in this subsection.
Thus we often have $O(50)$ different operators for any given
quantum numbers. Of course we do not need to consider
all of these; many are dominated by uninteresting
ultraviolet excitations. How to choose the `best'
is the question we shall return to, after a brief detour
describing the slightly different problem of extracting
the string tension.

\subsubsection{operators for the string tension}

We can calculate the string tension by calculating
the energy of the lightest state composed of a static $q$ 
and $\bar q$ a distance $R$ apart. (Any fundamental charges 
will do; we use $q$ for quarks because they are so familiar.) 
If we have linear confinement then this energy, $E_{min}(R)$,
provides our definition of the string tension, $\sigma$,
as well as providing us with a definition of the
static quark ``potential'', $V_{q\bar{q}}(R)$, 
\begin{equation}
E_{min}(R) \equiv V_{q\bar{q}}(R) 
\stackrel{R\to \infty}{\simeq} \sigma R.
\label{B10}
\end{equation}
For large $R$ one thinks of this state as being composed
of the dressed static quarks with a confining flux tube
of length $\simeq R$ joining them.

We note that the usual potential that enters phenomenological 
discussions of the string tension 
\cite{Perkins}
is essentially based on the Schrodinger equation and 
the relationship with our definition is not a 
simple one; this is apparent if one considers, for example, 
the case of QCD. Vacuum quark fluctuations break the string,
so the potential as defined in eqn(\ref{B10}) will
flatten off for larger $R$. The phenomenological potential, on
the other hand, continues to rise, although it may acquire
a modest imaginary part to incorporate the decay of the
confining flux tube. Effectively it incorporates 
information about the time-scales associated with the
different dynamical processes that contribute. The 
two definitions differ most dramatically in the
large-$N_c$, narrow-width limit of QCD. (Less
so in the pure gauge theories of interest here; unless, for
example, one considers the potential between adjoint sources.)

To project onto this $q\bar q$ state we define the 
gauge-invariant operator
\begin{equation}
\phi(t) = \bar q(0) \prod U_l q(R)  
\label{B11}
\end{equation}
where we can suppose that the quarks are separated along the
$x$-direction and the the product of link matrices is along the
shortest path joining them. The correlation function of this
operator, taken from $t=0$ to $t=T$, will, for large enough
$T$, be $\propto \exp\{-E_{min}(R) T\}$. This correlation
function involves two quark propagators; one from $(x=R,t=0)$
to $(x=R,t=T)$ and the other from $(x=0,t=T)$ to $(x=0,t=0)$.
In the $m_q\to\infty$ limit (which is how one implements
static quarks dynamically) the quark hops along the shortest
available route: that is to say its propagator is equal to
the product of links along the straight line joining its 
end-points. Thus the correlation function is equal (up to
some irrelevant factor) to the expectation value of the 
Wilson loop, $\langle W(R,T)\rangle$. If we have linear 
confinement, as in eqn(\ref{B10}), then $\langle W(R,T) \rangle
\propto  \exp\{-E_{min}(R) T\} \propto \exp\{ \sigma R T\}$, 
the usual confining
area decay of Wilson loops. We can improve this calculation,
just as we have improved the glueball calculation,
using smeared link matrices in eqn(\ref{B11}). The timelike
link matrices will, of course, not be smeared; they arise
from the quark propagator calculation.

We shall use a modified version of the above that employs
Polyakov loops rather than Wilson loops. Construct
a product of link matrices that closes on itself
through a spatial boundary; for example
\begin{equation}
\phi_P(x,t) = Tr \prod^L_{n=1} U_y(x,y+n\hat{y},t)
\label{B12}
\end{equation}
on a $L\times L$ spatial lattice. This non-contractible
loop is what one gets if one stretches our operator in
eqn(\ref{B11}) till the $q$ and $\bar q$ meet and
annihilate. It couples to the corresponding state:
a flux tube of length $L$ that encircles the torus. Such an
operator has zero overlap onto any contractible
loop. One can readily prove this using the symmetry
of the action and measure under the transformation 
$U_y(x,y_0,t) \to z_N U_y(x,y_0,t), \forall t$ where 
$y_0$ is an arbitrarily chosen value of $y$
and $z_N$ is a non-trivial element of the centre.
A contractible loop is obviously invariant under this 
symmetry while the Polyakov loop is not. This argument
breaks down if the symmetry is spontaneously broken;
which occurs, for example, in the high temperature 
deconfining phase.

If we sum over $x$ to make $\phi$ translation invariant 
($\vec{p}=0$) and form the correlation function, we obtain
at large $t$ the mass, $m_P(L)$, of the lightest state containing
a periodic flux loop of length $aL$
\begin{eqnarray}
\langle \phi^{\dagger}_P(t) \phi_P(0) \rangle
& \stackrel{t\to \infty}{\propto} & 
e^{-m_P(L) t}  
\nonumber \\
& = & e^{-\{\sigma aL -{\pi\over{6aL}} + ...\}t}
\label{B13}
\end{eqnarray}
Here we have explicitly included the first correction
term which is the translation to Polyakov loops
\cite{polycor}
of the usual Luscher correction 
\cite{wilcor}
for Wilson loops. This correction is `universal', but obviously
one needs to test whether the physical flux tube does indeed
fall into this particular universality class. 

As we have seen above,
using Wilson loops produces a heavy-quark potential.
This contains a Coulomb term which is long range $\propto g^2\log r$
in $D=2+1$. This term will of course be screened, but having
to disentangle it from  the linear piece, at the intermediate
values of $r$ where the calculations are accurate, can decrease
the accuracy of the estimate of $\sigma$. In $D=3+1$ the Coulomb
term is $\propto 1/r$ and its presence makes it difficult
to identify the $\pi/12r$ universal string correction.
By contrast, in using as we do correlators of
$\vec{p}=0$ sums of $spatial$ 
Polyakov loops, we have completely dispensed
with any charges and have transformed the problem into
a standard mass calculation. Because there are no charges,
there is no longer a Coulomb contribution. This benefit
has of course been achieved at a price:
we no longer have a calculation of the heavy 
quark potential, but only of the string tension.

Just as for glueballs, the simplest operator is too ultraviolet 
to be useful as $a\to 0$. To remedy this we replace the product 
of elementary links in eqn(\ref{B12}) with a product of
blocked link matrices, as defined earlier in this Section.
As we shall see, there is always a blocking level for
which this smeared Polyakov loop is very close to the
wave-functional of the ground state of a flux tube that winds
around the torus.

Two technical asides. When using link matrices at a blocking
level $N_B$, the sites are spaced a distance $2^{N_B-1}a$
apart. A given product of blocked links, that starts at
say $y=1$, is not quite invariant under translations in
the $y$-direction because the blocked links themselves are not 
completely invariant. One can remedy this by summing products 
that start at $y=1,2,..,2^{N_B-1}-1$ respectively; and
this does in fact improve the operator overlap slightly.
A second point is that $L$ need  not be divisible
by the length of the blocked link. In that case we
include links of a lower blocking level, averaged with
staples that include transverse links blocked to the
level of interest. (In practice this extra smearing with
staples is of marginal utility in getting a good overlap.)

\subsubsection{variational criterion and excited states}

Our lattice action possesses the positivity properties  
that allow our lattice correlation functions to be
decomposed as in eqn(\ref{B7}). Let us define an effective
mass by:
\begin{equation}
am_{eff}(t) = -\ln \biggl\{ 
{{\langle \phi^{\dagger}(t) \phi(0) \rangle}
\over
{\langle \phi^{\dagger}(t-a) \phi(0) \rangle}}
\biggr\}
\label{B14}
\end{equation}
Then it is easy to see from the fact that all the 
coefficients in eqn(\ref{B7}) are positive that
\begin{equation}
am_{eff}(t) \geq am_{eff}(t+a)  \ \ \ \ \ \forall t
\label{B15}
\end{equation}
This is a very useful property; it tells us that
$m_{eff}(t)$ provides an upper bound for the mass, $m_G$, 
of the lightest state with the quantum numbers of the
operator $\phi$; whatever the value of $t$ and whatever
the actual operator used. Since the statistical errors
on $am_{eff}(t)$ increase with $t$, we can assume that
any apparent increase of the effective mass with $t$ is 
in fact a statistical fluctuation. 

Now we know from eqn(\ref{B7}) that
\begin{equation}
am_{eff}(t) 
\stackrel{t\to \infty}{\longrightarrow}
am_G
\label{B16}
\end{equation}
When is $t$ large enough for this limit to have been
effectively reached? Since we know that $m_{eff}(t)$
decreases with increasing $t$ then we can estimate
$m_G$ by the value of the effective mass
\begin{equation}
am_G\simeq am_{eff}(t_0) 
\label{B17a}
\end{equation}
where $t_0$ is the lowest value of $t$ for which
\begin{equation}
m_{eff}(t_0) \leq  m_{eff}(t>t_0)
\label{B17}
\end{equation}
within errors. 

Obviously this criterion becomes
convincing only if the errors are small enough
for the relation in eqn(\ref{B17}) to represent a
significant constraint. In practice that will only be
the case if $t_0$ is small, which will only happen if
we have a `good' operator; i.e. one which mainly
projects onto the lightest state. Thus it would
be useful to have a simple practical criterion 
to decide, early on in a calculation, which 
operator is the best. Such a criterion is immediately
suggested by considering the normalised correlation 
function:
\begin{equation}
C(t)
\equiv 
{{\langle \phi^{\dagger}(t) \phi(0) \rangle}
\over
{\langle \phi^{\dagger}(0) \phi(0) \rangle}}
=
{{\langle \phi^{\dagger} e^{-Ht} \phi \rangle}
\over
{\langle \phi^{\dagger} \phi \rangle}}
\label{B18}
\end{equation}
Clearly if we were using a complete basis of operators, 
then the best operator would be the one that maximised 
$C(t)$: it would be
the wave-functional of the lightest state and we would
have $C(t) = \exp\{-m_G t\}$. If the basis is not complete,
this suggests a variational criterion: the `best' 
operator, $\phi$, is the one which maximises $C(t)$.
In practice we shall use $t=a$. The reason is that one
obtains an accurate value of $C(a)$ in even a small 
calculation, and so can determine early on which are the
operators that one needs to calculate with. The value
of $C(a)$ provides us with an estimate of $\exp\{-am_G\}$
and hence $am_G$, which we know to be an upper bound
on the true mass. In practice we improve upon this 
estimate by calculating the correlation function of this
best operator and getting our mass estimate using the
first effective mass that satisfies eqn(\ref{B17}).

Our general strategy for obtaining estimates of the 
ground state and excited state masses is an extension
of the procedure we have just described. We start
with some set of, say, $N$ lattice operators, 
$\phi_i: i=1,..,N$, which we normalise so that 
$\langle {\phi_i}^\dagger \phi_i \rangle = 1$.
(These are chosen from the operators discusses
earlier in this section.) We then find the 
normalised linear combination of the $\phi_i$ that
maximises $C(a) = \langle \phi^\dagger(a) \phi(0) \rangle$.
Call this operator $\Phi_1$. This is our best estimate for
the ground state wave-functional within the space 
$\{ \phi_i \}$; and the associated
value of $C(a)$ provides us with a lower bound estimate for
$\exp\{-am_1\}$ where $m_1$ is the ground state mass.
We can find higher excited states just as simply.
First we construct a basis of operators,
$\phi_i^{\prime}: i=1,..,N-1$,
that spans the $(N-1)$-dimensional subspace of 
the space $\{ \phi_i \}$ which is orthogonal to  $\Phi_1$.
We now find  the linear combination of these $\phi_i^{\prime}$
that maximises 
$C(a) = \langle \phi^{\prime\dagger}(a) \phi^{\prime}(0) \rangle$.
Call this operator $\Phi_2$. This is our best estimate for
the wave-functional of the first excited state.
The associated value of $C(a)$ provides us with an estimate for
$\exp\{-am_2\}$ where $m_2$ is the mass of the excited state.
We can continue this procedure obtaining operators
$\Phi_3, \Phi_4, ...$ from which we can obtain
the energies of higher excited states.

Because our basis is finite the above mass estimates need 
not be very good. To improve upon them we calculate
correlations of our approximate wave-functionals, 
$\langle {\Phi_i}^\dagger(t) \Phi_i(0) \rangle$, and 
from these obtain effective masses for as large a range of
$t$ as our statistical errors (which grow with $t$) will allow.
For each correlation function we look for a 
`plateau' in the effective masses and use the first mass
along that plateau. For the lightest state we are, in
principle, looking
for a plateau that extends to $t=\infty$. For the 
excited states we expect, with our incomplete basis, to
have some admixture of lighter eigenstates, and
so the initial plateau should be finite and will
eventually drop to the masses of the lighter states.
That is to say, for excited states the mass estimate can 
be lower than the mass of the state whose mass is being 
estimated. This undoubtedly means that there is
a larger systematic error on our estimate of the mass of an
excited state than on that of
a ground state. We do not know how to estimate this
error (for either type of state) but the reader should
be aware of its existence.

We have not yet said how we calculate the $\Phi_i$. We use
the following standard procedure
\cite{matrix_corr}.
Define the $N \times N$ correlation matrix $C(t)$ by
\begin{equation}
C_{ij}(t) =  \langle {\phi_i}^\dagger(t) \phi_j(0) \rangle\; .
\label{B19}
\end{equation}
Let the eigenvectors of the matrix $C^{-1}(0) C(a)$ be
$\vec{v}^i ; i=1,\ldots ,N$. Then
\begin{equation}
\Phi_i = c_i \sum_{k=1}^N v_k^i \phi_k 
\equiv \sum_{k=1}^N a_{ik}\phi_k
\label{B20}
\end{equation}
where the constant $c_i$ is chosen so that $\Phi_i$ is
normalised to unity. 
There are of course
many variations possible on the above procedure. 

We return now to the choice of our original basis
of $N$ operators, $\phi_i ; I=1,...,N$. What we do is to 
carry out a short preliminary calculation with typically
5 blocking levels of perhaps 6 to 12 different operators.
We calculate only the diagonal correlation functions.
Comparing the values at $t=a$ we identify the best operator
and a few which are almost as good. We also take a number
which are significantly worse, since, after all, we want 
our basis to contain a reasonable overlap onto some excited 
states. The sort of basis that we were easily able to
accommodate (in terms of memory) had $\sim 15$ operators.
In those cases where we had more we split the basis into
two and worked with both bases separately. Ideally of
course one wants to work with a single basis.
The smallest basis was for the string tension; but here
we were only interested in the ground state because
by using operators that are translation invariant along
the Polyakov loop, we automatically exclude any 
significant overlap onto the interesting string excitations
of the basic flux loop.

\subsubsection{lattice and continuum J}

Suppose we have an operator $\phi$ obtained by multiplying
the (blocked) link matrices around some closed curve $C$. 
The rotation of this curve 
by an angle $\theta$ gives the operator $\phi_{\theta}$.
We can then form an operator of spin J 
\begin{equation}
\phi(J) = \int d\theta e^{iJ\theta} \phi_{\theta}
\label{BC1}
\end{equation}
in the usual way. This assumes we are in the continuum of course.
On our square lattice we only use rotations of $\pi/2$:
\begin{equation}
\phi_L(J) = \sum_n  e^{iJn{\pi\over 2}} \phi_{n{\pi\over 2}}.
\label{BC2}
\end{equation}
As we remarked earlier, $\phi_L(J)$ is not just spin $J$ but
will obviously contain all spins $J\pm4N, \forall N$, since all
these spins provide identical phases at $\theta=n\pi/2$.
It is nonetheless customary to label the lowest energy state 
by the lowest possible spin, in the expectation that higher
spin states will naturally be more massive. This is quite
unsatisfactory: for example it is really not at all obvious 
that a J=3 glueball must be heavier than a J=1 glueball
(these are ambiguous since J=-1 and J=1 are degenerate
parity transforms and 3-(-1)=4). Which one is heavier can
be an important issue in any given dynamical model (as, for
example, in
\cite{rjmt}).

In fact the situation is significantly better than this
\cite{tepmor},
in the case where one uses smeared 
operators with large overlaps onto the ground state.
We shall now show this.

We note that the smeared operators that we construct and,
which we then insert into eqn(\ref{BC2}), spread substantially 
in all spatial directions. We are here only interested in 
the fact that this also involves an angular spread. We might
imagine modelling this qualitative feature using some function like 
$\sim \exp\{- \theta^2/\alpha^2\}$, with the value of $\alpha$ 
determining the angular spread of the operator. This would be
the amplitude to find $|\theta\rangle$ in $\phi|vac\rangle$.
The amplitude would change to 
$\sim \exp\{- (\theta-\theta_0)^2/\alpha^2\}$
if we rotate $\phi$ through an angle $\theta_0$. 
Of course, we cannot
use precisely this form because it does not reflect the
periodic nature of the angular variable. However we can modify it
slightly so that it does,
\begin{equation}
\phi = \sum_n  e^{-{{(\theta-2\pi n)^2}\over{\alpha^2}}},
\label{BC3}
\end{equation}
and in that case it possesses the Fourier expansion 
\begin{equation}
\phi = {1\over{2\alpha\surd\pi}}
\sum_{m=-\infty}^{m=+\infty} e^{-{{\alpha^2 m^2}\over 4}}
e^{im\theta}.
\label{BC3b}
\end{equation}
Suppose we now insert this in eqn(\ref{BC2}) 
with, for example, $J=0$. We obtain
\begin{eqnarray}
\phi_L(J=0) 
& = & {1\over{2\alpha\surd\pi}}
\sum_{m=-\infty}^{m=+\infty} e^{-{{\alpha^2 m^2}\over 4}}
e^{im\theta} \{ 1 + i^m + (-1)^m  + (-i)^m\} \nonumber \\
& = & {2\over{\alpha\surd\pi}}
\sum_{N=-\infty}^{N=+\infty} e^{-4\alpha^2 N^2}
e^{i4N\theta}.
\label{BC4}
\end{eqnarray}
We see, as expected, that we not only have $J=0$ but
that states with $J=4N, \forall N$ also contribute. However
what is interesting is their overlap, which is
$\propto \exp\{-4\alpha^2 N^2\}$. We see from this that
if $\alpha$ is not small, then these higher spin contributions
are severely suppressed. This should be no surprise: in the 
extreme limit where our operator is smeared uniformly over
all angles it is obvious that only $J=0$ can contribute.

We see from this argument that smeared operators will
generically have the largest overlap onto the lowest $|J|$.
The argument relies on the operator being smooth over some
finite angular region. This is true of our elementary smeared
operators, but is not necessarily true of linear combinations
of these. Since our variational calculation produces such
linear combinations, we need to continue the argument a little 
further.

What eqn(\ref{BC4}) tells us is that states of larger than
minimal $J$ will have a suppressed coupling to an
elementary smeared operator. Thus while it is certainly
possible that the lightest state with ``$J=0$'' actually
possesses $J=4$ and that it has a large overlap onto
the variationally selected linear combination of 
elementary smeared operators, its overlap onto any 
individual smeared operator should be visibly suppressed.
In practice we have found this not to be the case in
any of the channels: typically we can find a smeared operator
for which the overlap is $\geq 80\%$. Thus we can confidently
state that the lightest states with $J=0,1,2$ do indeed
have those spins, for all values of $P,C$. We have not
attempted to perform a similar check for the excited states
in these channels.

Clearly one should use the approximate rotational 
invariance on scales $\xi \gg a$ to construct operators
that, to a good approximation, have any value of $J$ that
one desires. Such a calculation is in progress
\cite{rjmtJ}.
\section{Confinement and the string tension}

In the previous section we described how to calculate the
mass, $m_P(L)$, of a flux tube that winds around our $L\times L$ 
spatial torus. Whether such a flux tube actually exists, that is
to say whether we have linear confinement, will be revealed by
how $m_P(L)$ varies with $L$. This is the first question we
address. 

Having shown that we do have linear confinement, we turn to the
problem of extracting continuum values of the string tension in
units of the mass scale provided by $g^2$. This turns out to
be much less ambiguous than the corresponding $D=3+1$ 
calculations where the scale is provided by, say, $\Lambda_{mom}$.
Nonetheless we shall see that using `improved' couplings does
indeed enable us to produce more accurate extrapolations.
An explicit demonstration of the extent of the improvement 
is provided in Appendix C.

Having obtained the continuum values of $\surd\sigma/g^2$
for the SU(2), SU(3), SU(4), SU(5) gauge theories, we
then test certain expectations concerning the large-$N_c$ limit:

{\noindent}$\bullet$ is $SU(\infty)$ confining?

{\noindent}$\bullet$ is the $N_c \to \infty$ limit reached
by varying $g^2 \propto 1/N_c$?
 
{\noindent}$\bullet$ is the leading correction $O(1/N_c^2)$?

{\noindent}We also get our first indication of how small we can
make $N_c$ and still be close to the $N_c=\infty$ limit.

\subsection{Testing for linear confinement}

When we are  using eqn(\ref{B13}) to extract $m_P(L)$, it
is not $t$ that we know but $n_t$ where $t=an_t$:
so what we actually extract is $a m_P(L)$. If we have
linear confinement with a string tension $\sigma$ then 
we should find 
\begin{equation}
a m_P(L) = a^2\sigma L - {\pi\over{6L}} + ...
\label{C1}
\end{equation}
for large enough $L$. Here we have also included the ``universal''
string correction. Its presence is also something we would like 
to test.

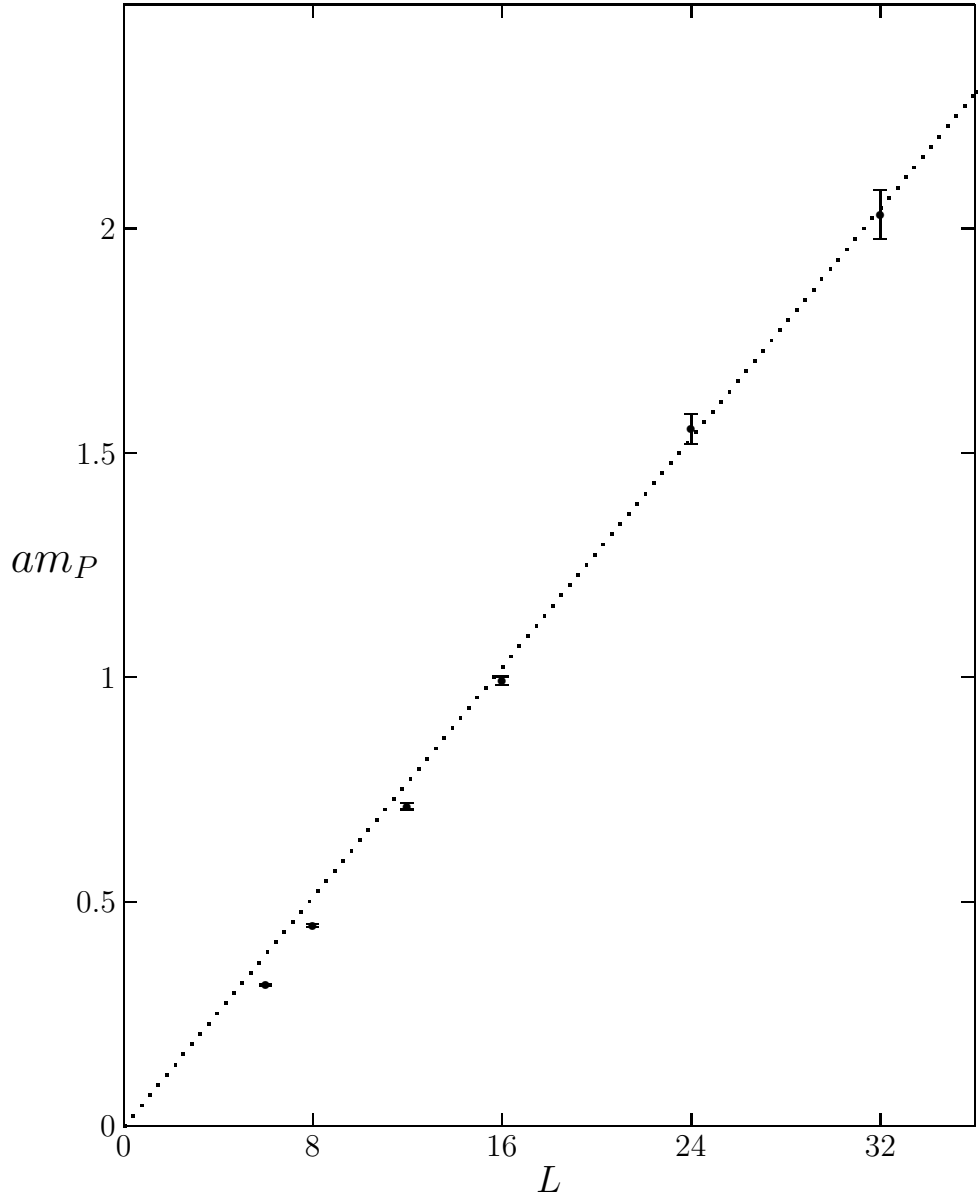
\begin	{figure}[p]
\begin	{center}
\leavevmode
\setlength{\unitlength}{0.240900pt}
\ifx\plotpoint\undefined\newsavebox{\plotpoint}\fi
\sbox{\plotpoint}{\rule[-0.200pt]{0.400pt}{0.400pt}}%
\begin{picture}(1500,1800)(0,0)
\font\gnuplot=cmr10 at 12pt
\gnuplot
\sbox{\plotpoint}{\rule[-0.200pt]{0.400pt}{0.400pt}}%
\put(120.0,31.0){\rule[-0.200pt]{4.818pt}{0.400pt}}
\put(108,31){\makebox(0,0)[r]{{$0$}}}
\put(1436.0,31.0){\rule[-0.200pt]{4.818pt}{0.400pt}}
\put(120.0,383.0){\rule[-0.200pt]{4.818pt}{0.400pt}}
\put(108,383){\makebox(0,0)[r]{{$0.5$}}}
\put(1436.0,383.0){\rule[-0.200pt]{4.818pt}{0.400pt}}
\put(120.0,736.0){\rule[-0.200pt]{4.818pt}{0.400pt}}
\put(108,736){\makebox(0,0)[r]{{$1$}}}
\put(1436.0,736.0){\rule[-0.200pt]{4.818pt}{0.400pt}}
\put(120.0,1088.0){\rule[-0.200pt]{4.818pt}{0.400pt}}
\put(108,1088){\makebox(0,0)[r]{{$1.5$}}}
\put(1436.0,1088.0){\rule[-0.200pt]{4.818pt}{0.400pt}}
\put(120.0,1441.0){\rule[-0.200pt]{4.818pt}{0.400pt}}
\put(108,1441){\makebox(0,0)[r]{{$2$}}}
\put(1436.0,1441.0){\rule[-0.200pt]{4.818pt}{0.400pt}}
\put(120.0,31.0){\rule[-0.200pt]{0.400pt}{4.818pt}}
\put(120,19){\makebox(0,0){\shortstack{\\ \\ \\ {$0$}}}}
\put(120.0,1773.0){\rule[-0.200pt]{0.400pt}{4.818pt}}
\put(417.0,31.0){\rule[-0.200pt]{0.400pt}{4.818pt}}
\put(417,19){\makebox(0,0){\shortstack{\\ \\ \\ {$8$}}}}
\put(417.0,1773.0){\rule[-0.200pt]{0.400pt}{4.818pt}}
\put(714.0,31.0){\rule[-0.200pt]{0.400pt}{4.818pt}}
\put(714,19){\makebox(0,0){\shortstack{\\ \\ \\ {$16$}}}}
\put(714.0,1773.0){\rule[-0.200pt]{0.400pt}{4.818pt}}
\put(1011.0,31.0){\rule[-0.200pt]{0.400pt}{4.818pt}}
\put(1011,19){\makebox(0,0){\shortstack{\\ \\ \\ {$24$}}}}
\put(1011.0,1773.0){\rule[-0.200pt]{0.400pt}{4.818pt}}
\put(1308.0,31.0){\rule[-0.200pt]{0.400pt}{4.818pt}}
\put(1308,19){\makebox(0,0){\shortstack{\\ \\ \\ {$32$}}}}
\put(1308.0,1773.0){\rule[-0.200pt]{0.400pt}{4.818pt}}
\put(120.0,31.0){\rule[-0.200pt]{321.842pt}{0.400pt}}
\put(1456.0,31.0){\rule[-0.200pt]{0.400pt}{424.466pt}}
\put(120.0,1793.0){\rule[-0.200pt]{321.842pt}{0.400pt}}
\put(12,912){\makebox(0,0){{\Large{$am_P$}}}}
\put(788,-53){\makebox(0,0){{\large{$L$}}}}
\put(120.0,31.0){\rule[-0.200pt]{0.400pt}{424.466pt}}
\put(1308,1462){\circle*{12}}
\put(1011,1126){\circle*{12}}
\put(714,730){\circle*{12}}
\put(565,533){\circle*{12}}
\put(417,346){\circle*{12}}
\put(343,253){\circle*{12}}
\put(1308.0,1424.0){\rule[-0.200pt]{0.400pt}{18.549pt}}
\put(1298.0,1424.0){\rule[-0.200pt]{4.818pt}{0.400pt}}
\put(1298.0,1501.0){\rule[-0.200pt]{4.818pt}{0.400pt}}
\put(1011.0,1102.0){\rule[-0.200pt]{0.400pt}{11.322pt}}
\put(1001.0,1102.0){\rule[-0.200pt]{4.818pt}{0.400pt}}
\put(1001.0,1149.0){\rule[-0.200pt]{4.818pt}{0.400pt}}
\put(714.0,723.0){\rule[-0.200pt]{0.400pt}{3.373pt}}
\put(704.0,723.0){\rule[-0.200pt]{4.818pt}{0.400pt}}
\put(704.0,737.0){\rule[-0.200pt]{4.818pt}{0.400pt}}
\put(565.0,528.0){\rule[-0.200pt]{0.400pt}{2.409pt}}
\put(555.0,528.0){\rule[-0.200pt]{4.818pt}{0.400pt}}
\put(555.0,538.0){\rule[-0.200pt]{4.818pt}{0.400pt}}
\put(417.0,343.0){\rule[-0.200pt]{0.400pt}{1.204pt}}
\put(407.0,343.0){\rule[-0.200pt]{4.818pt}{0.400pt}}
\put(407.0,348.0){\rule[-0.200pt]{4.818pt}{0.400pt}}
\put(343.0,251.0){\rule[-0.200pt]{0.400pt}{0.723pt}}
\put(333.0,251.0){\rule[-0.200pt]{4.818pt}{0.400pt}}
\put(333.0,254.0){\rule[-0.200pt]{4.818pt}{0.400pt}}
\sbox{\plotpoint}{\rule[-0.500pt]{1.000pt}{1.000pt}}%
\put(120,31){\usebox{\plotpoint}}
\put(120.00,31.00){\usebox{\plotpoint}}
\multiput(133,47)(13.194,16.022){2}{\usebox{\plotpoint}}
\put(159.38,79.23){\usebox{\plotpoint}}
\put(172.57,95.26){\usebox{\plotpoint}}
\put(185.67,111.36){\usebox{\plotpoint}}
\put(199.28,127.03){\usebox{\plotpoint}}
\put(212.02,143.41){\usebox{\plotpoint}}
\put(225.52,159.16){\usebox{\plotpoint}}
\put(238.32,175.49){\usebox{\plotpoint}}
\put(251.76,191.30){\usebox{\plotpoint}}
\put(264.99,207.29){\usebox{\plotpoint}}
\put(278.16,223.33){\usebox{\plotpoint}}
\put(291.28,239.42){\usebox{\plotpoint}}
\put(304.44,255.46){\usebox{\plotpoint}}
\put(317.56,271.54){\usebox{\plotpoint}}
\put(331.04,287.33){\usebox{\plotpoint}}
\put(344.03,303.50){\usebox{\plotpoint}}
\put(357.28,319.46){\usebox{\plotpoint}}
\put(370.33,335.58){\usebox{\plotpoint}}
\put(383.52,351.60){\usebox{\plotpoint}}
\put(396.88,367.47){\usebox{\plotpoint}}
\put(410.03,383.53){\usebox{\plotpoint}}
\put(423.17,399.60){\usebox{\plotpoint}}
\put(436.31,415.66){\usebox{\plotpoint}}
\put(449.46,431.72){\usebox{\plotpoint}}
\put(462.80,447.62){\usebox{\plotpoint}}
\put(476.04,463.59){\usebox{\plotpoint}}
\put(489.04,479.76){\usebox{\plotpoint}}
\put(502.34,495.68){\usebox{\plotpoint}}
\put(515.28,511.89){\usebox{\plotpoint}}
\put(528.78,527.65){\usebox{\plotpoint}}
\put(541.90,543.74){\usebox{\plotpoint}}
\put(555.07,559.78){\usebox{\plotpoint}}
\put(568.18,575.87){\usebox{\plotpoint}}
\put(581.36,591.90){\usebox{\plotpoint}}
\put(594.56,607.92){\usebox{\plotpoint}}
\put(608.05,623.68){\usebox{\plotpoint}}
\put(620.80,640.06){\usebox{\plotpoint}}
\put(634.35,655.77){\usebox{\plotpoint}}
\put(647.04,672.19){\usebox{\plotpoint}}
\put(660.65,687.86){\usebox{\plotpoint}}
\multiput(673,704)(13.668,15.620){2}{\usebox{\plotpoint}}
\multiput(687,720)(13.088,16.109){0}{\usebox{\plotpoint}}
\multiput(700,736)(13.194,16.022){2}{\usebox{\plotpoint}}
\put(726.33,768.18){\usebox{\plotpoint}}
\put(739.52,784.20){\usebox{\plotpoint}}
\put(752.62,800.30){\usebox{\plotpoint}}
\put(766.23,815.97){\usebox{\plotpoint}}
\put(778.97,832.35){\usebox{\plotpoint}}
\put(792.47,848.11){\usebox{\plotpoint}}
\put(805.27,864.43){\usebox{\plotpoint}}
\put(818.71,880.24){\usebox{\plotpoint}}
\put(831.94,896.23){\usebox{\plotpoint}}
\put(845.11,912.27){\usebox{\plotpoint}}
\put(858.23,928.36){\usebox{\plotpoint}}
\put(871.39,944.40){\usebox{\plotpoint}}
\put(884.52,960.48){\usebox{\plotpoint}}
\put(897.99,976.27){\usebox{\plotpoint}}
\put(910.98,992.44){\usebox{\plotpoint}}
\put(924.23,1008.40){\usebox{\plotpoint}}
\put(937.28,1024.53){\usebox{\plotpoint}}
\put(950.47,1040.54){\usebox{\plotpoint}}
\put(963.84,1056.41){\usebox{\plotpoint}}
\put(976.98,1072.48){\usebox{\plotpoint}}
\put(990.13,1088.54){\usebox{\plotpoint}}
\put(1003.26,1104.61){\usebox{\plotpoint}}
\put(1016.42,1120.66){\usebox{\plotpoint}}
\put(1029.75,1136.57){\usebox{\plotpoint}}
\put(1042.99,1152.53){\usebox{\plotpoint}}
\put(1055.99,1168.70){\usebox{\plotpoint}}
\put(1069.30,1184.62){\usebox{\plotpoint}}
\put(1082.23,1200.84){\usebox{\plotpoint}}
\put(1095.73,1216.60){\usebox{\plotpoint}}
\put(1108.85,1232.68){\usebox{\plotpoint}}
\put(1122.02,1248.72){\usebox{\plotpoint}}
\put(1135.14,1264.81){\usebox{\plotpoint}}
\put(1148.31,1280.85){\usebox{\plotpoint}}
\put(1161.51,1296.87){\usebox{\plotpoint}}
\put(1175.01,1312.62){\usebox{\plotpoint}}
\put(1187.75,1329.00){\usebox{\plotpoint}}
\put(1201.31,1344.71){\usebox{\plotpoint}}
\put(1213.99,1361.13){\usebox{\plotpoint}}
\put(1227.63,1376.78){\usebox{\plotpoint}}
\multiput(1240,1392)(13.194,16.022){2}{\usebox{\plotpoint}}
\multiput(1254,1409)(13.088,16.109){0}{\usebox{\plotpoint}}
\multiput(1267,1425)(13.194,16.022){2}{\usebox{\plotpoint}}
\put(1293.30,1457.14){\usebox{\plotpoint}}
\put(1306.94,1472.78){\usebox{\plotpoint}}
\put(1319.63,1489.20){\usebox{\plotpoint}}
\put(1333.18,1504.92){\usebox{\plotpoint}}
\put(1345.93,1521.29){\usebox{\plotpoint}}
\put(1359.42,1537.05){\usebox{\plotpoint}}
\put(1372.62,1553.07){\usebox{\plotpoint}}
\put(1385.79,1569.11){\usebox{\plotpoint}}
\put(1398.91,1585.19){\usebox{\plotpoint}}
\put(1412.08,1601.24){\usebox{\plotpoint}}
\put(1425.20,1617.32){\usebox{\plotpoint}}
\put(1438.70,1633.08){\usebox{\plotpoint}}
\put(1451.64,1649.30){\usebox{\plotpoint}}
\put(1456,1655){\usebox{\plotpoint}}
\end{picture}
\end	{center}
\vskip 0.15in
\caption{Mass of periodic flux loop, $am_P$, against its length, $L$, 
at $\beta=6$. The straight line is to guide the eye.}
\label{fig_linear6}
\end 	{figure}

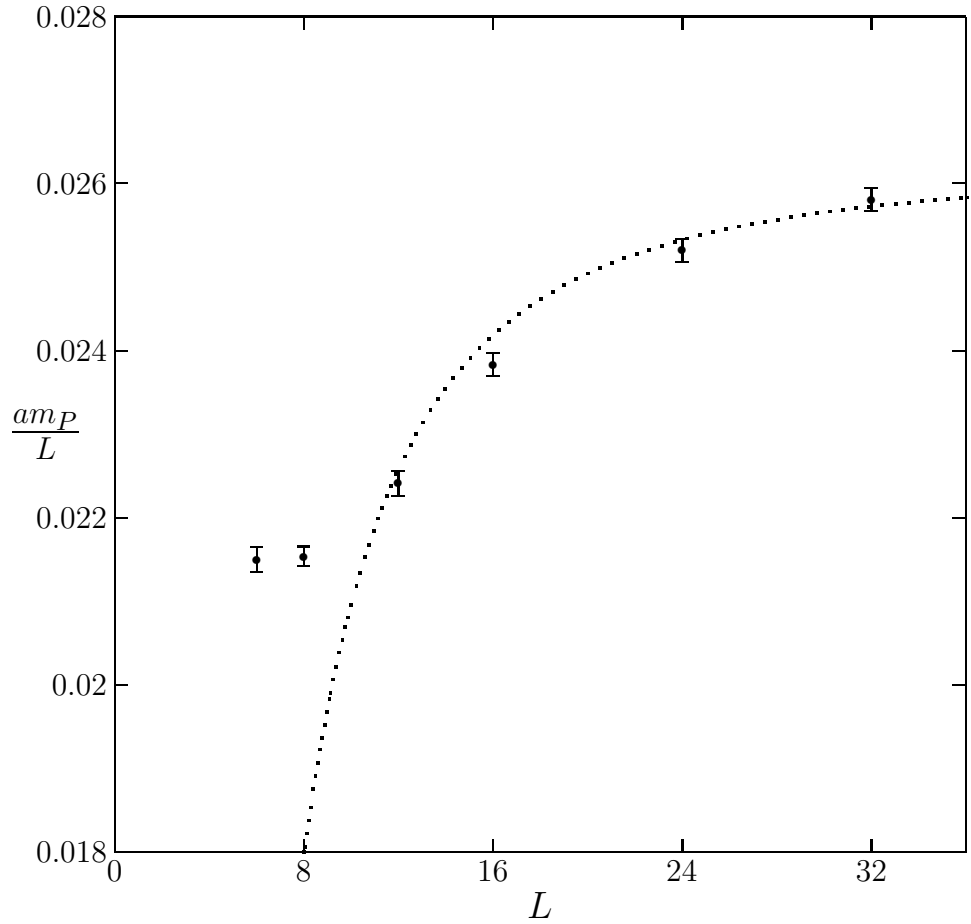
\begin	{figure}[p]
\begin	{center}
\leavevmode
\setlength{\unitlength}{0.240900pt}
\ifx\plotpoint\undefined\newsavebox{\plotpoint}\fi
\sbox{\plotpoint}{\rule[-0.200pt]{0.400pt}{0.400pt}}%
\begin{picture}(1500,1350)(0,0)
\font\gnuplot=cmr10 at 12pt
\gnuplot
\sbox{\plotpoint}{\rule[-0.200pt]{0.400pt}{0.400pt}}%
\put(120.0,31.0){\rule[-0.200pt]{4.818pt}{0.400pt}}
\put(108,31){\makebox(0,0)[r]{{$0.018$}}}
\put(1436.0,31.0){\rule[-0.200pt]{4.818pt}{0.400pt}}
\put(120.0,293.0){\rule[-0.200pt]{4.818pt}{0.400pt}}
\put(108,293){\makebox(0,0)[r]{{$0.02$}}}
\put(1436.0,293.0){\rule[-0.200pt]{4.818pt}{0.400pt}}
\put(120.0,556.0){\rule[-0.200pt]{4.818pt}{0.400pt}}
\put(108,556){\makebox(0,0)[r]{{$0.022$}}}
\put(1436.0,556.0){\rule[-0.200pt]{4.818pt}{0.400pt}}
\put(120.0,818.0){\rule[-0.200pt]{4.818pt}{0.400pt}}
\put(108,818){\makebox(0,0)[r]{{$0.024$}}}
\put(1436.0,818.0){\rule[-0.200pt]{4.818pt}{0.400pt}}
\put(120.0,1081.0){\rule[-0.200pt]{4.818pt}{0.400pt}}
\put(108,1081){\makebox(0,0)[r]{{$0.026$}}}
\put(1436.0,1081.0){\rule[-0.200pt]{4.818pt}{0.400pt}}
\put(120.0,1343.0){\rule[-0.200pt]{4.818pt}{0.400pt}}
\put(108,1343){\makebox(0,0)[r]{{$0.028$}}}
\put(1436.0,1343.0){\rule[-0.200pt]{4.818pt}{0.400pt}}
\put(120.0,31.0){\rule[-0.200pt]{0.400pt}{4.818pt}}
\put(120,19){\makebox(0,0){\shortstack{\\ \\ \\ {$0$}}}}
\put(120.0,1323.0){\rule[-0.200pt]{0.400pt}{4.818pt}}
\put(417.0,31.0){\rule[-0.200pt]{0.400pt}{4.818pt}}
\put(417,19){\makebox(0,0){\shortstack{\\ \\ \\ {$8$}}}}
\put(417.0,1323.0){\rule[-0.200pt]{0.400pt}{4.818pt}}
\put(714.0,31.0){\rule[-0.200pt]{0.400pt}{4.818pt}}
\put(714,19){\makebox(0,0){\shortstack{\\ \\ \\ {$16$}}}}
\put(714.0,1323.0){\rule[-0.200pt]{0.400pt}{4.818pt}}
\put(1011.0,31.0){\rule[-0.200pt]{0.400pt}{4.818pt}}
\put(1011,19){\makebox(0,0){\shortstack{\\ \\ \\ {$24$}}}}
\put(1011.0,1323.0){\rule[-0.200pt]{0.400pt}{4.818pt}}
\put(1308.0,31.0){\rule[-0.200pt]{0.400pt}{4.818pt}}
\put(1308,19){\makebox(0,0){\shortstack{\\ \\ \\ {$32$}}}}
\put(1308.0,1323.0){\rule[-0.200pt]{0.400pt}{4.818pt}}
\put(120.0,31.0){\rule[-0.200pt]{321.842pt}{0.400pt}}
\put(1456.0,31.0){\rule[-0.200pt]{0.400pt}{316.061pt}}
\put(120.0,1343.0){\rule[-0.200pt]{321.842pt}{0.400pt}}
\put(12,687){\makebox(0,0){{\Large{${{am_P}\over{L}}$}}}}
\put(788,-53){\makebox(0,0){{\large{$L$}}}}
\put(120.0,31.0){\rule[-0.200pt]{0.400pt}{316.061pt}}
\put(1308,1056){\circle*{12}}
\put(1011,976){\circle*{12}}
\put(714,796){\circle*{12}}
\put(565,610){\circle*{12}}
\put(417,495){\circle*{12}}
\put(343,490){\circle*{12}}
\put(1308.0,1037.0){\rule[-0.200pt]{0.400pt}{8.913pt}}
\put(1298.0,1037.0){\rule[-0.200pt]{4.818pt}{0.400pt}}
\put(1298.0,1074.0){\rule[-0.200pt]{4.818pt}{0.400pt}}
\put(1011.0,957.0){\rule[-0.200pt]{0.400pt}{8.913pt}}
\put(1001.0,957.0){\rule[-0.200pt]{4.818pt}{0.400pt}}
\put(1001.0,994.0){\rule[-0.200pt]{4.818pt}{0.400pt}}
\put(714.0,778.0){\rule[-0.200pt]{0.400pt}{8.672pt}}
\put(704.0,778.0){\rule[-0.200pt]{4.818pt}{0.400pt}}
\put(704.0,814.0){\rule[-0.200pt]{4.818pt}{0.400pt}}
\put(565.0,590.0){\rule[-0.200pt]{0.400pt}{9.395pt}}
\put(555.0,590.0){\rule[-0.200pt]{4.818pt}{0.400pt}}
\put(555.0,629.0){\rule[-0.200pt]{4.818pt}{0.400pt}}
\put(417.0,480.0){\rule[-0.200pt]{0.400pt}{7.468pt}}
\put(407.0,480.0){\rule[-0.200pt]{4.818pt}{0.400pt}}
\put(407.0,511.0){\rule[-0.200pt]{4.818pt}{0.400pt}}
\put(343.0,471.0){\rule[-0.200pt]{0.400pt}{9.395pt}}
\put(333.0,471.0){\rule[-0.200pt]{4.818pt}{0.400pt}}
\put(333.0,510.0){\rule[-0.200pt]{4.818pt}{0.400pt}}
\sbox{\plotpoint}{\rule[-0.500pt]{1.000pt}{1.000pt}}%
\put(416.00,31.00){\usebox{\plotpoint}}
\multiput(417,39)(2.935,20.547){4}{\usebox{\plotpoint}}
\multiput(430,130)(3.578,20.445){4}{\usebox{\plotpoint}}
\multiput(444,210)(3.738,20.416){4}{\usebox{\plotpoint}}
\multiput(457,281)(4.503,20.261){3}{\usebox{\plotpoint}}
\multiput(471,344)(4.774,20.199){3}{\usebox{\plotpoint}}
\multiput(484,399)(5.596,19.987){2}{\usebox{\plotpoint}}
\multiput(498,449)(5.760,19.940){2}{\usebox{\plotpoint}}
\multiput(511,494)(6.707,19.642){2}{\usebox{\plotpoint}}
\multiput(525,535)(7.049,19.522){2}{\usebox{\plotpoint}}
\multiput(538,571)(7.903,19.192){2}{\usebox{\plotpoint}}
\multiput(552,605)(8.253,19.044){2}{\usebox{\plotpoint}}
\put(573.57,652.14){\usebox{\plotpoint}}
\put(582.97,670.64){\usebox{\plotpoint}}
\multiput(592,688)(10.792,17.729){2}{\usebox{\plotpoint}}
\put(614.02,724.58){\usebox{\plotpoint}}
\put(625.51,741.83){\usebox{\plotpoint}}
\put(637.59,758.71){\usebox{\plotpoint}}
\put(649.86,775.41){\usebox{\plotpoint}}
\put(663.38,791.16){\usebox{\plotpoint}}
\put(676.75,807.02){\usebox{\plotpoint}}
\put(691.05,822.05){\usebox{\plotpoint}}
\put(705.94,836.52){\usebox{\plotpoint}}
\put(721.17,850.62){\usebox{\plotpoint}}
\put(737.08,863.92){\usebox{\plotpoint}}
\put(753.50,876.62){\usebox{\plotpoint}}
\multiput(754,877)(16.889,12.064){0}{\usebox{\plotpoint}}
\put(770.40,888.66){\usebox{\plotpoint}}
\put(787.83,899.90){\usebox{\plotpoint}}
\put(805.64,910.55){\usebox{\plotpoint}}
\multiput(808,912)(18.021,10.298){0}{\usebox{\plotpoint}}
\put(823.64,920.88){\usebox{\plotpoint}}
\put(842.02,930.51){\usebox{\plotpoint}}
\put(860.76,939.43){\usebox{\plotpoint}}
\multiput(862,940)(19.077,8.176){0}{\usebox{\plotpoint}}
\put(879.78,947.74){\usebox{\plotpoint}}
\put(898.74,956.18){\usebox{\plotpoint}}
\multiput(903,958)(19.372,7.451){0}{\usebox{\plotpoint}}
\put(918.07,963.74){\usebox{\plotpoint}}
\put(937.73,970.38){\usebox{\plotpoint}}
\multiput(943,972)(19.546,6.981){0}{\usebox{\plotpoint}}
\put(957.36,977.11){\usebox{\plotpoint}}
\put(977.24,983.07){\usebox{\plotpoint}}
\multiput(984,985)(19.838,6.104){0}{\usebox{\plotpoint}}
\put(997.12,989.03){\usebox{\plotpoint}}
\put(1017.15,994.42){\usebox{\plotpoint}}
\put(1037.20,999.77){\usebox{\plotpoint}}
\multiput(1038,1000)(20.224,4.667){0}{\usebox{\plotpoint}}
\put(1057.44,1004.38){\usebox{\plotpoint}}
\put(1077.69,1008.93){\usebox{\plotpoint}}
\multiput(1078,1009)(20.295,4.349){0}{\usebox{\plotpoint}}
\put(1097.96,1013.38){\usebox{\plotpoint}}
\put(1118.39,1016.91){\usebox{\plotpoint}}
\multiput(1119,1017)(20.224,4.667){0}{\usebox{\plotpoint}}
\put(1138.73,1020.96){\usebox{\plotpoint}}
\multiput(1146,1022)(20.514,3.156){0}{\usebox{\plotpoint}}
\put(1159.26,1024.05){\usebox{\plotpoint}}
\put(1179.62,1028.02){\usebox{\plotpoint}}
\multiput(1186,1029)(20.547,2.935){0}{\usebox{\plotpoint}}
\put(1200.16,1031.02){\usebox{\plotpoint}}
\put(1220.68,1034.10){\usebox{\plotpoint}}
\multiput(1227,1035)(20.514,3.156){0}{\usebox{\plotpoint}}
\put(1241.22,1037.09){\usebox{\plotpoint}}
\put(1261.85,1039.21){\usebox{\plotpoint}}
\multiput(1267,1040)(20.547,2.935){0}{\usebox{\plotpoint}}
\put(1282.40,1042.11){\usebox{\plotpoint}}
\put(1303.03,1044.29){\usebox{\plotpoint}}
\multiput(1308,1045)(20.514,3.156){0}{\usebox{\plotpoint}}
\put(1323.57,1047.18){\usebox{\plotpoint}}
\put(1344.27,1048.71){\usebox{\plotpoint}}
\multiput(1348,1049)(20.547,2.935){0}{\usebox{\plotpoint}}
\put(1364.87,1051.22){\usebox{\plotpoint}}
\put(1385.57,1052.75){\usebox{\plotpoint}}
\multiput(1389,1053)(20.514,3.156){0}{\usebox{\plotpoint}}
\put(1406.15,1055.30){\usebox{\plotpoint}}
\put(1426.85,1056.83){\usebox{\plotpoint}}
\multiput(1429,1057)(20.703,1.479){0}{\usebox{\plotpoint}}
\put(1447.55,1058.35){\usebox{\plotpoint}}
\put(1456,1059){\usebox{\plotpoint}}
\end{picture}
\end	{center}
\vskip 0.15in
\caption{Mass of periodic flux loop of length, $L$, 
at $\beta=9$; divided by $L$ to expose the correction to
the linear rise. Curve is fit using eqn(\ref{C1}).}
\label{fig_linear9}
\end 	{figure}

Since the numerical calculations are fastest in SU(2) that is 
where we have performed our most detailed tests. In 
Fig.\ref{fig_linear6} and Fig.\ref{fig_linear9} we
show how $a m_P(L)$ varies with $L$ for $\beta=6.0$
and $\beta=9.0$ respectively. The first thing we note is that 
there is indeed an approximate linear dependence of
$a m_P(L)$ on $L$, with an apparent trend towards exact
linearity at large $L$. Ideally we would
like to see this rise continue to $L=\infty$. This is not
possible to test in a numerical calculation, but 
what we can ask is whether the linear rise extends to 
physically large values of the string length, $aL$, or not.
Now a convenient  physical length scale is given by 
$\xi_s \equiv 1/\surd\sigma$ where we can get $a^2\sigma$ 
from the asymptotic linear rise. Doing so we find that our 
largest lengths correspond to $aL\sim 8 \xi_s$ and 
$aL \sim 5 \xi_s$ at $\beta=6$ and $\beta=9$ respectively.
These, we claim, are large distances. For example, they
would correspond to $\simeq 4 fm$ and $2.5 fm$ respectively in 
the real world where $\xi_s \equiv 1/\surd\sigma \simeq 0.5 fm$.

We can see from Fig.\ref{fig_linear6} and Fig.\ref{fig_linear9}
that the dependence of the mass on $L$ is not exactly linear; 
indeed in the latter figure we plot the ratio $a m_P(L)/L$ 
precisely in order to expose the deviations from linearity.
We note that the approach is from below, i.e. the leading 
correction must have a negative sign, just as it does in 
eqn(\ref{C1}). We have plotted a fit of this form in
Fig.\ref{fig_linear9} and we see that it appears to be
compatible with the observed variation. Can we test
the correction term in  eqn(\ref{C1}) more precisely?
Suppose the lattice sizes $L_i$ are ordered so that
$L_{i+1} > L_i$. Let us parameterise the corresponding
loop masses by $am_P(L_i) = a^2 \sigma L_i - c_{eff}/L_i$
and the same for $am_P(L_{i+1})$. Then we obtain
\begin{equation}
c_{eff} = 
{ { {{am_P(L_{i+1})}\over{L_{i+1}}} - {{am_P(L_{i})}\over{L_{i}}} }
\over
{ {1\over {L_{i}^2}} - {1\over {L_{i+1}^2} } } }
\label{C2}
\end{equation}
What are we looking for? At small $L$ higher order corrections
in $1/L$ will be important and so $c_{eff}$ will vary as
we increase $L$. If however $c_{eff}\to c$ as $L \to \infty$
then this tells us that the functional form of the leading 
correction is indeed $c/L$. If the value of $c$ is compatible
with $\pi/6 \simeq 0.52$ then we have some evidence that
the correction is of the universal form. How much evidence
depends on the precision of the comparison of course.

In Table~\ref{table_univcorr} we list the values of 
$c_{eff}$ for various ranges of $L_i,L_{i+1}$.
We also show a single value obtained at $\beta=12$.
In comparing the distances at various values of $\beta$,
we can use the fact that $\lim_{a\to 0} \beta = 4/ag^2$
which tells us that, roughly, $a \propto 1/\beta$. So
$L=32-48$ at $\beta=12$ corresponds roughly to
$L=24-36$ at $\beta=9$ and to $L=16-24$ at $\beta=6$.
As expected we see a strong variation of $c_{eff}$ at
the small values of $L$ where our calculations are
most accurate. As the length, $L$, of the flux 
loop increases its mass also increases and so the relative
error on $c_{eff}$ increases quite rapidly. So while there is
good evidence that for larger $L$ $c_{eff}$ grows to be
at least as large as the theoretical value of $\pi/6 \simeq 0.52$,
there is only a little direct evidence, from the $\beta=9$
values, that this is indeed the asymptotic $L \to \infty$ value.
Taken as a whole, we read the results in    
Table~\ref{table_univcorr} as providing significant 
support for the applicability of the Luscher
universal string correction to the confining flux tube.
We remark that in contrast to D=3+1 studies
using Wilson loops, the present analysis has the advantage
of there being no confusion with a Coulomb term
of the same functional form as the Luscher term.

Since this is our first serious mass calculation in this
paper, it might be worth discussing the extraction
of those masses in a little more detail. By way of example
we list in Table~\ref{table_massloop} the effective masses,
as defined in eqn(\ref{B14}), for the $\beta=9, 12$ and $14.5$
calculations. (The last corresponds to our smallest lattice
spacing.) We show not only the masses obtained
using $\vec{p}=0$ operators, but also those obtained
using operators with the lowest non-zero momentum,
$ap=2\pi/L$ on an $L\times L$ spatial lattice. From the latter
we obtain effective energies, $aE(p)$, which we have translated
into effective masses using the continuum dispersion relation 
$m^2 = E^2 - p^2$. As $L$ decreases, $ap=2\pi/L$ becomes larger
and at some point it should become sensitive to the cut-off
at which point this relation will break down. Comparing
the two sets of masses in Table~\ref{table_massloop}, we 
observe that they are compatible, within small errors, thus
demonstrating the restoration of continuum Lorentz invariance.
The dispersion relation does break down on the $L=6$ lattice,
but at this point $p = 2\pi/3a \sim 2/a$ which is certainly an
ultraviolet momentum.

It might seem remarkable how small we can make $L$ while
still retaining all the string-like properties of the
flux tube, which after all will have a width of the 
order of $\sim 1/a\surd\sigma$. In fact, as we have 
argued elsewhere 
\cite{MT2K},
this is not surprising if the fluctuations of the
tube are not too rough, and the transverse volume is
periodic. 

We return to the masses. Our criterion is that we
choose $m(t_0)$ as our mass estimate if, within 
errors, $m(t_0) = m(t) \forall t \geq t_0$. In most
of the cases shown in Table~\ref{table_massloop}
that is straightforward; the choice of $m(t=2a)$ 
seems appropriate. In some cases there is a downward drift
in the value of $m(t)$ at larger $t$. For example
on the $L=48$ lattice at $\beta=12$. In this case
however the mass from the $\vec{p}\not=0$ operator
shows no such effect: indeed it shows a slight upward
drift. This suggests that this drift is a statistical
fluctuation. Indeed the large drop seen in going from 
$m(3a)$ to $m(4a)$ cannot be accommodated in any
realistic decomposition of the correlator 
that respects positivity. In the case of the $L=32$
masses at the same $\beta$ there is a marginal hint
that $m(2a)$ is not asymptotic, but it is difficult
to see why it should not be if $t=2a$ is asymptotic
at $\beta=9$ and at $\beta=14.5$ (as it appears to be).
The fact that correlators can drift
\cite{CMT}
and indeed oscillate
\cite{hadosc}
outside their apparent errors introduces some subjective
bias into our analysis. But, as we have seen, it can often be 
resolved either by performing simultaneous $\vec{p}\not=0$ 
calculations, or by considering other calculations
at nearby values of $\beta$ and $L$. This renders the
problem a minor one in practice, although it may well induce
a systematic bias at the level of the statistical errors.
For this reason we do not take seriously the fact that
the mean values of $c_{eff}$ at larger $L$ tend to be above
$\pi/6$ albeit within statistical errors.

We turn now to the SU(3), SU(4) and SU(5) theories.
In each case we have performed an explicit test of 
linear confinement at one value of $\beta$ at least.
These are listed in Table~\ref{table_linearsun}.
We see that in every case there is an approximately
linear increase of the mass with the length of the
loop. That is to say, we have linear confinement.
We also extract and list the values of $c_{eff}$
as defined in eqn(\ref{C2}). We see that we have some
evidence for the validity of the string correction in 
eqn(\ref{C1}) for all our values of $N_c$.

On the basis of this evidence we shall assume that
we have linear confinement at all other values of $\beta$
so that we need calculate $m_P(L)$ for only one value of
$L$ and can then use eqn(\ref{C1}) to extract $a^2\sigma$.
Of course $aL$ has to be chosen large enough for the
leading correction to be the dominant one. We shall use 
lattices that are about as large as $L=24$ at $\beta=9.0$ 
in SU(2). As we from Table~\ref{table_univcorr},
this should certainly be large enough.

\subsection{Extrapolating to the continuum limit}

In Tables~\ref{table_Ksu2}, ~\ref{table_Ksu3} and 
~\ref{table_Ksu45} we list the values of the string
tension that we shall use. These have been obtained
from the calculated flux loop masses using eqn(\ref{C1}). 
The lengths of these loops are also shown in the tables. 

The flux loop masses have been obtained from the
cross-correlation matrix, as described earlier.
The exceptions are all in SU(2): the $\beta=6.56$
and $L=16,\beta=6.0$ calculations and those at such strong
coupling, $\beta \leq 3.47$, that the eigenvalue 
calculation becomes error-driven and breaks down. In these
cases we applied the simplified variational calculation
where one chooses, from the original basis of operators,
 the single smeared Polyakov loop that
maximises $am_{eff}(t=a)$. By comparing how such a procedure
differs from the full one at neighbouring values of $L$ and
$\beta$ we believe that any bias induced is within the
statistical errors. Once we have chosen the `best'
operator, we extract $\sigma$ from  $am_{eff}(t=2a)$. There
are a few cases where the naive application of our
`effective mass plateau' criterion would lead us to use
$am_{eff}(t=3a)$ (or larger $t$). However these are typically
two standard deviation effects that occur infrequently enough 
that they can be fluctuations. And in practice if we were
to use them it would make no material difference to the
calculations we now describe.

We now wish to use these values to obtain the continuum
string tension. Since the only explicit mass scale is
provided by $g^2$, we expect that $\surd\sigma$ should be
some multiple of it. We can obtain this ratio from our
tabulated values of $a\surd\sigma$:
\begin{equation}
\lim_{\beta\to\infty} \beta a \surd \sigma
= 2 N_c {{\surd\sigma}\over{g^2}}
\label{C3}
\end{equation}
using eqn(\ref{B2}). 

To perform the limit in eqn(\ref{C3}) we can add a
correction term 
\begin{equation}
\beta a \surd \sigma = c_0 + {{c_1}\over\beta}
\label{C4}
\end{equation}
and fit the unknown constants, $ c_0=  2 N_c
{{\surd\sigma}\over{g^2}}$ and $c_1$, to the values of
$\beta a \surd \sigma$ that we obtain from our Tables.
In practice higher order corrections will be important
at small $\beta$ and so we will need to systematically drop 
off the lowest-$\beta$ points until we get a fit with an
acceptable $\chi^2$. Although this is a workable approach,
we recall, from the $D=3+1$ case, that the lattice bare 
coupling provides a poor definition of a running coupling.
The basic problems are similar in 2+1 and 3+1 dimensions
and so we might expect that the higher order corrections
to eqn(\ref{C4}) will be much larger than if we were
to use a physically motivated coupling. A very simple
such coupling 
\cite{MFI}
is the mean field improved coupling
\begin{equation}
\beta_I  =  \beta \times  \langle {1\over{N_c}} Tr U_p \rangle.
\label{C5}
\end{equation}
To define $\beta_I$ we need the values of the average plaquettes,
$\langle {1\over{N_c}} Tr U_p \rangle$.
These are provided in Tables~\ref{table_plaqsu2},
~\ref{table_plaqsu3} and ~\ref{table_plaqsu45}.
In Appendix C we compare extrapolations in $\beta$ and $\beta_I$
in the cases of SU(2) and SU(3), where we have calculations
over a wide range of $\beta$ values. We are able to demonstrate 
that the mean field improved  
coupling does indeed provide a much better expansion
parameter. Thus we shall extrapolate to the continuum limit using
\begin{equation}
\beta_I a \surd \sigma = c_0
+ {{c_1}\over\beta_I}
\label{C6}
\end{equation}
in all cases.

The results of these extrapolations are listed in 
Table~\ref{table_Kcont} together with the confidence 
levels of the fits and the fitted range. Having obtained
the continuum string tensions
\begin{equation}
{ {\surd\sigma} \over {g^2} } =
\left\{ \begin{array}{ll}
0.3353(18) & \ \ \ \mbox{SU(2)} \\
0.5530(20) & \ \ \ \mbox{SU(3)} \\
0.7581(40) & \ \ \ \mbox{SU(4)} \\
0.9657(54) & \ \ \ \mbox{SU(5)}
\end{array}
\right.
\label{C7}
\end{equation}
we turn now to an analysis of their $N_c$ dependence.

\subsection{Confinement at large N}

\begin	{figure}[p]
\begin	{center}
\leavevmode
\setlength{\unitlength}{0.240900pt}
\ifx\plotpoint\undefined\newsavebox{\plotpoint}\fi
\sbox{\plotpoint}{\rule[-0.200pt]{0.400pt}{0.400pt}}%
\begin{picture}(1500,1800)(0,0)
\font\gnuplot=cmr10 at 12pt
\gnuplot
\sbox{\plotpoint}{\rule[-0.200pt]{0.400pt}{0.400pt}}%
\put(120.0,31.0){\rule[-0.200pt]{4.818pt}{0.400pt}}
\put(108,31){\makebox(0,0)[r]{{$0$}}}
\put(1436.0,31.0){\rule[-0.200pt]{4.818pt}{0.400pt}}
\put(120.0,337.0){\rule[-0.200pt]{4.818pt}{0.400pt}}
\put(108,337){\makebox(0,0)[r]{{$0.2$}}}
\put(1436.0,337.0){\rule[-0.200pt]{4.818pt}{0.400pt}}
\put(120.0,644.0){\rule[-0.200pt]{4.818pt}{0.400pt}}
\put(108,644){\makebox(0,0)[r]{{$0.4$}}}
\put(1436.0,644.0){\rule[-0.200pt]{4.818pt}{0.400pt}}
\put(120.0,950.0){\rule[-0.200pt]{4.818pt}{0.400pt}}
\put(108,950){\makebox(0,0)[r]{{$0.6$}}}
\put(1436.0,950.0){\rule[-0.200pt]{4.818pt}{0.400pt}}
\put(120.0,1257.0){\rule[-0.200pt]{4.818pt}{0.400pt}}
\put(108,1257){\makebox(0,0)[r]{{$0.8$}}}
\put(1436.0,1257.0){\rule[-0.200pt]{4.818pt}{0.400pt}}
\put(120.0,1563.0){\rule[-0.200pt]{4.818pt}{0.400pt}}
\put(108,1563){\makebox(0,0)[r]{{$1$}}}
\put(1436.0,1563.0){\rule[-0.200pt]{4.818pt}{0.400pt}}
\put(343.0,31.0){\rule[-0.200pt]{0.400pt}{4.818pt}}
\put(343,19){\makebox(0,0){\shortstack{\\ \\ \\ {$1$}}}}
\put(343.0,1773.0){\rule[-0.200pt]{0.400pt}{4.818pt}}
\put(565.0,31.0){\rule[-0.200pt]{0.400pt}{4.818pt}}
\put(565,19){\makebox(0,0){\shortstack{\\ \\ \\ {$2$}}}}
\put(565.0,1773.0){\rule[-0.200pt]{0.400pt}{4.818pt}}
\put(788.0,31.0){\rule[-0.200pt]{0.400pt}{4.818pt}}
\put(788,19){\makebox(0,0){\shortstack{\\ \\ \\ {$3$}}}}
\put(788.0,1773.0){\rule[-0.200pt]{0.400pt}{4.818pt}}
\put(1011.0,31.0){\rule[-0.200pt]{0.400pt}{4.818pt}}
\put(1011,19){\makebox(0,0){\shortstack{\\ \\ \\ {$4$}}}}
\put(1011.0,1773.0){\rule[-0.200pt]{0.400pt}{4.818pt}}
\put(1233.0,31.0){\rule[-0.200pt]{0.400pt}{4.818pt}}
\put(1233,19){\makebox(0,0){\shortstack{\\ \\ \\ {$5$}}}}
\put(1233.0,1773.0){\rule[-0.200pt]{0.400pt}{4.818pt}}
\put(120.0,31.0){\rule[-0.200pt]{321.842pt}{0.400pt}}
\put(1456.0,31.0){\rule[-0.200pt]{0.400pt}{424.466pt}}
\put(120.0,1793.0){\rule[-0.200pt]{321.842pt}{0.400pt}}
\put(12,1056){\makebox(0,0){{\Large{${{\surd\sigma} \over g^2}$}}}}
\put(788,-53){\makebox(0,0){{\large{$N_c$}}}}
\put(120.0,31.0){\rule[-0.200pt]{0.400pt}{424.466pt}}
\put(565,545){\circle*{12}}
\put(788,878){\circle*{12}}
\put(1011,1193){\circle*{12}}
\put(1233,1511){\circle*{12}}
\put(565.0,542.0){\rule[-0.200pt]{0.400pt}{1.204pt}}
\put(555.0,542.0){\rule[-0.200pt]{4.818pt}{0.400pt}}
\put(555.0,547.0){\rule[-0.200pt]{4.818pt}{0.400pt}}
\put(788.0,875.0){\rule[-0.200pt]{0.400pt}{1.445pt}}
\put(778.0,875.0){\rule[-0.200pt]{4.818pt}{0.400pt}}
\put(778.0,881.0){\rule[-0.200pt]{4.818pt}{0.400pt}}
\put(1011.0,1186.0){\rule[-0.200pt]{0.400pt}{3.132pt}}
\put(1001.0,1186.0){\rule[-0.200pt]{4.818pt}{0.400pt}}
\put(1001.0,1199.0){\rule[-0.200pt]{4.818pt}{0.400pt}}
\put(1233.0,1502.0){\rule[-0.200pt]{0.400pt}{4.095pt}}
\put(1223.0,1502.0){\rule[-0.200pt]{4.818pt}{0.400pt}}
\put(1223.0,1519.0){\rule[-0.200pt]{4.818pt}{0.400pt}}
\sbox{\plotpoint}{\rule[-0.500pt]{1.000pt}{1.000pt}}%
\put(293.00,31.00){\usebox{\plotpoint}}
\multiput(295,38)(7.708,19.271){2}{\usebox{\plotpoint}}
\put(315.34,89.09){\usebox{\plotpoint}}
\multiput(322,106)(8.777,18.808){2}{\usebox{\plotpoint}}
\multiput(336,136)(8.253,19.044){2}{\usebox{\plotpoint}}
\put(357.85,183.71){\usebox{\plotpoint}}
\put(367.01,202.33){\usebox{\plotpoint}}
\multiput(376,221)(9.840,18.275){2}{\usebox{\plotpoint}}
\put(395.37,257.75){\usebox{\plotpoint}}
\multiput(403,273)(10.141,18.109){2}{\usebox{\plotpoint}}
\put(424.89,312.56){\usebox{\plotpoint}}
\put(435.05,330.66){\usebox{\plotpoint}}
\multiput(444,346)(10.213,18.069){2}{\usebox{\plotpoint}}
\put(466.40,384.45){\usebox{\plotpoint}}
\put(476.86,402.37){\usebox{\plotpoint}}
\put(487.36,420.27){\usebox{\plotpoint}}
\multiput(498,437)(10.559,17.869){2}{\usebox{\plotpoint}}
\put(520.07,473.25){\usebox{\plotpoint}}
\put(531.09,490.83){\usebox{\plotpoint}}
\put(542.09,508.43){\usebox{\plotpoint}}
\multiput(552,524)(10.925,17.648){2}{\usebox{\plotpoint}}
\put(575.60,560.90){\usebox{\plotpoint}}
\put(586.70,578.44){\usebox{\plotpoint}}
\put(597.93,595.89){\usebox{\plotpoint}}
\put(609.38,613.20){\usebox{\plotpoint}}
\multiput(619,628)(11.513,17.270){2}{\usebox{\plotpoint}}
\put(643.56,665.25){\usebox{\plotpoint}}
\put(655.34,682.34){\usebox{\plotpoint}}
\put(666.64,699.73){\usebox{\plotpoint}}
\put(677.98,717.11){\usebox{\plotpoint}}
\put(689.74,734.21){\usebox{\plotpoint}}
\multiput(700,750)(11.902,17.004){2}{\usebox{\plotpoint}}
\put(724.74,785.70){\usebox{\plotpoint}}
\put(736.61,802.73){\usebox{\plotpoint}}
\put(748.14,819.98){\usebox{\plotpoint}}
\put(759.74,837.19){\usebox{\plotpoint}}
\put(771.58,854.24){\usebox{\plotpoint}}
\put(783.34,871.34){\usebox{\plotpoint}}
\multiput(795,888)(11.720,17.130){2}{\usebox{\plotpoint}}
\put(818.84,922.49){\usebox{\plotpoint}}
\put(830.61,939.59){\usebox{\plotpoint}}
\put(842.45,956.64){\usebox{\plotpoint}}
\put(854.27,973.70){\usebox{\plotpoint}}
\put(866.19,990.68){\usebox{\plotpoint}}
\put(878.30,1007.53){\usebox{\plotpoint}}
\multiput(889,1024)(12.312,16.709){2}{\usebox{\plotpoint}}
\put(913.74,1058.70){\usebox{\plotpoint}}
\put(925.94,1075.49){\usebox{\plotpoint}}
\put(937.86,1092.48){\usebox{\plotpoint}}
\put(949.68,1109.54){\usebox{\plotpoint}}
\put(961.51,1126.60){\usebox{\plotpoint}}
\put(973.40,1143.61){\usebox{\plotpoint}}
\put(985.63,1160.38){\usebox{\plotpoint}}
\multiput(997,1177)(12.312,16.709){2}{\usebox{\plotpoint}}
\put(1021.46,1211.29){\usebox{\plotpoint}}
\put(1033.64,1228.09){\usebox{\plotpoint}}
\put(1045.57,1245.07){\usebox{\plotpoint}}
\put(1057.61,1261.97){\usebox{\plotpoint}}
\put(1069.69,1278.85){\usebox{\plotpoint}}
\put(1081.58,1295.86){\usebox{\plotpoint}}
\put(1093.80,1312.63){\usebox{\plotpoint}}
\multiput(1105,1329)(12.743,16.383){2}{\usebox{\plotpoint}}
\put(1130.09,1363.20){\usebox{\plotpoint}}
\put(1142.30,1379.98){\usebox{\plotpoint}}
\put(1154.20,1396.98){\usebox{\plotpoint}}
\put(1166.27,1413.86){\usebox{\plotpoint}}
\put(1178.31,1430.76){\usebox{\plotpoint}}
\put(1190.24,1447.75){\usebox{\plotpoint}}
\put(1202.52,1464.48){\usebox{\plotpoint}}
\put(1214.69,1481.29){\usebox{\plotpoint}}
\multiput(1227,1498)(11.720,17.130){2}{\usebox{\plotpoint}}
\put(1250.97,1531.89){\usebox{\plotpoint}}
\put(1263.16,1548.68){\usebox{\plotpoint}}
\put(1275.42,1565.43){\usebox{\plotpoint}}
\put(1287.41,1582.37){\usebox{\plotpoint}}
\put(1299.39,1599.32){\usebox{\plotpoint}}
\put(1311.65,1616.06){\usebox{\plotpoint}}
\put(1323.84,1632.86){\usebox{\plotpoint}}
\multiput(1335,1648)(11.720,17.130){2}{\usebox{\plotpoint}}
\put(1360.12,1683.45){\usebox{\plotpoint}}
\put(1372.30,1700.26){\usebox{\plotpoint}}
\put(1384.58,1717.00){\usebox{\plotpoint}}
\put(1396.51,1733.98){\usebox{\plotpoint}}
\put(1408.77,1750.71){\usebox{\plotpoint}}
\put(1421.07,1767.41){\usebox{\plotpoint}}
\put(1432.90,1784.47){\usebox{\plotpoint}}
\put(1439,1793){\usebox{\plotpoint}}
\end{picture}
\end	{center}
\vskip 0.15in
\caption{The value of $\surd\sigma/g^2$ as a function
of $N_c$. The line shows the fit in eqn(\ref{C12}).}
\label{fig_stringN}
\end 	{figure}
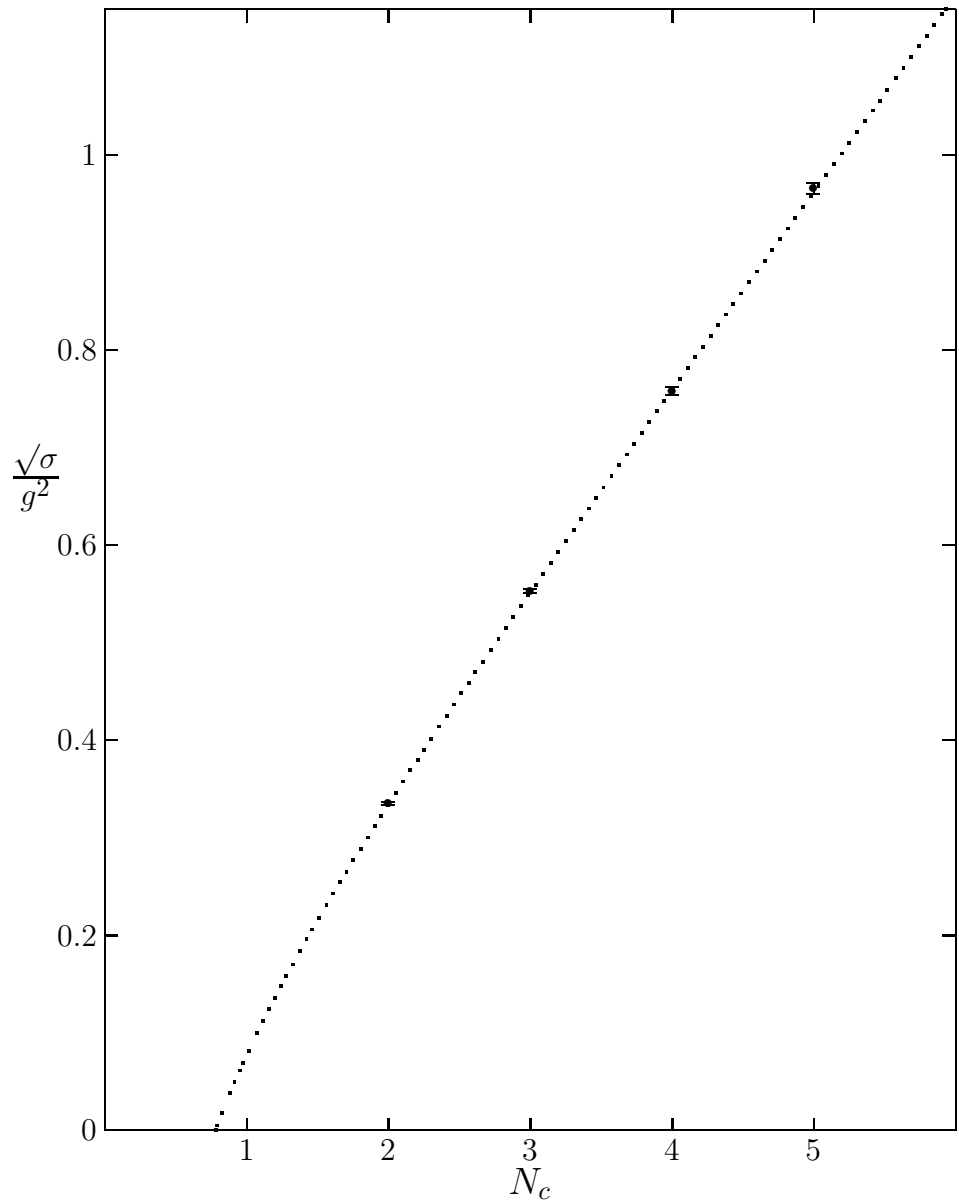

In Fig.\ref{fig_stringN} we plot our calculated values of
$\surd\sigma/g^2$ against $N_c$. We immediately observe that 
the variation approaches a linear form for larger $N_c$
\begin{equation}
{ {\surd\sigma} \over {g^2} } \propto N_c.
\label{C8}
\end{equation}
and indeed is nearly linear even down to $N_c=2$. Now, if our
$SU(N_c)$ gauge theories are to have a smooth $N_c\to\infty$
limit, then in that limit they will have some fixed physical 
mass scale that we shall call $\mu$. If this limit is to be 
confining we must have
\begin{equation}
{ {\surd\sigma} \over {\mu} } \to const \ \ \ \ \ ; \ N_c\to\infty
\label{C9}
\end{equation}
From eqn(\ref{C8}) and eqn(\ref{C9}) we immediately infer that
\begin{equation}
g^2  \propto {{\mu} \over {N_c}}.
\label{C10}
\end{equation}
We recall that the usual all-order diagrammatic analysis demands 
that $g^2\propto 1/N_c$ for a smooth large-$N_c$ limit. 
Eqn(\ref{C10}) embodies precisely that requirement and so provides
a fully non-perturbative confirmation of those arguments.

To complete our demonstration that the theory is 
confining in the $N_c\to\infty$ limit we need to show that
$\lim_{N_c\to\infty} m_G/\surd\sigma$ is finite and non-zero
for the lightest glueball masses. That this is in fact the case
is something that we shall demonstrate in the next Section;
for now we shall assume it to be so. 

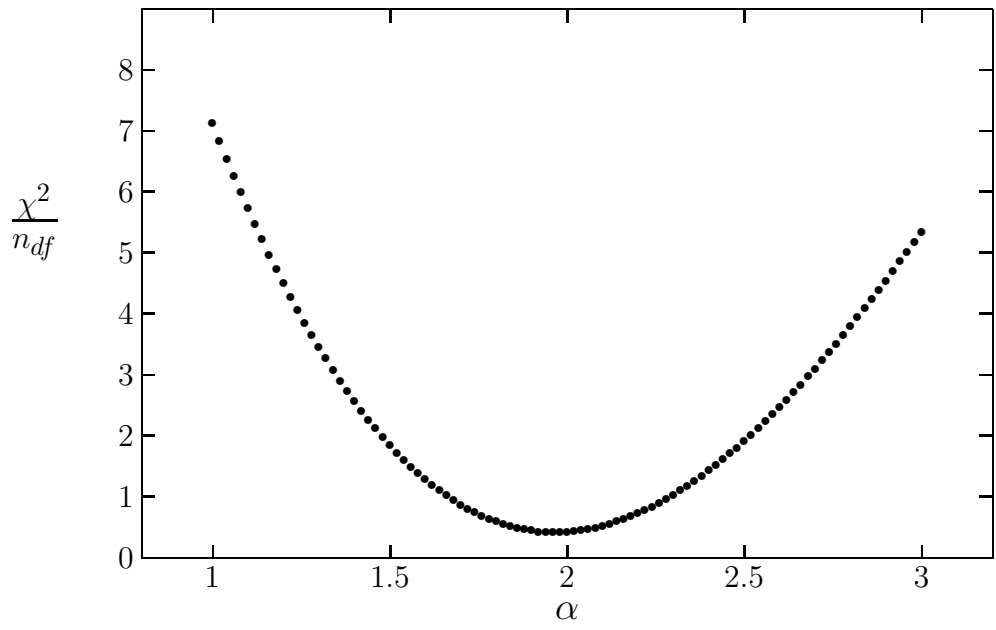
\begin	{figure}[p]
\begin	{center}
\leavevmode
\setlength{\unitlength}{0.240900pt}
\ifx\plotpoint\undefined\newsavebox{\plotpoint}\fi
\sbox{\plotpoint}{\rule[-0.200pt]{0.400pt}{0.400pt}}%
\begin{picture}(1500,900)(0,0)
\font\gnuplot=cmr10 at 12pt
\gnuplot
\sbox{\plotpoint}{\rule[-0.200pt]{0.400pt}{0.400pt}}%
\put(120.0,31.0){\rule[-0.200pt]{4.818pt}{0.400pt}}
\put(108,31){\makebox(0,0)[r]{{$0$}}}
\put(1436.0,31.0){\rule[-0.200pt]{4.818pt}{0.400pt}}
\put(120.0,127.0){\rule[-0.200pt]{4.818pt}{0.400pt}}
\put(108,127){\makebox(0,0)[r]{{$1$}}}
\put(1436.0,127.0){\rule[-0.200pt]{4.818pt}{0.400pt}}
\put(120.0,223.0){\rule[-0.200pt]{4.818pt}{0.400pt}}
\put(108,223){\makebox(0,0)[r]{{$2$}}}
\put(1436.0,223.0){\rule[-0.200pt]{4.818pt}{0.400pt}}
\put(120.0,318.0){\rule[-0.200pt]{4.818pt}{0.400pt}}
\put(108,318){\makebox(0,0)[r]{{$3$}}}
\put(1436.0,318.0){\rule[-0.200pt]{4.818pt}{0.400pt}}
\put(120.0,414.0){\rule[-0.200pt]{4.818pt}{0.400pt}}
\put(108,414){\makebox(0,0)[r]{{$4$}}}
\put(1436.0,414.0){\rule[-0.200pt]{4.818pt}{0.400pt}}
\put(120.0,510.0){\rule[-0.200pt]{4.818pt}{0.400pt}}
\put(108,510){\makebox(0,0)[r]{{$5$}}}
\put(1436.0,510.0){\rule[-0.200pt]{4.818pt}{0.400pt}}
\put(120.0,606.0){\rule[-0.200pt]{4.818pt}{0.400pt}}
\put(108,606){\makebox(0,0)[r]{{$6$}}}
\put(1436.0,606.0){\rule[-0.200pt]{4.818pt}{0.400pt}}
\put(120.0,701.0){\rule[-0.200pt]{4.818pt}{0.400pt}}
\put(108,701){\makebox(0,0)[r]{{$7$}}}
\put(1436.0,701.0){\rule[-0.200pt]{4.818pt}{0.400pt}}
\put(120.0,797.0){\rule[-0.200pt]{4.818pt}{0.400pt}}
\put(108,797){\makebox(0,0)[r]{{$8$}}}
\put(1436.0,797.0){\rule[-0.200pt]{4.818pt}{0.400pt}}
\put(231.0,31.0){\rule[-0.200pt]{0.400pt}{4.818pt}}
\put(231,19){\makebox(0,0){\shortstack{\\ \\ \\ {$1$}}}}
\put(231.0,873.0){\rule[-0.200pt]{0.400pt}{4.818pt}}
\put(510.0,31.0){\rule[-0.200pt]{0.400pt}{4.818pt}}
\put(510,19){\makebox(0,0){\shortstack{\\ \\ \\ {$1.5$}}}}
\put(510.0,873.0){\rule[-0.200pt]{0.400pt}{4.818pt}}
\put(788.0,31.0){\rule[-0.200pt]{0.400pt}{4.818pt}}
\put(788,19){\makebox(0,0){\shortstack{\\ \\ \\ {$2$}}}}
\put(788.0,873.0){\rule[-0.200pt]{0.400pt}{4.818pt}}
\put(1066.0,31.0){\rule[-0.200pt]{0.400pt}{4.818pt}}
\put(1066,19){\makebox(0,0){\shortstack{\\ \\ \\ {$2.5$}}}}
\put(1066.0,873.0){\rule[-0.200pt]{0.400pt}{4.818pt}}
\put(1345.0,31.0){\rule[-0.200pt]{0.400pt}{4.818pt}}
\put(1345,19){\makebox(0,0){\shortstack{\\ \\ \\ {$3$}}}}
\put(1345.0,873.0){\rule[-0.200pt]{0.400pt}{4.818pt}}
\put(120.0,31.0){\rule[-0.200pt]{321.842pt}{0.400pt}}
\put(1456.0,31.0){\rule[-0.200pt]{0.400pt}{207.656pt}}
\put(120.0,893.0){\rule[-0.200pt]{321.842pt}{0.400pt}}
\put(-48,558){\makebox(0,0){{\Large{${{\chi^2}\over{n_{df}}}$}}}}
\put(788,-53){\makebox(0,0){{\large{$\alpha$}}}}
\put(120.0,31.0){\rule[-0.200pt]{0.400pt}{207.656pt}}
\put(231,714){\circle*{12}}
\put(242,686){\circle*{12}}
\put(254,658){\circle*{12}}
\put(265,631){\circle*{12}}
\put(276,605){\circle*{12}}
\put(287,580){\circle*{12}}
\put(298,555){\circle*{12}}
\put(309,531){\circle*{12}}
\put(320,507){\circle*{12}}
\put(332,484){\circle*{12}}
\put(343,462){\circle*{12}}
\put(354,441){\circle*{12}}
\put(365,420){\circle*{12}}
\put(376,400){\circle*{12}}
\put(387,381){\circle*{12}}
\put(398,362){\circle*{12}}
\put(409,344){\circle*{12}}
\put(421,326){\circle*{12}}
\put(432,309){\circle*{12}}
\put(443,293){\circle*{12}}
\put(454,277){\circle*{12}}
\put(465,262){\circle*{12}}
\put(476,248){\circle*{12}}
\put(487,234){\circle*{12}}
\put(499,221){\circle*{12}}
\put(510,208){\circle*{12}}
\put(521,196){\circle*{12}}
\put(532,185){\circle*{12}}
\put(543,174){\circle*{12}}
\put(554,164){\circle*{12}}
\put(565,154){\circle*{12}}
\put(576,145){\circle*{12}}
\put(588,137){\circle*{12}}
\put(599,129){\circle*{12}}
\put(610,121){\circle*{12}}
\put(621,114){\circle*{12}}
\put(632,108){\circle*{12}}
\put(643,102){\circle*{12}}
\put(654,97){\circle*{12}}
\put(666,92){\circle*{12}}
\put(677,88){\circle*{12}}
\put(688,84){\circle*{12}}
\put(699,81){\circle*{12}}
\put(710,78){\circle*{12}}
\put(721,76){\circle*{12}}
\put(732,74){\circle*{12}}
\put(743,72){\circle*{12}}
\put(755,72){\circle*{12}}
\put(766,71){\circle*{12}}
\put(777,71){\circle*{12}}
\put(788,72){\circle*{12}}
\put(799,73){\circle*{12}}
\put(810,74){\circle*{12}}
\put(821,76){\circle*{12}}
\put(833,78){\circle*{12}}
\put(844,81){\circle*{12}}
\put(855,84){\circle*{12}}
\put(866,88){\circle*{12}}
\put(877,92){\circle*{12}}
\put(888,96){\circle*{12}}
\put(899,101){\circle*{12}}
\put(910,106){\circle*{12}}
\put(922,111){\circle*{12}}
\put(933,117){\circle*{12}}
\put(944,123){\circle*{12}}
\put(955,130){\circle*{12}}
\put(966,137){\circle*{12}}
\put(977,144){\circle*{12}}
\put(988,152){\circle*{12}}
\put(1000,160){\circle*{12}}
\put(1011,168){\circle*{12}}
\put(1022,177){\circle*{12}}
\put(1033,186){\circle*{12}}
\put(1044,195){\circle*{12}}
\put(1055,204){\circle*{12}}
\put(1066,214){\circle*{12}}
\put(1077,224){\circle*{12}}
\put(1089,235){\circle*{12}}
\put(1100,246){\circle*{12}}
\put(1111,257){\circle*{12}}
\put(1122,268){\circle*{12}}
\put(1133,279){\circle*{12}}
\put(1144,291){\circle*{12}}
\put(1155,303){\circle*{12}}
\put(1167,316){\circle*{12}}
\put(1178,328){\circle*{12}}
\put(1189,341){\circle*{12}}
\put(1200,354){\circle*{12}}
\put(1211,367){\circle*{12}}
\put(1222,381){\circle*{12}}
\put(1233,395){\circle*{12}}
\put(1244,409){\circle*{12}}
\put(1256,423){\circle*{12}}
\put(1267,437){\circle*{12}}
\put(1278,452){\circle*{12}}
\put(1289,466){\circle*{12}}
\put(1300,481){\circle*{12}}
\put(1311,497){\circle*{12}}
\put(1322,512){\circle*{12}}
\put(1334,527){\circle*{12}}
\put(1345,543){\circle*{12}}
\end{picture}
\end	{center}
\vskip 0.15in
\caption{The $\chi^2$ per degree of freedom against
the power, $\alpha$, of the leading large-$N_c$ correction
when fitting $\surd\sigma/g^2N_c$.}
\label{fig_corrN}
\end 	{figure}

In addition to predicting that $g^2\propto 1/N_c$, the usual
diagrammatic analysis also predicts that the leading correction
should be $O(1/N_c^2)$. To test this we fit our string tensions
with the functional form
\begin{equation}
{ {\surd\sigma} \over {g^2 N_c} } = 
c_0 + {{c_1} \over {N_c^{\alpha}}}
\label{C11}
\end{equation}
In Fig.\ref{fig_corrN} we show how the goodness of fit varies
with the power $\alpha$. From this we can infer that 
\begin{equation}
\alpha = 1.96 \pm 0.45.
\label{C11B}
\end{equation}
If we assume, in addition, that 
the power should be an integer, then only one value is
allowed: $\alpha=2$. Thus we conclude that we have rather 
strong evidence that the leading correction is also in agreement 
with the usual diagrammatic expectations.

Fitting our calculated values, we obtain
\begin{equation}
{ {\surd\sigma} \over {g^2 N_c} } = 
0.1975(10) - {{0.119(8)} \over {N_c^2}}
\label{C12}
\end{equation}
This fit has a good confidence level, $\sim 80 \%$. We note
that this tells us something interesting: we can describe
the physics of $SU(N_c)$ gauge theories, all the way down to
$SU(2)$, by that of the $SU(\infty)$ theory supplemented
by the leading correction with a modest coefficient. Of course,
so far we have only shown this for the string tension: in the
next Section we shall see that this is also the case for the 
mass spectrum.

Before moving on from our result for the string tension, it is
interesting to ask whether it is possible to quantify the 
potential error associated with keeping only the leading
correction in eqn(\ref{C12}). There is no unique way to do 
this, of course, but a first step would be to include a higher
order correction and see what difference it makes. When we
do so we obtain the following range of fits:
\begin{equation}
{ {\surd\sigma} \over {g^2 N_c} } = 
0.1976(22) - {{0.121(43)} \over {N_c^2}} 
- {{0.01(14)} \over {N_c^4}}.
\label{C12B}
\end{equation}
We observe that our result for $\lim_{N_c\to\infty} 
{\surd\sigma}/{g^2 N_c}$ is robust under the inclusion of
the higher order correction. Our calculations
constrain the coefficient of this 
higher-order correction to be small,
and the only significant effect from including it 
is to double the error on the extrapolated value of the
string tension.

The discussion so far has concerned the continuum limit, which
is of course what we are mainly interested in. However the
large-$N_c$ expectations will also apply to lattice corrections,
and we can ask if they are fulfilled. What we would expect is
that the $O(ag^2)$ correction should also be a function of 
$g^2N_c$, i.e. 
\begin{equation}
{ {a\surd\sigma} \over {ag^2 N_c} } = 
b_0 + b_1 a g^2N_c  \ \ \ \ \ \ : \ N_c\to\infty.
\label{C13}
\end{equation}
In terms of our fit in eqn(\ref{C6}) this implies that the 
lattice correction, $c_1$, should be given by
\begin{equation}
c_1 = 4N_c^4 b_1  \ \ \ \ \ \ : \ N_c\to\infty.
\label{C14}
\end{equation}
We note that our calculated values of $c_1$, as listed in 
Table~\ref{table_Kcont}, are entirely consistent with
this being the leading large-$N_c$ behaviour. Indeed, if we
fit these values with a functional form $c_1=cN^{\alpha}$
we find a good fit with $\alpha = 4.2 \pm 0.6$.

\section{The Mass Spectrum}

Having seen that all our gauge theories are linearly
confining, we infer that the asymptotic states are 
colour singlet and so we can calculate the mass spectrum
using the operators described earlier on in this paper.

We shall first indicate the quality of the lattice mass 
calculations. We then investigate the finite volume
dependence of these masses so as to establish control
over this potential source of systematic error. 
We shall then carry out the extrapolation to the
continuum limit. Finally we turn to a study of the
dependence of the mass spectrum on $N_c$.
We finish with a discussion of some features of the
calculated mass spectrum.

\subsection{Calculating the masses}

We shall focus on the lightest states because the
correlations mediated by heavier states decrease so
rapidly with $t$ that it becomes hard to know whether
we have indeed isolated the asymptotic exponential
decay. Moreover, glueballs that are heavy enough will 
decay into lighter glueballs and this may require
more careful analysis.

Clearly we want to obtain the ground state in each
$J^{PC}$ channel, and in those cases where the
ground state is light enough we can estimate one
or two excited masses as well. So the states whose
masses we shall calculate are those of the 
$0^{++}$, $0^{++\ast}$, $0^{++\ast\ast}$, 
$0^{--}$, $0^{--\ast}$, $0^{--\ast\ast}$,
$0^{-+}$, $0^{+-}$, 
$2^{++}$, $2^{++\ast}$, $2^{-+}$, $2^{-+\ast}$,
$2^{--}$, $2^{--\ast}$, $2^{+-}$, $2^{+-\ast}$,
$1^{++}$, $1^{-+}$, $1^{--}$ and $1^{+-}$ glueballs.

We shall calculate the masses, as described earlier,
for SU(2), SU(3), SU(4) and SU(5) gauge groups and, in
each case, for a range of $\beta$ values  sufficient
to allow a continuum extrapolation. In practice
this means for most, but not all, of the values of $\beta$
at which we calculated the string tension.

For any state, the first question must be: how confident are 
we that we have indeed calculated the mass? That is to say,
do we have evidence for an effective mass plateau? 

To address this question we analyse, by way of an example, 
our SU(5) calculation at the highest value of $\beta$. We 
show in Table~\ref{table_meff5} the effective masses we obtain 
there. Since the highest $\beta$ corresponds to the smallest
$a$, these calculations are the closest to the continuum limit 
(and the closest to $N_c=\infty$) and so are the ones which are the
most interesting. We note that it is when the value of $a$ is 
smallest that the correlation functions drop most slowly and we
can extract effective masses to larger $t$. At smaller $\beta$,
further away from the continuum limit, it will be harder to confirm 
that we are seeing mass plateaux.

From this Table we infer that a good estimate of the 
mass is provided by $m_{eff}(2a)$ in each case. That is to say,
within errors the effective mass is on a plateau for $t\geq 2a$.
This is self-evident in most cases. In some cases, e.g.
for the $0^{--\ast}$, one sees a drop in $m_{eff}$
of over one standard deviation when going from $t=2a$ to $t=3a$. 
However that is to be expected, just statistically, given the 
large number of correlation functions that we consider. 
Positivity can be useful in such cases. If $m_{eff}(a)$
and  $m_{eff}(2a)$ are sufficiently close, then one can argue
that it is not possible for $m_{eff}(3a)$ to be very much lower.
At the margins, this allows us to make choices about
what is, or is not, likely to be a statistical fluctuation.
At this level there is some subjective element in the
analysis, although it should be evident from 
Table~\ref{table_meff5} that this will not be an
important problem in our calculation. To test this we
have performed continuum extrapolations using $m_{eff}(3a)$
whenever a blind application of our criterion for
identifying mass plateaux so dictated. We found it makes
no significant difference although the fits are often 
worse. The reason for the latter fact
is that we discount any $rise$ in $m_{eff}(t)$ with $t$
simply because we know from positivity that the
effective masses must decrease monotonically with $t$.
From the statistical point of view, this is a bias in the 
procedure which undermines the statistical analysis.
 
It is apparent from Table~\ref{table_meff5} that as we
go to heavier states, the evidence for effective mass 
plateaux becomes less significant simply because
the statistical errors will overwhelm the signal at smaller
values of $t$. As we go to smaller $\beta$, and so larger
$a$, this becomes very much worse and we will often
not have a useful effective mass beyond $t=2a$. In these
cases we simply assume that $m_{eff}(2a)$ provides a good 
mass estimate. This is reasonable. If at a high value of
$\beta$ a particular operator gives us a mass plateau
from $t=2a$ then at a larger lattice spacing,
e.g. $a^{\prime} = 2a$, an operator that is one
blocking level down, and hence half the size,
should surely give us a mass plateau from
$t=2a^{\prime}=4a$. 

While the above argument is plausible, it cannot replace
a direct demonstration. This can be provided by allowing 
the spacelike, $a_s$, and timelike, $a_t$, lattice
spacings to differ. We then choose $a_t$ small enough
that the correlation functions fall slowly enough over
several (temporal) lattice spacings for us to obtain
several accurate effective masses. This is an old
idea that was used precisely for this purpose
\cite{asymold}
in the early days of glueball calculations. More
recently it has been used very successfully
\cite{asymnew}
as part of the action improvement program. Since this
is a somewhat different type of calculation to the one
in this paper, we leave its discussion to Appendix D.
The reader will find there an explicit demonstration
that even for a coarse spatial discretisation, 
using effective masses at distances between $a_s$
and $2a_s$ is an accurate way to estimate the masses.

In summary, we have taken all the masses that we use in the
spectrum calculations of this paper from $m_{eff}(t=2a)$.
We have checked that using $m_{eff}(t=3a)$ in the few
cases that are ambiguous makes no significant difference.

\subsection{Finite volume effects}

In a theory with a mass gap, $m$, and on a periodic spatial 
volume that is $L$ lattice units across, the leading finite 
size corrections to masses are typically $O(e^{-camL})$ where 
the constant $c=O(1)$ will depend on the details of the 
theory being considered
\cite{LuscherV}.
Of course this correction will only be relevant 
once $aL$ is significantly larger than the typical hadronic
length scale, $\xi$. In that case $amL$ will be large, since 
hadron Compton wavelengths are usually $\ll \xi$,
and so the correction will be small:
usually too small to be observed with the kind of
accuracy we possess. This means that we cannot
expect to obtain a reliable estimate of the coefficient
of this correction term. This correction is interesting because
it is proportional to a triple-glueball effective 
coupling. However, if what we are interested in is
controlling finite-volume corrections, then the
known functional form of this correction has a very
useful consequence. Essentially it tells us that
if we calculate a mass on volumes $aL_1$ and $aL_2$
which are both significantly larger than $\xi$,
and if we find that the change in mass is small
when we go from $L_1$ to $L_2$, then we can be
confident that any mass shift in going from
$L=L_2$ to $L=\infty$ will be small compared to the 
observed change in the mass (as long as  $a(L_2-L_1) \sim \xi$).
This is important: if the leading correction were
power-like rather than exponential then this would
not be true and controlling finite-volume corrections
would be appreciably more difficult.

So our strategy to control finite volume corrections is as 
follows. We calculate masses on a range of lattice volumes.
We include volumes that satisfy the conditions of the
previous paragraph. And once we observe very small 
changes on our larger volumes, we can be confident that
the mass calculated on the very largest volume
is identical, within errors, to the $L=\infty$ mass.

In practice it would be wasteful to perform such an
analysis at each value of $\beta$. Instead we choose
a couple of values of $\beta$ where we perform an
extensive analysis, including very large volumes in
order to make sure there are no unpleasant surprises.
This allows us to establish what volumes are large
enough that any change in mass becomes invisible
within the typical statistical errors of our calculations.
We then use scaling to infer how this translates
to other values of $\beta$. 

We perform these calculations in SU(2), simply because
that consumes much less computer time. Having
determined how large the volume has to be in, say,
units of $1/a\surd\sigma$ we can take this criterion
over to SU(3) etc. Of course there is some danger in doing
this and so we perform at least a modest finite-volume
check for each of our non-Abelian groups.

\begin	{figure}[p]
\begin	{center}
\leavevmode
\input	{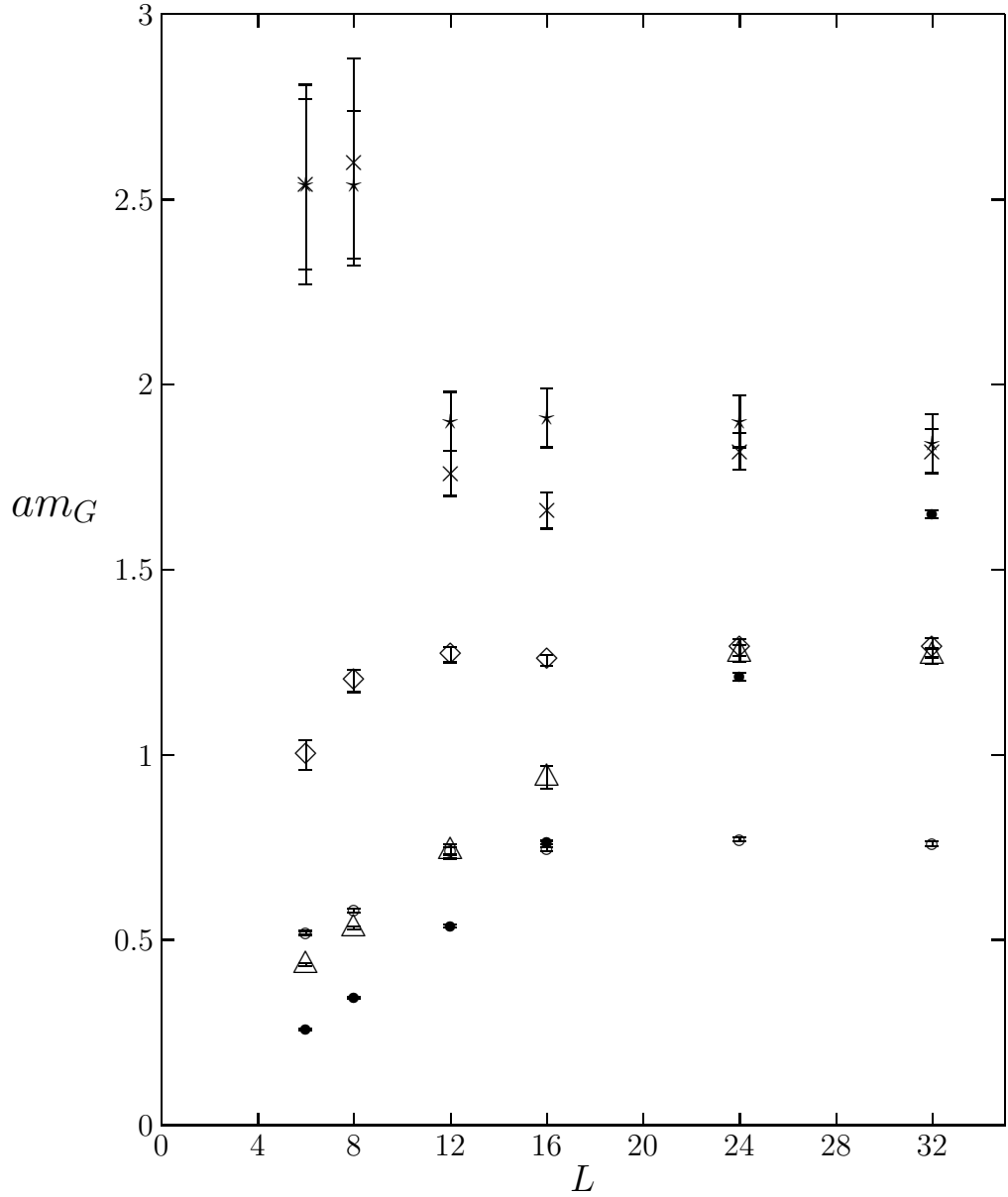}
\end	{center}
\vskip 0.15in
\caption{How the lightest SU(2) masses vary with the spatial
volume, $L^2$, at $\beta=9$. 
States are  the $0^{++}$($\circ$), the $2^{++}$($\bigtriangleup$), 
the $2^{-+}$($\diamond$), the $1^{++}$($\times$) and the 
$1^{-+}$($\star$). Also shown is twice the mass of the periodic
flux loop ($\bullet$).}
\label{fig_Vsu2}
\end 	{figure}

As a first step, we show in Fig.\ref{fig_Vsu2}
how some of the lightest masses vary with the size, $L$,
of the spatial volume, in the case of SU(2) and at
a coupling of $\beta=9.0$. This, as we see from 
Table~\ref{table_Ksu2}, corresponds to quite a small lattice
spacing. Our spatial length varies from $L=6$ to
$L=32$ which corresponds to a variation of $\sim 1$ to
$\sim 5$ in units of $1/a\surd\sigma$ (about 0.5 to 2.5
fermi if we were in QCD); a range of sizes that satisfies the
conditions laid out above. Note that since this is SU(2),
we have no $C=-$ states; these shall appear in our
(less extensive) SU(3) study. So what we show in 
Fig.\ref{fig_Vsu2} are the masses of the lightest $0^{++}$, 
$0^{-+}$, $2^{++}$, $2^{-+}$, $1^{++}$ and $1^{-+}$ glueballs.
We also show twice the mass of the periodic flux loop,
$2am_P$, for reasons that shall soon be apparent.

There are several observations we can make from
Fig.\ref{fig_Vsu2} and the calculations on which it is based.

\vskip 0.1in

{\noindent}$\bullet$ 
As we decrease $L$ we do indeed observe the onset of
substantial finite size effects.

\vskip 0.1in

{\noindent}$\bullet$ 
The parity doubling that we see at large volumes
is badly broken by these finite-volume corrections. This
is not unexpected: the toroidal boundary conditions break the
effective rotational symmetry from the 
(dynamically restored) continuous one 
down to rotations of $\pi/2$. As discussed earlier,
this undermines the argument for $2^{\pm}$ degeneracy
although not $1^{\pm}$ degeneracy. This is precisely
what we observe in  Fig.\ref{fig_Vsu2}. 
Thus the observed degeneracy of the $2^{\pm}$ states 
can serve as a criterion for the lack of finite volume
effects.

\vskip0.1in

{\noindent}$\bullet$ 
We observe that the value of $L$ at which
the $2^{+}$ begins to show finite-volume corrections
is roughly where the asymptotic glueball mass equals
twice the flux loop mass:
\begin{equation}
 2am_P(L) \simeq am_{2^+}{\big /}_{L=\infty}.
\label{D1}
\end{equation}
The same is true for the $0^+$. Since the latter is lighter,
and since $m_P \uparrow$ as $L\uparrow$, the scalar becomes
volume-independent at smaller volumes than the
tensor. 

This correspondence with $2m_P$ is easy to understand. 
Suppose we denote  by $l_x$ the $\vec{p} = 0$ smeared
Polyakov loop in the x-direction which has the best overlap
onto the $x$-periodic flux tube. Typically this overlap
will be $\sim 90-100\%$. Consider now the operator 
$l_x l_x^{\dagger}$. This 
will also be $\vec{p} = 0$ and colour singlet, but it
falls into the sector of contractible loops and so can
couple to glueball states. If the transverse
spatial size, $L_y$, were very large, then this operator would
mainly couple to a state that consists of two periodic flux 
loops whose energy would be $2m_P(L)$. We shall refer to
such states as `torelons'. On our lattices
$L_x = L_y = L$ and as $L$ becomes small these flux loops
will necessarily interact; thus the lowest energy will
deviate somewhat from $2m_P$. We can form $0^{++}$ and $2^{++}$
combinations, $l_x l_x^{\dagger}+l_y l_y^{\dagger}$
and  $l_x l_x^{\dagger}-l_y l_y^{\dagger}$ repectively.
Again, on large volumes these will mainly couple to
states with two flux tubes and mass $2m_P(L)$. On
smaller volumes, $aL \leq \xi$, the interaction between the 
flux tubes will split these states away from each other and
from this mass. While we cannot predict the precise 
variation of these masses with $L$, one would expect them to
decrease, at least until $aL \ll \xi$. Thus the fact that
the $2^{++}$ mass begins to decrease with decreasing $L$ just
when eqn(\ref{D1}) is satisfied, would seem to simply reflect
the fact that for smaller $L$ than this the $2^{++}$ state
which is composed of a pair of flux loops becomes the lightest state
in that sector. And the same for the $0^{++}$. 

We have explicitly confirmed this scenario. For large $L$
at least one of our usual glueball operators has a large,
$\sim 90-100\%$, overlap onto the lightest $0^{++}$ or 
$2^{++}$ state. By contrast the double flux loop operators 
have poor overlaps. The value of $L$ at which the lightest
mass begins to decrease as $L$ decreases, marks the point
at which things reverse. For smaller $L$ it is one of
the double flux loop operators that has a very high
overlap onto the lightest state and the usual operators
all become poor. 

We remark that similar finite volume effects are observed in 
D=3+1 gauge theories
\cite{torelon};
but because $m_G/\surd\sigma$ is slightly
higher in D=2+1, the effects occur on somewhat larger
volumes, and so their interpretation is that much less
ambiguous.   

\vskip 0.1in

{\noindent}$\bullet$ 
Naively we would expect the spatial size at which we begin to
encounter large finite size
effects to be related to the size of the glueball.
However in the case of the $0^{++}$ and $2^{++}$ glueballs,
we have seen that the onset of finite size effects is simply
determined by the (infinite volume) mass and the string tension.
Thus the fact that we can go to smaller spatial volumes for
the scalar than for the tensor, before encountering large finite 
size effects, is not telling us that the size of the scalar is
less than that of the tensor. The same holds true in the
case of four dimensions.

\vskip 0.1in

{\noindent}$\bullet$ In addition to the above, there are finite 
size effects, visible in the $0^-,2^-,1^{\pm}$ states,
whose onset appears at much smaller
values of $L$, and which does not appear to be linked
to mixing with torelon states. (There are no simple
torelon states with these quantum numbers.) 

\vskip 0.1in

{\noindent}$\bullet$ We infer from Fig.\ref{fig_Vsu2} that at 
$\beta=9$ a spatial size of $L=24$ is large enough 
for the lightest glueballs to be free
of finite size effects within
our statistical errors. Assuming scaling and
$\beta = 4/ag^2$ this implies that at a general value
of $\beta$ a safe size is $L \geq 24\beta/9$. One 
can do better by using the calculated value of
$a\surd\sigma$ to set the scale. Doing so, one can then 
extend the criterion to SU(3) etc.

\vskip 0.1in

In practice we are more cautious than this and have performed 
an extensive finite volume analysis at $\beta=6$ as well.
This is for a larger value of $a$, and we include larger
lattices ; up to more than 8 in units of $1/a\surd\sigma$.
This masses are  shown in Table~\ref{table_Vsu2}. We also 
have a more limited study, on $L=32$ and $L=48$ lattices, 
at $\beta=12$ where $a$ is smaller. These masses will
be displayed later, in Table~\ref{table_msu2b}, where
we display our ``$V=\infty$'' SU(2) mass values. All these
results confirm our criterion for what constitutes a
safe volume. 

In Table~\ref{table_Vsu3} we show our SU(3) study. 
The presentation here is slightly different to that
in Table~\ref{table_Vsu2} in that we show estimates
of the $0^{++}$ and $2^{++}$ torelon masses using
the operators described earlier on in this Section.
The glueball masses have been obtained from the
usual glueball operators based on contractible loops.
(Of course, the torelon and glueball operators do mix and
at $t \to \infty$ we would always find the same effective 
mass. But if the mixing is small one will, in general, find 
different effective mass plateux at small values of $t$.) Just 
as we saw in the case of SU(2), it is clear that the onset of
large finite size effects for the scalar and tensor glueball
masses is linked to the mass of the corresponding torelon state.
We are also
now able to see what happens to the $C=-$ states and
we show the $0^{--}$ which is the lightest of these.
Since our $a$ is not small, we do not have accurate
values of $m_{eff}(t=2a)$ for the heavier states.
(As usual our quoted masses are extracted at $t\geq 2a$.)
We can of course look at  $m_{eff}(t=a)$ which has the
disadvantage of having some excited state component, but
which is accurately calculated even for the heaviest
states. Although we do not show the values here, we
remark that they show no sign of any finite size effects
that violate our above criterion, either for the $C=+$
or for the $C=-$ states. The same is true for our
modest SU(4) and SU(5) finite volume studies, which
appear in Table~\ref{table_Vsu45}.

This establishes the level of our control of finite volume
effects. There is one further important point. In the
case of the $2^{++}$ the torelon appears to exist as
a bound state in the mass spectrum for larger $L$.
This makes it difficult to extract a consistent
picture of the excited $2^{++}$ states. Since there is
no such difficulty for the excited $2^{-+}$ states, and 
since these states should be degenerate with the non-torelon 
$2^{++}$ excitations, we do not try to overcome this
difficulty. So the reader should not be surprised to find
no masses being quoted for the excited $2^{++}$ states later on.
As an addendum to this, we remark that this `difficulty'
appears to disappear for larger values of $N_c$.
We assume that this is a manifestation of the suppression
of all mixings at large $N_c$.

\subsection{The lattice mass spectra}

In the previous two subsections we discussed our criteria
first for minimising the systematic error associated  with
the extraction of masses on a given lattice, and second for
controlling the finite volume corrections to such masses. 
We now use those criteria to extract our ``infinite volume''
lattice mass spectra. 

In Tables~\ref{table_msu2a},~\ref{table_msu2b},
~\ref{table_msu3}, ~\ref{table_msu4} 
and ~\ref{table_msu5} we list some of the masses that we have
extracted in the SU(2), SU(3), SU(4) and SU(5) calculations
respectively. All are in lattice units. In addition, all
these masses have been extracted from $am_{eff}(t=2a)$.
There are a few exceptions to this. It sometimes occurs,
particularly at the smaller values of $\beta$, that
$am_{eff}(t=a) < am_{eff}(t=2a)$. On the other hand, we
know from positivity that  $am_{eff}(t)$ must decrease
as $t$ increases. Since the error at $t=a$ is smaller than 
at $t=2a$, it is clear that the $t=a$ effective mass is
the better mass estimate in these cases, and that is
the value we list. However the error quoted is the larger
one which is associated with $t=2a$. In principle
when we come to extrapolating to the continuum limit
we should use the $t=2a$ effective masses in these cases
since otherwise there is a systematic downward bias in
the statistical analysis. (That is: we correct some
large upward statistical fluctuations in $am_{eff}(t=2a)$,
but none of the ones that are large and downwards.)
However so few values are affected, and these are usually
at the lowest values of $\beta$, that we choose not to 
complicate the analysis by doing so. We also note that
the volumes used here are always at least as large as the
minimum necessary, as indicated by our earlier finite volume
studies.

We begin with a brief technical aside.
Our calculations, at each value of $\beta$ and $L$, typically
involved between 80000 and 200000 Monte Carlo sweeps with
calculations of glueball correlators being made every
5 sweeps. Typically we would have 3 or 5 over-relaxed sweeps
for each heat bath sweep. By comparing the values of
$a\surd\sigma$ the reader can see that the SU(4) and 
SU(5) $\beta$ values are more-or-less equivalent. They
are also nearly equivalent to some of the SU(3) and
SU(2) $\beta$ values. 

Some comments now, starting with the SU(5) masses listed in 
Table~\ref{table_msu5}. Here we focus on features that
might affect the reliability of the calculations; we
leave a discussion of the physics till later. We first note
that the lightest $J=1$ and $J=2$ states display 
parity doubling within errors. In contrast to the 
marked lack of doubling in the $J=0$ sector. This confirms 
that we have made our ultraviolet cut-off small enough, and our
infrared cutoff large enough. The same is true of
the $J=2$ excited states. However it is less clear what is 
going on amongst the excited states in 
the $J=1$ sector; there appears to be a
near-degenaracy between the ground and excited states in
some cases. And the expected degeneracy between the $1^{++\ast}$ 
and the $1^{-+\ast}$ appears to be broken. This may indicate 
the presence of a $J=3$ state which is nearly degenerate with 
the $J=1$ state; or it may be that some of these states are
multiglueball scattering states; or there may be finite volume
corrections. We are not well placed to distinguish amongst these
possibilities in our present calculations. For example
to investigate the last possibility, we need to do a
finite volume study for a small value of $a$ where these
very heavy masses can be accurately calculated. 
The only calculation of this kind is in SU(2) at $\beta=12$
(see Table~\ref{table_msu2b}).
We do not see any trend for the $1^{++\ast}$ and  $1^{-+\ast}$ 
masses to converge as we increase $L$ from $L=32$ to $L=48$.
So it does not seem to be a finite volume effect. If we
compare different $\beta$ values there appears to be no
trend for this effect to decrease; so it would not seem to
be a finite-$a$ effect. This is a puzzle. As far as the
$1^{\pm -}$ states are concerned, we have even less to go on,
because we have no $C=-$ in our SU(2) studies. These oddities
need to be resolved but they afflict the very heaviest of the
states we study and so we shall not attempt to resolve the issue
here.

Although we have carefully chosen the volumes so that the ground 
state $J=0$ and $J=2$ masses are essentially infinite volume,
this is not necessarily the case for the $J=2$ excited states.
Indeed we observe in  Table~\ref{table_msu2b} that the mass
of the $2^{++\ast}$ is volume dependent and is only degenerate
with the $2^{-+\ast}$ on the largest volume, and then only 
in those cases,
$\beta=9$ and $12$, where this is exceptionally large. Explicit
calculations with the double Polyakov loop operators
described in the previous subsection indicate that this is
an artifact of the presence of a corresponding scattering
(or bound?) state whose mass increases approximately 
linearly with $L$. So as $L\uparrow$ the mass moves out of the
range of masses we probe. We remark that we have no clear evidence 
of a corresponding $J=0$ state. (It would, in any case,
not interfere with the lowest $0^{++}$ excitations because
these are so light.)  It is interesting that this
problem appears to disappear for the larger SU(N) groups. 
A possibility is that, as expected, the mixing between
the double flux loops and our ordinary `local' glueballs
is suppressed by powers of $1/N$. To investigate this
properly one needs to include both double flux loops
and our normal contractible loops within a single
basis for our cross-correlation matrix (as has been
done in 
\cite{PWT}
for precisely this purpose.) We have not been able to do this 
here because of the very large storage costs this would have 
entailed. Since we cannot resolve the states unambiguously,
we shall not attempt a continuum extrapolation of the SU(2)
$2^{++\ast}$. For larger groups it seems that this problem
is not there and so we shall attempt to obtain the corresponding
continuum masses.

\subsection{The continuum mass spectrum}

A lattice spectrum is only interesting insofar as it
can lead us to the spectrum of the corresponding continuum 
theory. To obtain the continuum spectrum we need to extrapolate 
our lattice masses to $a = 0$. 
The first step is to take ratios of masses so that the scale, $a$,
in which they are expressed cancels. We choose to take ratios of 
the glueball masses, $am_G$, to $a\surd\sigma$ since the string
tension is our most accurately calculated quantity. 

The second step is motivated by the observation
\cite{Sym}
that in pure lattice gauge theories the leading 
lattice correction to dimensionless
ratios of physical quantities, such as $m_G/\surd\sigma$ 
is $O(a^2)$. So for small enough $a$ we expect
the $a$-dependence to be given, just as in four dimensions, by 
\begin{equation}
 {{m_G(a)} \over {\surd\sigma(a)}}
 = 
 {{m_G(a=0)} \over {\surd\sigma(a=0)}} + ca^2\sigma.
\label{D2}
\end{equation}
Of course, instead of using the correction term  $ca^2\sigma$
we could use ${\tilde c}a^2{\tilde m}_G^2$ where $a{\tilde m}_G$
is any calculated glueball mass. The difference is formally
$O(a^4)$. The reason for choosing $\sigma$ is simply
that it is so accurately determined. 
An alternative way of extrapolating such a mass ratio to $a=0$
is to use the fact that $g^2$ has dimensions of mass, and
that $\lim_{\beta\to\infty} \beta = 2N_c/ag^2$. Thus for small
enough $a$
\begin{equation}
 {{m_G(a)} \over {\surd\sigma(a)}}
 = 
 {{m_G(a=0)} \over {\surd\sigma(a=0)}} + { c \over {\beta^2}}.
\label{D3}
\end{equation}
What we do in practice is to choose one of the above forms and 
attempt to fit all the mass values of some state with it.
If a good fit is not possible we assume that this is because the
largest value of $a$ used is too large for the $O(a^2)$
correction to be adequate. So we drop the mass corresponding
to the largest value of $a$ and try again. We keep doing
this until we get a good fit.
  
\begin	{figure}[p]
\begin	{center}
\leavevmode
\input	{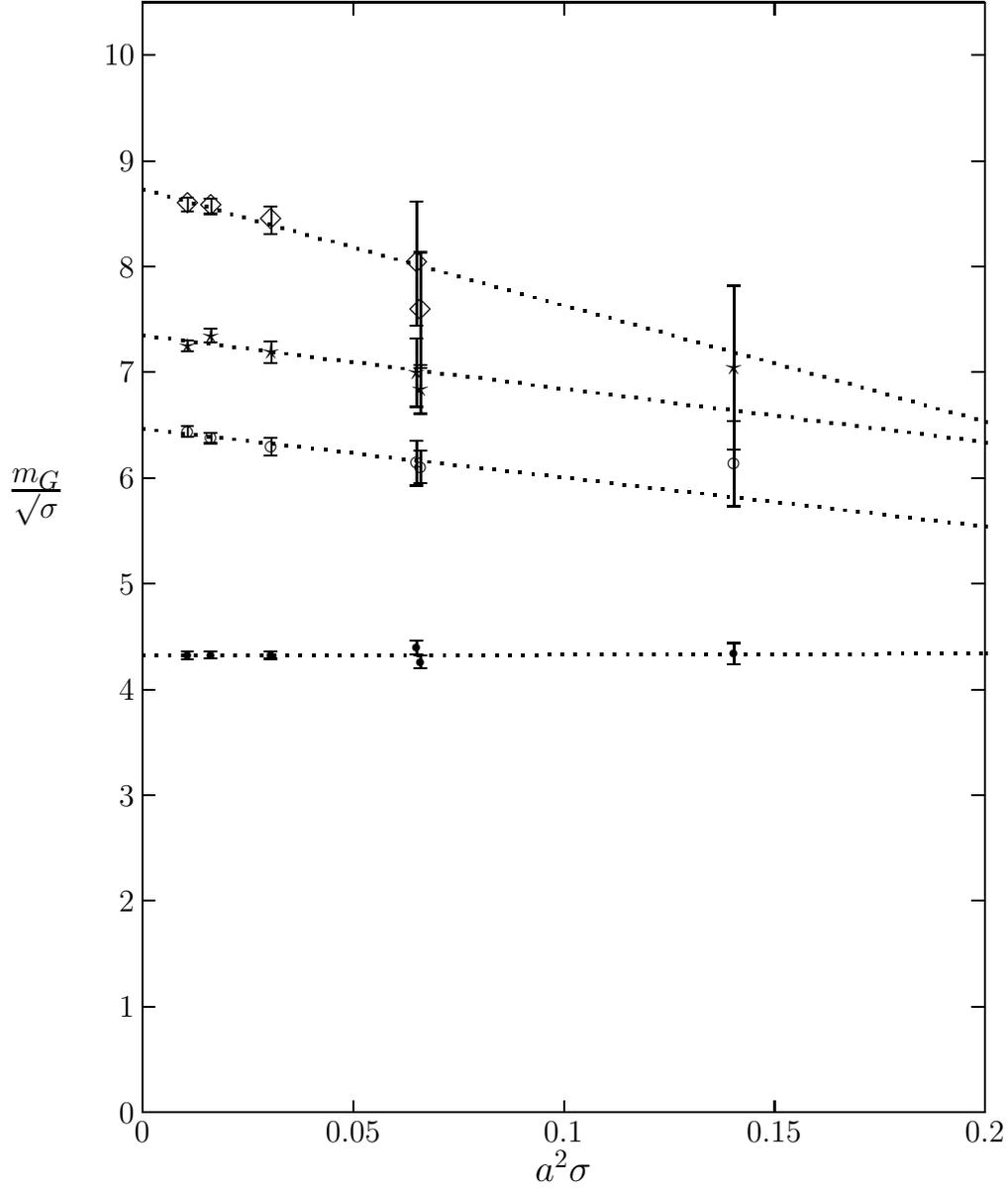}
\end	{center}
\vskip 0.15in
\caption{The ratio of some SU(3) masses to 
$a\surd\sigma$, plotted against $a^2\sigma$ to
show how they vary with $a$: the $0^{++}$($\bullet$),
the $0^{--}$($\circ$), the $2^{-+}$($\star$) and the 
$2^{--}$($\diamond$). Extrapolations
to the continuum limit are shown as straight lines.}
\label{fig_msu3}
\end 	{figure}

In Fig.\ref{fig_msu3} we show some examples drawn from
the SU(3) calculation. Since the mass ratios are plotted against
$a^2\sigma$, the continuum extrapolations, in 
eqn(\ref{D2}), are simple straight lines. The really striking
feature of this plot is how little variation with $a$ there
is. This will make for unambiguous and very accurate
continuum extrapolations.

Our continuum extrapolations for our various theories
are displayed in  Table~\ref{table_mcont}. They have been
obtained by fitting the form in
eqn(\ref{D3}) to the masses listed
in Tables~\ref{table_msu2a},~\ref{table_msu2b},
~\ref{table_msu3},~\ref{table_msu4},~\ref{table_msu5}
and the string tensions listed in Tables~\ref{table_Ksu2},
~\ref{table_Ksu3},~\ref{table_Ksu45}. The quality 
of each fit, as given by the confidence level, is 
given in Table~\ref{table_clcont}. We have also
performed extrapolations using  eqn(\ref{D2}); these
give essentially identical results, with any differences
being much smaller than our quoted errors. 

The reader will note that the mass of the $2^{++\ast}$ 
is missing from the SU(2) and SU(3) columns. This is
because we could find no acceptable fits. We believe
this is related to mixing with torelons, as discussed 
previously. The mass of the latter depends sensitively on
the volume and so will not provide a consistent set of 
masses at different values of $\beta$ (since the volumes
are not exactly the same). This is no longer a problem
with SU(4) and  SU(5) and we assume that this is because
any such mixing becomes suppressed at large $N_c$.
A calculation including overlaps between torelon
and glueball operators would resolve this question,
but we have not carried this out. Why we cannot get
an acceptable fit for the SU(2) $2^{-+\ast}$ is
less clear. The reason might be that our SU(2)
calculations in the $2^{-+}$ channel had a smaller 
basis (4 operators) than in the later calculations
with larger groups. This meant a very small basis for
the excited states.

In addition to these spectra we have performed in
Appendix D some calculations with a very asymmetric
lattice action, $a_t \simeq a_s/4$. This may be thought
of as being close to the `Hamiltonian' limit, and it is
interesting, as a test of universality, to confront
this spectrum with the SU(2) spectrum that we have
obtained in this section. This we do in 
Table~\ref{table_compasym} and we observe good agreement
within errors. For a detailed discussion of our
calculations with the asymmetric lattice action we
refer the reader to Appendix D.

We now have all our continuum spectra and can turn to their
dependence on $N_c$.

\subsection{The ${\rm N_c}$ dependence of the mass spectrum}

We can already see from Table~\ref{table_mcont} that the
variation of our mass ratios with $N_c$ is weak; and that
it appears to become weaker with increasing $N_c$. 

To illustrate this we plot in Fig.\ref{fig_mcpsuN} 
and Fig.\ref{fig_mcnsuN} the quantity 
$m_G/g^2N_c$, which is obtained from the ratios in 
Table~\ref{table_mcont} and the string tensions in eqn(\ref{C7}).
We choose to plot against $1/N_c^2$ because the usual
diagrammatic analysis predicts that at large enough $N_c$ we
should expect 
\begin{equation}
 {{m_G} \over {g^2N_c}}
 = 
 R_{\infty} + {{R_1} \over {N_c^2}}
\label{D4}
\end{equation}
where $R_{\infty} = \lim_{N_c\to\infty} m_G/g^2N_c$.
So on our plot this will be a simple straight line.
We observe that the dependence of our masses on $N_c$
is really very weak indeed, all the way down to $N_c=2$. 
This indicates once again that the mass scale of the 
SU($N_c$) theory is $\propto g^2N_c$, as expected from
the diagrammatic analysis.

In Table~\ref{table_massgN} we list the $N_c \to \infty$
limits and the slopes, $R_1$, that result from fitting 
our continuum masses with eqn(\ref{D4}). We can, of
course, perform a similar analysis using $m_G/\surd\sigma$
instead. The results of the corresponding extrapolations
are presented in Table~\ref{table_masskN}.

Is there anything we can add to our previous result,
in eqn(\ref{C11B}), on the power of the leading correction?
The only mass that is accurate enough to be potentially useful
is $m_{0^{++}}/g^2N_c$. However, as we see from
Fig.\ref{fig_mcpsuN}, this
varies almost not at all with $N_c$ and so provides us with 
no useful information on the power of this correction.
(The stronger variation in $m_{0^{++}}/\surd\sigma$ simply
reflects the variation of $\surd\sigma$ which we have
already studied.)

For purposes of comparison, it would be interesting to 
provide an example of a mass spectrum that is quite
different to the one we have calculated here and
yet comes from a theory with a dynamics that is not
so dissimilar as to make the comparison meaningless.
A natural possibility is to consider the U(1) theory.
Since the leading-order large-$N_c$ 
arguments are in fact for U($N_c$), this theory belongs
naturally to the sequence of theories we have considered.
And yet it is so far from $N_c = \infty$ that we would not
expect it to fit into the pattern we have observed so
far. As far as dynamics goes, it is a lattice gauge theory which
is linearly confining and free at short distances.
We discuss our results for the U(1) mass spectrum (listed
in Table~\ref{table_massU1}) and some
peculiarities of the theory, in Appendix E.
Here we merely note that in the U(1) theory the $0^{++}$ is no
longer the lightest state; the $0^{--}$ is about half its mass.
We also note that the mass ratio  $m_{0^{++}}/\surd\sigma$ 
is much lower than in SU(2) (or in SU($\infty$) for that
matter). So this spectrum is indeed quite different; and
the comparison enhances, by contrast, 
our claim that SU(2) $\simeq$ SU($\infty$).

The lack of any visible $N_c$ dependence in our most
accurately calculated mass, $m_{0^{++}}/g^2N_c$, is
quite striking and provides strong evidence that
there is a smooth nontrivial large-$N_c$ limit,
with a physical mass scale $\propto g^2N_c$.
Coupled with our previous analysis of the string
tension, this also tells us that the $N_c = \infty$
limit possesses linear confinement. (As we see 
immediately from the behaviour of $m_G/\surd\sigma$.)
From the intercepts and slopes listed in 
Tables~\ref{table_massgN} and ~\ref{table_masskN}
and from eqn(\ref{C12}), we can obtain the mass spectrum
for any value of $N_c$. In this very concrete sense we
can say that SU($N_c$) theories are close to SU($\infty$) 
all the way down to SU(2). Thus the large-$N_c$ 
analysis unifies our understanding of all SU($N_c$)
theories in a compact and elegant fashion.

\subsection{Features of the mass spectrum}

The purpose of this paper is to calculate the mass spectrum.
Extracting interesting physics from the detailed features
of that spectrum
is something that belongs elsewhere. However it would be 
churlish of us not to make a few comments. These will be
brief and incomplete.

\vskip 0.1in

{\noindent}$\bullet$ There is clear evidence for the expected parity 
doubling in the cases of the
$2^{\pm +}$, $2^{\pm -}$, $2^{\pm -\ast}$,
$1^{\pm +}$ and $1^{\pm -}$. In contrast, for the
$J=0$ states, where we do not expect parity doubling,
the splitting between the $P=+$ and $P=-$ states is huge.

\vskip 0.1in

{\noindent}$\bullet$ Our lightest glueball state is the $0^{++}$; just
as it is in D=3+1 gauge theories. Moreover its mass, in units
of $\surd\sigma$ is not that different. 
If we take the SU(3) continuum extrapolation in 
\cite{MTnewton},
which uses the D=3+1 lattice glueball mass calculations in
\cite{CMT,massSU3},
and if we perform a corresponding SU(2) continuum extrapolation
using the D=3+1 lattice calculations in
\cite{MTunpub,massSU2},
then we find
%
\begin{equation}
{{m_{0^{++}}} \over {\surd\sigma}} =
\left\{ \begin{array}{ll}
3.87(12) & \ \ \ \mbox{SU(2) \ ; \ D=3+1} \\
3.65(11) & \ \ \ \mbox{SU(3) \ ; \ D=3+1}
\end{array}
\right.
\label{D5}
\end{equation}
The fact that the D=3+1 mass ratio is smaller
than the one in D=2+1 follows naturally
\cite{tepmor,rjmt}
in flux tube models of gluonic states
\cite{patisg}.
(It does so from the fact that the closed flux loop 
has more transverse
dimensions in which to oscillate; this increases the
corresponding ``Casimir energy'', and so decreases
the mass of the loop, for a given loop length.) 
We also note from
eqn(\ref{D5}) that the D=3+1 $N_c$ dependence has 
the same sign as in D=2+1.

\vskip 0.1in

{\noindent}$\bullet$ Just as in D=3+1
\cite{MTnewton,CMT,massSU3},
the next heaviest state 
in the $C=+$ sector is the $2^{++}$ (ignoring excitations 
of the $0^{++}$ since these have not been
calculated in 4 dimensions). The scalar-tensor
mass splitting is not dissimilar: e.g.
\begin{equation}
{{m_{2^{++}}} \over {m_{0^{++}}}} =
\left\{ \begin{array}{ll}
1.65(3) & \ \ \ \mbox{SU(3) \ ; \ D=2+1} \\
1.41(7) & \ \ \ \mbox{SU(3) \ ; \ D=3+1}
\end{array}
\right.
\label{D6}
\end{equation}

\vskip 0.1in

{\noindent}$\bullet$ Unlike the $C=+$ sector, the $C=-$ sector 
is very different from its D=3+1 counterpart. For example, we 
have a light $0^{--}$, while there are no light $C=-$ states 
in 4 dimensions. This may arise from the fact that 
in 3 space dimensions there is an interplay between $C$ 
and $J$ that does not exist in 2 space dimensions. 
Consider, for example, a circular flux string. It will
have an arrow on it, for $N_c\geq 3$. Under $C$ the
direction of the arrow flips. In 3 (but not 2) space dimensions
we can rotate the circle by $\pi$ around a diameter and
this also flips the direction of the arrow. Note that
this means that a rotationally symmetric linear
combination of such circular loops cannot be $C=-$.
One needs a fluctuation away from a circle to allow
$J=0$ and $C=-$ and this raises the energy. Of course
we have gradually incorporated some dynamical
assumptions as we moved through the last few sentences.
One needs to make the argument within a specific
model framework and that belongs elsewhere
\cite{patisg,tepmor,rjmt}.

\vskip 0.1in

{\noindent}$\bullet$ We observe that whatever splits the $C=+$ and
$C=-$ states is weakly dependent on $N_c$; and survives
the $N_c\to\infty$ limit. On the other hand, in SU(2),
where we have no $C=-$, the spectrum is clearly a
smooth continuation of the $N_c\geq 3$ $C=+$ spectrum
(since our simple mass fit encompasses $2 \leq N_c \leq 5$).
This provides a constraint on dynamical mechanisms for
the $C=\pm$ splitting.

\vskip 0.1in

{\noindent}$\bullet$ There are some striking approximate degeneracies
in the spectrum. The typical pattern is:
$m_{0^{++\ast}} \simeq m_{0^{--}}$, 
$m_{0^{++\ast\ast}} \simeq m_{0^{--\ast}}$
and similarly for the $J=2$ states. Again, if this is not 
an accident, it does suggests some simplicity in the dynamics.

\vskip 0.1in

To go further requires confronting specific models with
the spectrum we have calculated here. That goes well beyond the
scope of this paper.

\section{Conclusions}

In this paper we presented our calculations of the mass 
spectra and string tensions in three dimensional SU(2), SU(3), 
SU(4) and SU(5) lattice gauge theories. From these we obtained 
the corresponding continuum spectra. The accuracy of these
continuum results reflects the large range in the lattice
spacing, $a$, over which we performed our lattice
calculations. We can compare this range to that in the
more familiar D=4 SU(3) gauge theory by using the
calculated values of the string tension, $a^2\sigma$.
Doing so we observe that the (useful) range of our D=3 
calculations would correspond to $5.50 \leq \beta \leq 6.55$ for
the case of SU(2), $5.50 \leq \beta \leq 6.50$ for SU(3), and
$5.70 \leq \beta \leq 6.35$ for both SU(4) and SU(5).
This range, and the statistics of our calculations, is
the primary reason why our D=3 calculations are so much
more accurate than what is available in four dimensions. 
We also gain something from the fact that our best
operators are slightly better in three than in four dimensions.

We noted some strong similarities, in the $C=+$ sector, between
the 2+1 and 3+1 dimensional spectra. This should provide an
interesting test for models of glueballs. Indeed one of the
main motivations for our calculations is to provide a
detailed spectrum against which models and analytic approaches
can test themselves.   

At the more technical lattice level, we studied, during the
course of our calculations, the effectiveness of over-relaxation,
the use of asymmetric lattice actions, how good are operators with
baryonic vertices, the efficiency of our `blocking' algorithm,
and the extent to which the mean-field/tadpole improvement
of the coupling really represents an improvement. In this
last case we were aided by the super-renormalisability of the
theory; this allowed us to compare directly extrapolations
using the improved and bare lattice couplings, in a way
which is not at present possible in four dimensions.

The primary purpose of our calculations was to study the
large-$N_c$ limit of SU($N_c$) gauge theories in 2+1
dimensions, and to compare the results of our fully 
non-perturbative calculations with the standard expectations 
obtained from all-order perturbation theory. We found that
there does appear to be a smooth $N_c \to \infty$ limit
and that it is obtained, as expected, by varying 
$g^2 \propto 1/N_c$. The leading correction is $O(1/N^2_c)$,
again as expected. We found that confinement -- the crucial
ingredient for the usual phenomenology -- does indeed survive 
the large-$N_c$ limit. And we obtained the detailed mass
spectrum in that limit. Finally, we observed that even
SU(2) is close to SU($\infty$), in the sense that the
difference between the mass spectra can be described by
just the leading $O(1/N^2_c)$ correction.

Thus all $D=2+1$ SU($N_c$) gauge theories can be
described by the  SU($\infty$) theory with a
modest  $O(1/N^2_c)$ correction. This provides a very 
elegant way to unify and understand all these 
potentially quite different theories.

There is a wealth of large-$N_c$ expectations that we
have not explored. For example those involving decays,
$G \to GG$, and, more generally, the $N_c$-dependence of
matrix elements
involving various products of singlet operators,
as well as their factorisation properties.
Neither have we attempted to expose the existence
of Witten's Master field
\cite{EWmaster}
or to determine its properties. All these topics should be
readily accessible in three dimensional calculations
of the kind presented in this paper. The reason we have not
addressed them in this paper is not because we find
them less interesting than the questions we have addressed,
but because the SU(2) and SU(3) calculations were 
completed before we realised that we might have something
interesting to contribute concerning the large-$N_c$ limit.

These calculations also need to be extended by the
inclusion of matter fields in the fundamental representation.
In this case the leading corrections are expected to be
larger, $O(1/N_c)$, and so it is an interesting open
question whether SU(2) or even SU(3) will remain close
to  SU($\infty$). Needless to say, all the above questions need
to be addressed in four dimensions; and what we can say there
is that the first indications 
\cite{MTN}
are quite promising.

\vfill
\eject
\noindent {\large {\bf Acknowledgements}}

\vspace{0.25cm}

This work has been supported by the following
grants from PPARC for time on the RAL Cray J90:
GR/J21408, GR/K95338 and PPA/G/S/1997/00643. 
It has also been supported under PPARC grant GR/K55752.
The hospitality of the
Newton Institute during the last part of this work is
gratefully acknowledged. 
I have benefitted from many discussions, with many
people, during the course of this work. Particular
thanks to Simon Dalley for asking 
questions which provided my original motivation
for extending the SU(2) and SU(3) calculations
into a study of SU($N_c$).

\vspace{1.0cm}
\newpage

\noindent{\Large{\bf Appendix A : 
Testing the benefits of Monte Carlo over-relaxation }}

\vspace{0.25cm}

Although there is no systematic procedure known for reducing
the exponents associated with critical slowing down in
$D=4$, or $D=3$, non-Abelian gauge theories 
a method that appears to have some effectiveness, and
which is now in common use, is to
mix heatbath and overrelaxation
\cite{ORadler,ORcreutz}
sweeps during the update (for reviews see 
\cite{ORrevs}).
As far as testing the efficiency of this method
is concerned, what is available are studies
of the decorrelation of blocked Wilson loops in SU(2) 
and SU(3) 
\cite{ORtests}
which show that there is a strong reduction in fluctuations 
when most heatbath sweeps are replaced with overrelaxation. 
This has helped to motivate the widespread use of 
overrelaxation. 

However, useful as these tests are,
what one would like to see is how the statistical errors
on the physical quantities of interest (glueball masses,
string tension, ...) are reduced when some fraction of the
heatbath sweeps are replaced with overrelaxation sweeps. In D=3+1
such an exercise would be prohibitively expensive for the small
lattice spacings where the answer is interesting; and
so, as far as we know, no study of this kind has been
published. In $D=2+1$, such a study becomes possible and
this is what we shall present in this Appendix.
Because the D=2+1 and D=3+1 theories have so much in common
-- in particular they both become free at short distances --
we can hope that what we find has some relevance to
four dimensions as well.

Our heatbath and over-relaxation algorithms have been
described in Section 3.1 . We note that
in both SU(2) and SU(3) the overrelaxation algorithm
explores phase space at a constant value of the total action.
We shall characterise the update pattern by the ratio, $R_o$, of the
number of overrelaxation sweeps to the number of heatbath sweeps.
Since we use a pipelined CPU, all our sweeps employ a variation
of a chequerboard update.

Our study of SU(3) is the more extensive of the two and so
this is where we shall begin. We have performed
comparisons at three values of $\beta$: at $\beta=11$ and 15
on $12^{2}16$ lattices, 
and at $\beta=21$ on a $24^{3}$ lattice. If we use
the calculated string tension to set the scale of the lattice
spacing, then these three values of $\beta$ correspond to
$\beta \simeq 5.7, 5.9, 6.15$ respectively in the $D=4$ 
theory with which the reader is probably more familiar.
The lattices at $\beta=11$ and $\beta= 21.0$ are effectively of
infinite physical volume for the quantities we shall be
considering here. The lattice volume at $\beta=15$ is 
of an intermediate size, which mainly effects the
nature of the $2^{++}$ glueball.
At $\beta=11$ we performed calculations for $R_o=0$ (pure
heatbath) and for $R_o=5$. Each calculation involved 
80000 sweeps with the data split into 40 bins of 2000
sweeps each for the error analysis. 
At $\beta=15$ we performed 25000 sweeps at
each of $R_o=0,2,5,10,20$, with the data divided into
20 bins in each case. At $\beta=21$ we performed
20000 sweeps at each of $R_o=0,3,5,7,10$, with 20 bins
in each case. 

The quantities we use in our comparison are, firstly, the masses 
of the lightest glueballs: the $0^{++}, 0^{--}, 2^{++}$
and, where available, the $2^{-+}$ (which should be degenerate with
the $2^{++}$ in large volumes and for small lattice spacings). We   
also use the mass of the lightest flux loop that closes
through a periodic boundary. This provides us with our
estimate of the string tension, $\sigma$, since the mass of
this loop is $a^2\sigma L$, up to $O(1/L)$ finite size corrections,
where $L$ is the minimal length of the loop (in lattice units).
In addition to these masses we also calculate expectation values 
of the simplest closed  loops made out of our `blocked' 
links. At a `blocking' level of unity, $B_l=1$, we have the
simple plaquette. More generally these `superplaquettes' consist
of a square that is length $2^{B_l-1}$ in lattice units. The
simple plaquette is dominated by ultraviolet fluctuations and
is of relatively little physical interest. At higher smearing
levels, the expectation value is dominated by fluctuations
closer to physical length scales and how the accuracy of these
is affected by overrelaxation is a more interesting question.

We compare the errors on these quantities in the different runs
characterised by different values of $R_o$. The reference run
is the one with $R_o=0$, i.e. pure heat bath. If the $R_o=0$ error
is changed by a factor of, say, $\gamma$ in the mixed run
then the latter
is as good as a pure heatbath run whose length is $1/\gamma^2$
times that of the mixed run. This of course assumes that our
bins are large enough to be essentially independent. We have
performed a variety of checks to convince ourselves that this
is the case for the results we present here. For example, for
the $\beta=15$ pure heatbath run
we checked that the bins could be made a factor
of 10 smaller and still be negligibly correlated. (With a factor
of 20 the independence began to break down.) In order to
keep our bin sizes sufficiently large so that we could be
confident of their mutual independence, the number of bins
for each calculation could not be made very large. Hence there will
be substantial fluctuations on our error estimates. For this
reason the reader should be cautious about drawing conclusions
from any one error ratio, and in practice we will average the error 
comparisons over several quantities. 

In Tables~\ref{table_or11su3},~\ref{table_or15su3} and 
~\ref{table_or21su3} we show the statistical errors for 
the flux loop and glueball masses at the three different
values of $\beta$. There are 2 rows of numbers for each mass.
The second row contains the actual mass estimates. The first
row is the error on the effective mass extracted from the
same correlation function but from one time step earlier.
This contains an admixture (typically only a few percent)
of excited states. (See Section 3.2.3 for a discussion of
effective masses.) We display both because the individual
error estimates contain quite large fluctuations 
which appear to be largely independent and so can be averaged
to obtain more reliable error ratio estimates. Because we 
are equally interested in all these physical quantities, it
makes sense to construct a global average of these error ratios.
We attach to this average an `error' obtained by treating
the variations of the individual error ratios around the
global average as though they were statistical fluctuations.
This is intended to do no more than provide an $indication$
of the significance of the value of the average error ratio.
The reader can manipulate the numbers in the Tables in other
ways if he so prefers.  

Consider first $\beta=11.0$ (Table~\ref{table_or11su3}) 
and the ratio of errors in
the run with overrelaxation to the errors in the pure
heatbath run. Taking the ratio of corresponding
errors in the two columns we obtain an overall average error
ratio of 1.02(6). So in this case there is no improvement
in incorporating overrelaxation. At $\beta=15.0$ 
(Table~\ref{table_or15su3}) we
obtain average ratios 1.27(12), 0.96(7), 1.09(9), 1.09(10) for
$R_o=2,5,10,20$ respectively. Again there is no sign of a significant
improvement for any overrelaxation mix. The global error ratio
average, 1.10(5), confirms this.

We turn now to our calculation at the weakest coupling, $\beta=21.0$ 
(Table~\ref{table_or21su3}). 
We obtain average ratios of 0.91(8), 0.81(6), 
0.88(7), 0.94(7) for $R_o=3,5,7,10$ respectively. We observe
a clear reduction in the errors of the runs with overrelaxation:
the global error ratio average is 0.885(33). Although we cannot
be certain which mix is best, there is evidence that a ratio of 
around 5:1 to 7:1 is as good as any at this value of $\beta$
and that this leads to an error ratio of around 0.84. To this
improvement we should add the fact that an overrelaxed sweep,
for SU(3) in $D=2+1$, takes about $77\%$ of the time for a
heat bath sweep. Thus the gain in using over-relaxation
is about $40\%$ in the update time. Although
the gain in the total time will be reduced by the
inclusion of measurements (typically tuned to be
about half of the total time) there is no doubt that
this is a worthwhile gain. 

We turn now to the case of SU(2) where our tests are much more
limited. The lightest masses here are of the flux loop and
of the $0^+$ and $2^+$ glueballs.
We performed comparisons at $\beta=6.0$ and $\beta=9.0$.
Using the calculated string tension to set the physical scale,
these values of $\beta$ correspond to $\beta\simeq 2.4$ and 2.55, 
respectively, in the $D=4$ $SU(2)$ theory. At $\beta=6.0$ we see
no sign of any benefit from overrelaxation; albeit in a 
calculation of limited statistics. Our calculations at $\beta=9.0$
(Table~\ref{table_or9su2}) are with $R_o=0,5,9,49,249,\infty$ and
have much better statistics: 25000 sweeps, split into 25 bins,
for each value of $R_o$. The lattice is $12^2 24$ which is 
of small, but not very small, physical size: there are 
certainly some finite size effects involving 
the $2^+$. We see from Table~\ref{table_or9su2} 
that there is a significant benefit to using
overrelaxation. We do not show our results for $R_o=249$
and $R_o=\infty$ (all sweeps overrelaxed) which, while
amusing for various reasons, are not really relevant to 
this study. We obtain average error ratios of
0.76(8), 0.71(9), 0.80(11) for $R_o=5,9,49$ respectively. The
global error ratio average is 0.76(5). For SU(2) overrelaxation
is a simple operation and is much faster than the heatbath;
a run with $R_o=5$ or 9 is about twice as fast as a run
with only heatbath sweeps. Thus the overall saving is
a factor of $2 \times 1/0.76^2 \sim 4$. This is a large
reduction. Again, the inclusion of measurements will
reduce the gain somewhat.

We turn now to the smeared superplaquettes. The average error
ratios, for the runs described above,
are summarised in Table~\ref{table_orplaq}. 
We note that, not surprisingly,
the simple plaquette acquires a larger error if we
include overrelaxation. (These global averages mask the fact that
for small $R_o$ the error is often reduced, which $is$
surprising.) On the other hand we observe that the
errors on large superplaquettes are reduced and that, in
contrast to what we saw for masses, this effect is 
present at smaller couplings. This is similar to what has been 
found in 4 dimensions
\cite{ORtests}
for large and blocked Wilson loops.

In conclusion, we have seen that for sufficiently small 
couplings -- equivalent 
to $D=4$ values of $\beta \sim 6.15$  for SU(3) and 
$\beta \sim 2.55$ for SU(2) -- there is a substantial
increase in efficiency through mixing heatbath and
overrelaxation sweeps. The cpu saving is about $40\%$
for SU(3) and about $75\%$ for SU(2). The difference
is largely due to the fact that SU(2) overrelaxation
is a very simple and fast operation. In $D=4$ the
operation of calculating `staples' is a little lengthier and
so this effect will be somewhat weaker there.

\vspace{1.0cm}

\noindent{\Large{\bf Appendix B : 
Testing the efficiency of the operators.}}

\vspace{0.25cm}

In constructing a `good' basis of operators for our various
mass calculations, the use of spatial blocking is crucial. The
general motivation is that if one wants a good overlap onto
the lightest physical states then one needs to employ 
(combinations of) large smooth operators. There are 
obviously many possible variants on the particular
recipe we have used in this paper (which is the one that
has been used successfully in earlier $D=3+1$ 
mass calculations). In the first part of this Appendix we
will consider some variation in the blocking procedure
and we shall see that our choice is indeed an efficient one.

The range of operators we have used has been limited not
only in the type of blocking employed but also in the variety
of ways we put the blocked loops together to form colour
singlet operators. In practice we limited ourselves
to simple closed loops. However once we go beyond SU(2)
there is a whole new class of operators that we can
construct, and which take advantage of the fact that
one can tie together $N_c$ indices with a totally 
anti-symmetric tensor. We refer to this, for obvious reasons,
as a baryonic vertex. Such operators have not been
used in previous lattice glueball calculations as far
as we are aware. Our attention was drawn to them by
their possible role in splitting the $C=\pm$ sectors,
as pointed out in 
\cite{rjmt}.
For this reason we have carried out a small calculation
to check whether they encode some interesting new information.
This is described in the second part of this Appendix.

\vspace{0.25cm}

\noindent{\large{\bf Variations on the blocking procedure}}

\vspace{0.25cm}

In our construction of `blocked' link matrices 
the most obvious parameter is the weighting of the direct
path as compared to the staple-like paths. The choice
we made was to take an equal weighting for all the paths.
So, for say the $x$-direction, we would take 
$xx + yxxy^{\dagger} +  y^{\dagger}xxy$ using the notation
in Section 3.2.1.
In this section we shall perform some calculations using
a variable weighting $\gamma_d$. That is to say, we
use a blocking   
\begin{equation}
{\bar U}^{B}_x = \gamma_d U_x U_x
+ U_y U_x U_x U^{\dagger}_y
+ U^{\dagger}_y U_x U_x U_y
\label{APC1}
\end{equation}
%
where we have suppressed some obvious arguments etc. The
blocked link, $U^B_x$, is then obtained by projecting
${\bar U}^B_x$ back into the group.
We then see which value of $\gamma_d$ is most efficient
in the sense of producing the best operators.

Before doing so we briefly comment on the projection
back into the group and the resulting gauge
transformation properties of the $U^B$. We begin by
noting that if we perform a local gauge transformation 
on the fields then 
${\bar U}^{B} \to g_n {\bar U}^{B} g^\dagger_{n^\prime}$
where $g_n$ is the gauge transformation at site $n$
and the paths making up ${\bar U}^{B}$ start at the
site $n$ and end at the site $n^{\prime}$. If the
group is SU(2) we obtain $U^B$ by dividing  ${\bar U}^{B}$
by $\det\{{\bar U}^{B}\}$. Since the matrices $g$ are unitary
we have $\det g_n =1$ and so $\det\{{\bar U}^{B}\}$ is
gauge invariant. Thus in the case of SU(2) $U^B$ has
the gauge transformation properties of a product of links
from $n$ to $n^{\prime}$ and we can form colour singlet
operators out of closed loops in the usual way. For
SU($N_c \neq 2$) the situation is different:  
${\bar U}^{B}$ is not proportional to an SU($N_c$)
matrix and if we want $U^B$ to be in the group we need 
to define it some other way. The method we use is
to define $U^B$ as equal to the value of the SU($N_c$) 
matrix $U$ that maximises Tr$\{{\bar U}^{B}U^{\dagger}\}$.
It is easy to see that for SU(2) this reduces to 
the method we use there. 
It is also trivial to see (using the cyclic property of
the trace) that if  
${\bar U}^{B} \to g_n {\bar U}^{B} g^\dagger_{n^\prime}$
then, just as in SU(2),
${U}^{B} \to g_n {U}^{B} g^\dagger_{n^\prime}$ and we can form 
colour singlet closed loops in the usual way. However, 
in practice we maximise the trace by a simple iterative
procedure which we stop before complete
convergence in order to save computer time. This procedure
requires, as its starting point, some first guess, $U_s$, for 
the blocked matrix. In practice we construct $U_s$ from 
${\bar U}^{B}$ in such a way that it does not transform as
$U_s \to g_n U_s g^\dagger_{n^\prime}$ 
under a gauge transformation
$g$. This means that when we stop the algorithm prior to complete
convergence, the resulting $U^B$ only transforms approximately
as ${U}^{B} \to g_n {U}^{B} g^\dagger_{n^\prime}$. 
In principle this does not matter; averaging over
all field configurations in the Monte Carlo will lead to
a cancellation of the non-gauge invariant pieces in the
correlation functions. However, again in practice, this means we 
generate extra noise and this will increase our statistical
errors -- something to be avoided if possible.

We see from the discussion in the previous paragraph 
that there is more to `blocking' than  choosing
a sum of paths and a relative weighting. One can ask if
projecting back to the group produces better 
operators than not doing so (and perhaps using some
other form of normalisation). Studies in D=3+1 of several
alternative strategies in SU(2)
\cite{MTblock}
and SU(3) 
\cite{CMT}
suggested that
this was more-or-less so. Some tests then showed that
approximating the maximisation of the trace by one or two
iterations did not significantly worsen the operators
or increase the errors. However there has been no 
demonstration that this continues to be the case as we
increase the size of the group or that all this continues to
hold in D=2+1. These are studies that still need to be 
carried out.

We return now to our study of how the operators vary with the
choice of $\gamma_d$ in eqn(\ref{APC1}).
Our calculations are in SU(2) and are performed on
a $16^3$ lattice at $\beta = 7.5$ in a run consisting
of 10000 sweeps. On these configurations we performed
separate mass calculations, using our usual basis
of operators, for the 5 different blocking schemes
that used $\gamma_d = 0.25, 0.50, 1.0, 2.0$ and $4.0$. 

What we want to know is how efficient are the different
schemes; and in particular how  efficient is our usual 
choice, $\gamma_d =1$. We shall confine ourselves to the 
lightest states in each $J^{PC}$ channel. In that case our
usual variational criterion, as discussed in Section 3.2.3, 
provides us with a simple criterion for comparing operators 
of the same quantum numbers:
one calculates the effective mass at $t=a$ and the `best'
operator is the one that gives the smallest value of 
this effective mass. So what we have done here is to
find the best operator in our basis for each type
of blocking. The best form of blocking will then be the
one that produces the minimal value of the effective mass.

In Table~\ref{table_blockT} we present the value of
$a m_{eff}(t=a)$ for the best operator for each of
the 5 kinds of blocking we consider, and for the
various $J^{PC}$ quantum numbers. We observe that
$\gamma_d \in [1,2]$ seems to work best overall,
although $\gamma_d=0.5$ is virtually just as good if
we ignore the $1^{\pm}$ states. Note that since
the different calculations are performed on exactly 
the same sequence of field configurations, the errors
will be highly correlated.

We have therefore seen that with respect to variations
in this particular parameter, our choice of $\gamma_d = 1$
is about as good as any. Of course one can vary the algorithm 
in many other ways; for example by including other paths
than just the direct path and the `staples'. A systematic
study would be useful.

\vspace{0.25cm}

\noindent{\large{\bf Operators with `baryonic' vertices}}

\vspace{0.25cm}

We will consider the specific case of SU(3). Suppose we
have three curves $C_1,C_2,C_3$ each of which starts at 
some point $n$ and finish at some point $n^{\prime}$. 
Let us denote by $U^1,U^2,U^3$ the corresponding path
ordered products of (blocked) link matrices along
these three curves, running from $n$ to  $n^{\prime}$.
We can form singlet operators out of pairs of these
in the usual way, e.g.
\begin{equation}
\phi = Tr \ U^1 U^{2\dagger}.
\label{APC2}
\end{equation}
But in the case of SU(3) we also can form a colour singlet
out of all three of them:
\begin{equation}
\phi_Y =  
\epsilon_{ijk}
U^1_{ii^{\prime}} U^2_{jj^{\prime}} U^3_{kk^{\prime}}
\epsilon_{i^{\prime}j^{\prime}k^{\prime}}
\label{APC3}
\end{equation}
where we have exposed the matrix indices and $\epsilon_{ijk}$
is the usual totally anti-symmetric tensor. This extends
to $N_c > 3$ in the obvious way; we have $U^1,\ldots,U^{N_c}$
paths joined by the appropriate $N_c$-component $\epsilon$ 
tensor.

Since the operators in eqn(\ref{APC2}) and  eqn(\ref{APC3})
have the same quantum numbers they will have non-zero
overlaps and there is no {\it \`a priori} reason to think
that we have lost anything by excluding the latter. However it
might be that they constitute more efficient operators
for some states and if that is the case for one of the heavier
states, where in practice we cannot calculate correlators beyond 
small $t$, it might be that, in using them, we will expose a state
that we have not been able to see with operators of the type 
in eqn(\ref{APC2}).

In this Appendix we will describe a small exploratory
calculation designed to see if including such operators
might make a serious difference to our calculations.
We shall consider an operator, $\phi_Y$, of the form 
in eqn(\ref{APC3}) with
\begin{eqnarray}
U^1 & = & U_x U_y U^{\dagger}_x \nonumber \\
U^2 & = & U_y \\
U^3 & = & U^{\dagger}_x U_y U_x \nonumber
\label{APC4}
\end{eqnarray}
suppressing obvious arguments and indices. This is
a rectangle with a central link crossing the
rectangle. The path ordering is out from the same vertex
for all three curves. Under $C$ such an operator
reverses all three arrows on the curves; which for this
particular operator is equivalent to a rotation of $\pi$. 
So one can easily see that from this operator (and the one 
we obtain by $x \leftrightarrow y$) we can obtain $0^{++}$, 
$2^{++}$ and $1^{--}$ quantum numbers. 

We have performed a calculation on a $16^3$ lattice at 
$\beta=15$ with this operator. In Table~\ref{table_Y}
we list the effective $0^{++},2^{++}$ and $1^{--}$
masses obtained at $t=a$ using a basis that includes
the best two blocking levels of 
this operator. We compare it with what we obtain 
(on the same set of field configurations) if we 
do not include $\phi_Y$. We observe that there seems
to be nothing new in these channels 
when we include $\phi_Y$; at least not as far as the 
ground state and the first few excitations are 
concerned.
There is a slight improvement in some of the
overlaps, as indicated by a decrease in $m_{eff}(t=a)$,
but one would expect that just from an increase in the
size of the basis.

This, albeit minimal, study leads us to believe that the 
inclusion of operators incorporating baryonic vertices will
not alter our conclusions in any significant way. 
However such
operators  can be convenient in providing a simple
means for constructing $J=1$ operators; and they may well
be important in investigating some physics, e.g. the
$C=\pm$ splittings, lower order in $1/N_c$ corrections,
$\ldots$, so a more detailed investigation would
be useful.

\vspace{1.0cm}

\noindent{\Large{\bf Appendix C : 
Testing the benefits of mean-field improvement}}

\vspace{0.25cm}

In this Appendix we shall show that the
mean-field/tadpole improved inverse coupling
\cite{MFI}
\begin{equation}
 \beta_I = \beta \times 
\langle {1\over{N_c}} Tr(U_p) \rangle
\label{APB1}
\end{equation}
provides a much better expansion parameter than $\beta$
in our D=2+1 calculations. This both complements the
available D=3+1 evidence
\cite{MFI}
and provides us with a more accurate way to determine
$\lim_{a\to 0} \surd\sigma/g^2$.

Our strategy will be to compare directly various
extrapolations to $a=0$ using either $1/\beta$ or $1/\beta_I$
as expansion parameters. We shall perform these comparisons
using our calculations in the SU(2) and SU(3) theories,
since these cover large ranges in $\beta$.
Having found which extrapolation works best, we shall take
that information over to the SU(4) and SU(5) theories where
the range of our calculations is much more limited and
where the use of a good expansion parameter pays significant
dividends.

We have focussed upon the string tension because in practice this 
is the quantity that we calculate most accurately on the lattice.
Since $\lim_{a\to 0} \beta = 2N_{c}/ag^2$,
we know that
\begin{equation}
\lim_{\beta\to \infty} \beta a \surd\sigma
=
2N_c \lim_{a\to 0} {{\surd\sigma}\over {g^2}}
\label{APB2}
\end{equation}
The approach to the continuum limit will 
involve higher order corrections that are inverse powers of $\beta$ 
and hence vanish as powers of $a$. We thus expect that the
approach to the continuum limit will be under much better
control than in 4 dimensions, where the analogous
quantity that one would be calculating is 
$\lim_{a\to 0}  \surd\sigma/\Lambda_{mom}$ and where the corrections
would be inverse powers of $\log a$. This will allow us to make
a much more explicit and direct comparison than is possible in D=3+1.

\begin	{figure}[p]
\begin	{center}
\leavevmode
\input	{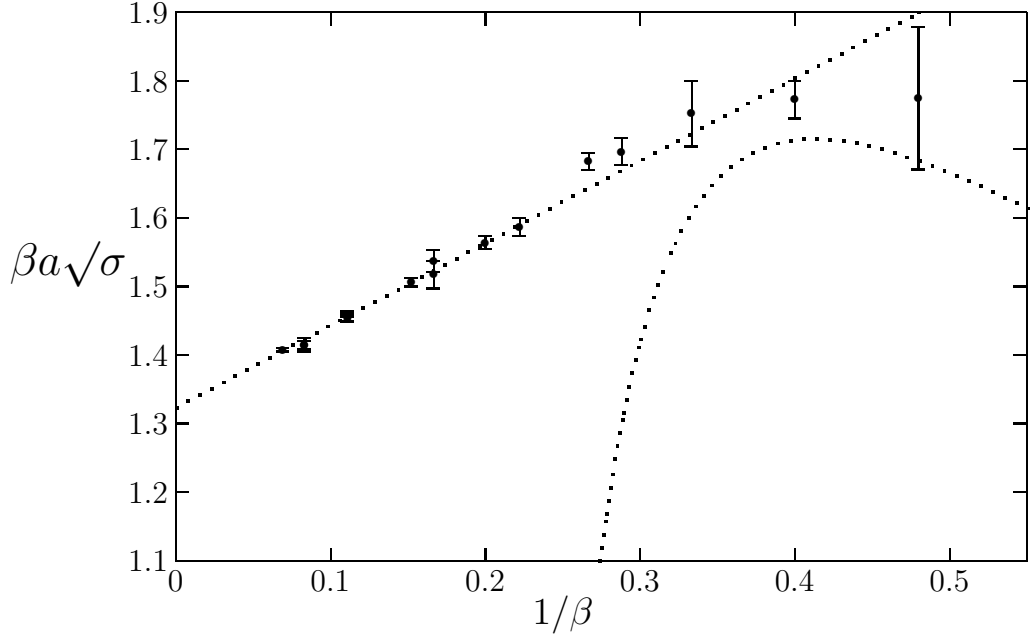}
\end	{center}
\vskip 0.15in
\caption{The values of $\beta a \surd\sigma$ plotted against
$1/\beta$ for SU(2). Also shown is the leading-order strong 
coupling prediction at low $\beta$, and a leading-order
continuum extrapolation at high $\beta$. }
\label{fig_Kbetasu2}
\end 	{figure}

\begin	{figure}[p]
\begin	{center}
\leavevmode
\input	{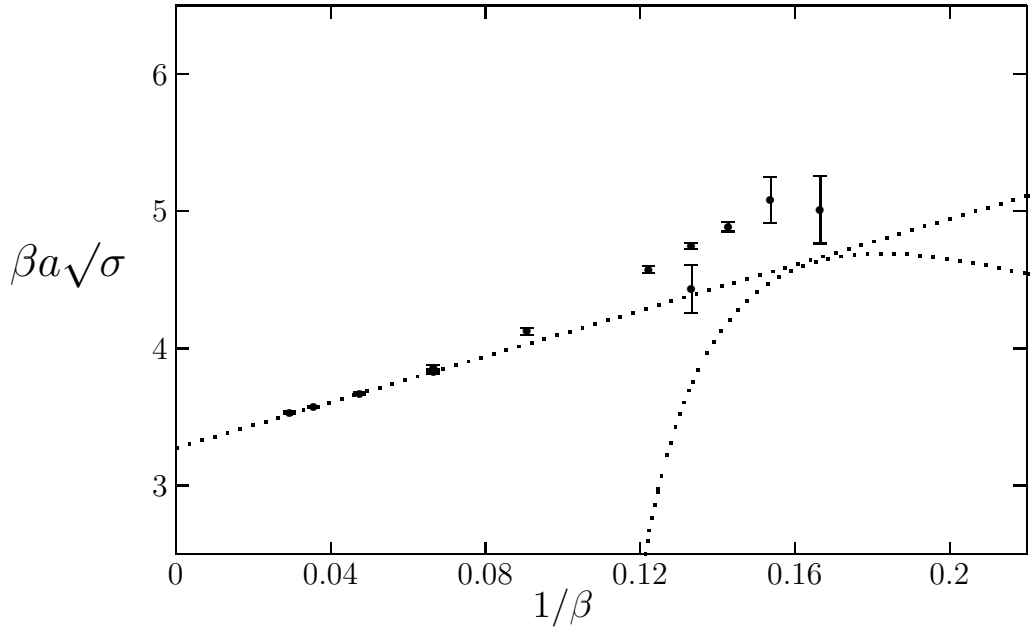}
\end	{center}
\vskip 0.15in
\caption{The values of $\beta a \surd\sigma$ plotted against
$1/\beta$ for SU(3). Also shown is the strong coupling 
prediction to $O(\beta)$ at low $\beta$, and a leading-order
continuum extrapolation at high $\beta$. }
\label{fig_Kbetasu3}
\end 	{figure}

Since we expect 
\begin{equation}
 \beta a \surd \sigma = c_0 + {c_1 \over \beta} +
{c_2 \over \beta^2} + \ldots
\label{APB3}
\end{equation}
for large $\beta$, it is useful to plot the values of
$\beta a \surd \sigma$ against $1/\beta$; we do this in 
Fig.\ref{fig_Kbetasu2} for $SU(2)$ and in 
Fig.\ref{fig_Kbetasu3} for $SU(3)$. For large enough $\beta$
the first two terms in eqn(\ref{APB3}) will dominate and so 
the values
should fall on a straight line as we approach the continuum 
limit. This we observe to be the case. For orientation
we also show in
Fig.\ref{fig_Kbetasu2} and Fig.\ref{fig_Kbetasu3} the
strong coupling predictions for the string tension up
to $O(\beta)$:
\begin{equation}
 a^2 \sigma = -\log({\beta \over 4}) + O(\beta^2) 
\label{APB3b}
\end{equation}
for SU(2), and 
\begin{equation}
 a^2 \sigma = -\log({\beta \over {18}}) - 
{\beta \over {12}} + O(\beta^2) 
\label{APB3c}
\end{equation}
in the case of SU(3). (The extra $O(\beta)$ term in the
case of SU(3) arises because in that case a product of 
two plaquettes can be used just as well as a single
plaquette in tiling the minimal surface spanning the
Wilson loop.) We see that our calculated values of
the string tension extend well into the strong-coupling
regime. In this region an expansion in $1/\beta$, such
as in eqn(\ref{APB3}) should no longer be valid.

The maximum range, more-or-less, over which we can perform linear 
fits with acceptable $\chi^2$, turns out to be  $\beta \geq 4.5$ 
in the case of $SU(2)$ and 
$\beta \geq 15.0$ for the case of $SU(3)$. These fits are
given by
\begin{equation}
\beta a \surd \sigma = 1.324(12) + {1.20(11) \over \beta}
 \ \ \ \ \ \ \ : SU(2) 
\label{APB4}
\end{equation}
and
\begin{equation}
\beta a \surd \sigma = 3.275(24) + {8.35(61) \over \beta}
 \ \ \ \ \ \ \ : SU(3) 
\label{APB5}
\end{equation}
(Note that the errors on the intercept and slope 
are anti-correlated.)

While such a linear extrapolation
is a perfectly acceptable procedure for
extracting the continuum value of the string tension, it must
suffer from some sytematic bias due to the neglect of higher
order terms. These, it is clear from the figures,
are certainly not negligible at intermediate values of
$\beta$. If we include $O(1/\beta^2)$ terms in our fits, we naturally
find larger acceptable ranges for the fits: $\beta \geq 3.0$
for $SU(2)$ and $\beta \geq 6.5$  for $SU(3)$. 
The fits are
\begin{equation}
\beta a \surd \sigma = 1.337(23) + {0.95(38) \over \beta}
+ {1.1(1.3) \over \beta^2}
 \ \ \ \ \ \ \ : SU(2) 
\label{APB6}
\end{equation}
and
\begin{equation}
\beta a \surd \sigma = 3.367(50) + {4.1(1.7) \over \beta}
+ {46.5(11.0) \over \beta^2}
 \ \ \ \ \ \ \ \ : SU(3)  
\label{APB7}
\end{equation}
We observe that in both cases the inclusion of the extra
$O(1/\beta^2)$ term has increased the value of the continuum 
limit by an amount that, while small in absolute units, is 
uncomfortably large when compared to the claimed errors,
especially so in the case of SU(3).
Moreover, in the case of SU(3) the coefficient of
the $1/\beta^2$ correction is so large that the value of this
correction is comparable to that of the $1/\beta$ correction
over much of our range. Under such circumstances one cannot
motivate the neglect of the next, $O(1/\beta^3)$, correction. However 
it is clear, from the large errors in eqn(\ref{APB7}), that our $SU(3)$ 
data will not be able to resolve these higher order terms with 
any useful accuracy. Moreover there is also the danger that
the $O(1/\beta^2)$ correction is being overly biased by the
values of $a\surd\sigma$ in the transition region between
weak and strong coupling, where the very validity of an 
expansion in $1/\beta$ is breaking down.
This leaves us with an intrinsic systematic
error on the SU(3) continuum limit that may well be larger than the 
quoted statistical error.

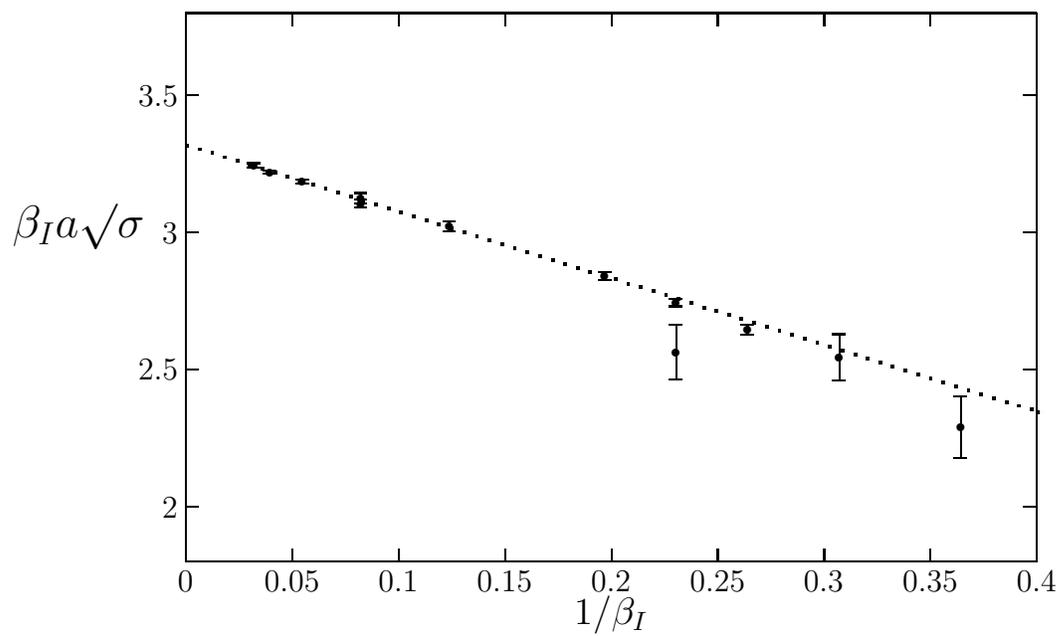
\begin	{figure}[p]
\begin	{center}
\leavevmode
\setlength{\unitlength}{0.240900pt}
\ifx\plotpoint\undefined\newsavebox{\plotpoint}\fi
\sbox{\plotpoint}{\rule[-0.200pt]{0.400pt}{0.400pt}}%
\begin{picture}(1500,900)(0,0)
\font\gnuplot=cmr10 at 12pt
\gnuplot
\sbox{\plotpoint}{\rule[-0.200pt]{0.400pt}{0.400pt}}%
\put(120.0,117.0){\rule[-0.200pt]{4.818pt}{0.400pt}}
\put(108,117){\makebox(0,0)[r]{{$2$}}}
\put(1436.0,117.0){\rule[-0.200pt]{4.818pt}{0.400pt}}
\put(120.0,333.0){\rule[-0.200pt]{4.818pt}{0.400pt}}
\put(108,333){\makebox(0,0)[r]{{$2.5$}}}
\put(1436.0,333.0){\rule[-0.200pt]{4.818pt}{0.400pt}}
\put(120.0,548.0){\rule[-0.200pt]{4.818pt}{0.400pt}}
\put(108,548){\makebox(0,0)[r]{{$3$}}}
\put(1436.0,548.0){\rule[-0.200pt]{4.818pt}{0.400pt}}
\put(120.0,764.0){\rule[-0.200pt]{4.818pt}{0.400pt}}
\put(108,764){\makebox(0,0)[r]{{$3.5$}}}
\put(1436.0,764.0){\rule[-0.200pt]{4.818pt}{0.400pt}}
\put(120.0,31.0){\rule[-0.200pt]{0.400pt}{4.818pt}}
\put(120,19){\makebox(0,0){\shortstack{\\ \\ \\ {$0$}}}}
\put(120.0,873.0){\rule[-0.200pt]{0.400pt}{4.818pt}}
\put(287.0,31.0){\rule[-0.200pt]{0.400pt}{4.818pt}}
\put(287,19){\makebox(0,0){\shortstack{\\ \\ \\ {$0.05$}}}}
\put(287.0,873.0){\rule[-0.200pt]{0.400pt}{4.818pt}}
\put(454.0,31.0){\rule[-0.200pt]{0.400pt}{4.818pt}}
\put(454,19){\makebox(0,0){\shortstack{\\ \\ \\ {$0.1$}}}}
\put(454.0,873.0){\rule[-0.200pt]{0.400pt}{4.818pt}}
\put(621.0,31.0){\rule[-0.200pt]{0.400pt}{4.818pt}}
\put(621,19){\makebox(0,0){\shortstack{\\ \\ \\ {$0.15$}}}}
\put(621.0,873.0){\rule[-0.200pt]{0.400pt}{4.818pt}}
\put(788.0,31.0){\rule[-0.200pt]{0.400pt}{4.818pt}}
\put(788,19){\makebox(0,0){\shortstack{\\ \\ \\ {$0.2$}}}}
\put(788.0,873.0){\rule[-0.200pt]{0.400pt}{4.818pt}}
\put(955.0,31.0){\rule[-0.200pt]{0.400pt}{4.818pt}}
\put(955,19){\makebox(0,0){\shortstack{\\ \\ \\ {$0.25$}}}}
\put(955.0,873.0){\rule[-0.200pt]{0.400pt}{4.818pt}}
\put(1122.0,31.0){\rule[-0.200pt]{0.400pt}{4.818pt}}
\put(1122,19){\makebox(0,0){\shortstack{\\ \\ \\ {$0.3$}}}}
\put(1122.0,873.0){\rule[-0.200pt]{0.400pt}{4.818pt}}
\put(1289.0,31.0){\rule[-0.200pt]{0.400pt}{4.818pt}}
\put(1289,19){\makebox(0,0){\shortstack{\\ \\ \\ {$0.35$}}}}
\put(1289.0,873.0){\rule[-0.200pt]{0.400pt}{4.818pt}}
\put(1456.0,31.0){\rule[-0.200pt]{0.400pt}{4.818pt}}
\put(1456,19){\makebox(0,0){\shortstack{\\ \\ \\ {$0.4$}}}}
\put(1456.0,873.0){\rule[-0.200pt]{0.400pt}{4.818pt}}
\put(120.0,31.0){\rule[-0.200pt]{321.842pt}{0.400pt}}
\put(1456.0,31.0){\rule[-0.200pt]{0.400pt}{207.656pt}}
\put(120.0,893.0){\rule[-0.200pt]{321.842pt}{0.400pt}}
\put(-48,558){\makebox(0,0){{\Large{$\beta_I a\surd\sigma$}}}}
\put(788,-53){\makebox(0,0){{\large{$1/\beta_I$}}}}
\put(120.0,31.0){\rule[-0.200pt]{0.400pt}{207.656pt}}
\put(227,654){\circle*{12}}
\put(252,643){\circle*{12}}
\put(303,628){\circle*{12}}
\put(395,602){\circle*{12}}
\put(395,593){\circle*{12}}
\put(534,558){\circle*{12}}
\put(778,480){\circle*{12}}
\put(890,360){\circle*{12}}
\put(890,438){\circle*{12}}
\put(1002,395){\circle*{12}}
\put(1146,352){\circle*{12}}
\put(1337,243){\circle*{12}}
\put(227.0,650.0){\rule[-0.200pt]{0.400pt}{1.686pt}}
\put(217.0,650.0){\rule[-0.200pt]{4.818pt}{0.400pt}}
\put(217.0,657.0){\rule[-0.200pt]{4.818pt}{0.400pt}}
\put(252.0,641.0){\rule[-0.200pt]{0.400pt}{0.964pt}}
\put(242.0,641.0){\rule[-0.200pt]{4.818pt}{0.400pt}}
\put(242.0,645.0){\rule[-0.200pt]{4.818pt}{0.400pt}}
\put(303.0,625.0){\rule[-0.200pt]{0.400pt}{1.445pt}}
\put(293.0,625.0){\rule[-0.200pt]{4.818pt}{0.400pt}}
\put(293.0,631.0){\rule[-0.200pt]{4.818pt}{0.400pt}}
\put(395.0,594.0){\rule[-0.200pt]{0.400pt}{3.854pt}}
\put(385.0,594.0){\rule[-0.200pt]{4.818pt}{0.400pt}}
\put(385.0,610.0){\rule[-0.200pt]{4.818pt}{0.400pt}}
\put(395.0,587.0){\rule[-0.200pt]{0.400pt}{3.132pt}}
\put(385.0,587.0){\rule[-0.200pt]{4.818pt}{0.400pt}}
\put(385.0,600.0){\rule[-0.200pt]{4.818pt}{0.400pt}}
\put(534.0,550.0){\rule[-0.200pt]{0.400pt}{3.854pt}}
\put(524.0,550.0){\rule[-0.200pt]{4.818pt}{0.400pt}}
\put(524.0,566.0){\rule[-0.200pt]{4.818pt}{0.400pt}}
\put(778.0,473.0){\rule[-0.200pt]{0.400pt}{3.132pt}}
\put(768.0,473.0){\rule[-0.200pt]{4.818pt}{0.400pt}}
\put(768.0,486.0){\rule[-0.200pt]{4.818pt}{0.400pt}}
\put(890.0,317.0){\rule[-0.200pt]{0.400pt}{20.717pt}}
\put(880.0,317.0){\rule[-0.200pt]{4.818pt}{0.400pt}}
\put(880.0,403.0){\rule[-0.200pt]{4.818pt}{0.400pt}}
\put(890.0,432.0){\rule[-0.200pt]{0.400pt}{2.891pt}}
\put(880.0,432.0){\rule[-0.200pt]{4.818pt}{0.400pt}}
\put(880.0,444.0){\rule[-0.200pt]{4.818pt}{0.400pt}}
\put(1002.0,387.0){\rule[-0.200pt]{0.400pt}{3.854pt}}
\put(992.0,387.0){\rule[-0.200pt]{4.818pt}{0.400pt}}
\put(992.0,403.0){\rule[-0.200pt]{4.818pt}{0.400pt}}
\put(1146.0,315.0){\rule[-0.200pt]{0.400pt}{17.586pt}}
\put(1136.0,315.0){\rule[-0.200pt]{4.818pt}{0.400pt}}
\put(1136.0,388.0){\rule[-0.200pt]{4.818pt}{0.400pt}}
\put(1337.0,194.0){\rule[-0.200pt]{0.400pt}{23.367pt}}
\put(1327.0,194.0){\rule[-0.200pt]{4.818pt}{0.400pt}}
\put(1327.0,291.0){\rule[-0.200pt]{4.818pt}{0.400pt}}
\sbox{\plotpoint}{\rule[-0.500pt]{1.000pt}{1.000pt}}%
\put(120,685){\usebox{\plotpoint}}
\put(120.00,685.00){\usebox{\plotpoint}}
\put(139.88,679.03){\usebox{\plotpoint}}
\put(159.76,673.07){\usebox{\plotpoint}}
\multiput(160,673)(19.546,-6.981){0}{\usebox{\plotpoint}}
\put(179.39,666.34){\usebox{\plotpoint}}
\put(199.30,660.49){\usebox{\plotpoint}}
\multiput(201,660)(19.838,-6.104){0}{\usebox{\plotpoint}}
\put(219.07,654.19){\usebox{\plotpoint}}
\put(238.78,647.68){\usebox{\plotpoint}}
\multiput(241,647)(19.957,-5.702){0}{\usebox{\plotpoint}}
\put(258.70,641.86){\usebox{\plotpoint}}
\put(278.38,635.29){\usebox{\plotpoint}}
\multiput(282,634)(19.838,-6.104){0}{\usebox{\plotpoint}}
\put(298.18,629.09){\usebox{\plotpoint}}
\put(318.09,623.20){\usebox{\plotpoint}}
\multiput(322,622)(19.957,-5.702){0}{\usebox{\plotpoint}}
\put(337.96,617.25){\usebox{\plotpoint}}
\put(357.58,610.55){\usebox{\plotpoint}}
\multiput(363,609)(19.838,-6.104){0}{\usebox{\plotpoint}}
\put(377.46,604.58){\usebox{\plotpoint}}
\put(397.20,598.23){\usebox{\plotpoint}}
\put(416.98,592.00){\usebox{\plotpoint}}
\multiput(417,592)(19.838,-6.104){0}{\usebox{\plotpoint}}
\put(436.86,586.04){\usebox{\plotpoint}}
\put(456.44,579.21){\usebox{\plotpoint}}
\multiput(457,579)(19.957,-5.702){0}{\usebox{\plotpoint}}
\put(476.35,573.35){\usebox{\plotpoint}}
\put(496.26,567.50){\usebox{\plotpoint}}
\multiput(498,567)(19.838,-6.104){0}{\usebox{\plotpoint}}
\put(516.04,561.20){\usebox{\plotpoint}}
\put(535.74,554.70){\usebox{\plotpoint}}
\multiput(538,554)(19.957,-5.702){0}{\usebox{\plotpoint}}
\put(555.66,548.87){\usebox{\plotpoint}}
\put(575.34,542.31){\usebox{\plotpoint}}
\multiput(579,541)(19.838,-6.104){0}{\usebox{\plotpoint}}
\put(595.15,536.10){\usebox{\plotpoint}}
\put(615.05,530.22){\usebox{\plotpoint}}
\multiput(619,529)(19.546,-6.981){0}{\usebox{\plotpoint}}
\put(634.68,523.48){\usebox{\plotpoint}}
\put(654.57,517.55){\usebox{\plotpoint}}
\multiput(660,516)(19.838,-6.104){0}{\usebox{\plotpoint}}
\put(674.45,511.59){\usebox{\plotpoint}}
\put(694.19,505.24){\usebox{\plotpoint}}
\put(713.97,499.01){\usebox{\plotpoint}}
\multiput(714,499)(19.838,-6.104){0}{\usebox{\plotpoint}}
\put(733.85,493.04){\usebox{\plotpoint}}
\put(753.43,486.22){\usebox{\plotpoint}}
\multiput(754,486)(19.957,-5.702){0}{\usebox{\plotpoint}}
\put(773.33,480.36){\usebox{\plotpoint}}
\put(793.25,474.50){\usebox{\plotpoint}}
\multiput(795,474)(19.372,-7.451){0}{\usebox{\plotpoint}}
\put(812.81,467.63){\usebox{\plotpoint}}
\put(832.70,461.71){\usebox{\plotpoint}}
\multiput(835,461)(19.957,-5.702){0}{\usebox{\plotpoint}}
\put(852.62,455.89){\usebox{\plotpoint}}
\put(872.31,449.32){\usebox{\plotpoint}}
\multiput(876,448)(19.838,-6.104){0}{\usebox{\plotpoint}}
\put(892.11,443.11){\usebox{\plotpoint}}
\put(912.01,437.23){\usebox{\plotpoint}}
\multiput(916,436)(19.546,-6.981){0}{\usebox{\plotpoint}}
\put(931.64,430.50){\usebox{\plotpoint}}
\put(951.53,424.56){\usebox{\plotpoint}}
\multiput(957,423)(19.838,-6.104){0}{\usebox{\plotpoint}}
\put(971.38,418.51){\usebox{\plotpoint}}
\put(991.03,411.84){\usebox{\plotpoint}}
\put(1010.95,406.01){\usebox{\plotpoint}}
\multiput(1011,406)(19.838,-6.104){0}{\usebox{\plotpoint}}
\put(1030.83,400.05){\usebox{\plotpoint}}
\put(1050.41,393.23){\usebox{\plotpoint}}
\multiput(1051,393)(19.957,-5.702){0}{\usebox{\plotpoint}}
\put(1070.32,387.36){\usebox{\plotpoint}}
\put(1090.23,381.51){\usebox{\plotpoint}}
\multiput(1092,381)(19.372,-7.451){0}{\usebox{\plotpoint}}
\put(1109.79,374.63){\usebox{\plotpoint}}
\put(1129.69,368.71){\usebox{\plotpoint}}
\multiput(1132,368)(19.957,-5.702){0}{\usebox{\plotpoint}}
\put(1149.52,362.65){\usebox{\plotpoint}}
\put(1169.19,356.09){\usebox{\plotpoint}}
\multiput(1173,355)(19.838,-6.104){0}{\usebox{\plotpoint}}
\put(1189.07,350.12){\usebox{\plotpoint}}
\put(1208.76,343.63){\usebox{\plotpoint}}
\multiput(1213,342)(19.957,-5.702){0}{\usebox{\plotpoint}}
\put(1228.58,337.51){\usebox{\plotpoint}}
\put(1248.47,331.58){\usebox{\plotpoint}}
\multiput(1254,330)(19.838,-6.104){0}{\usebox{\plotpoint}}
\put(1268.32,325.53){\usebox{\plotpoint}}
\put(1287.97,318.85){\usebox{\plotpoint}}
\put(1307.89,313.03){\usebox{\plotpoint}}
\multiput(1308,313)(19.838,-6.104){0}{\usebox{\plotpoint}}
\put(1327.63,306.63){\usebox{\plotpoint}}
\put(1347.36,300.20){\usebox{\plotpoint}}
\multiput(1348,300)(19.957,-5.702){0}{\usebox{\plotpoint}}
\put(1367.28,294.38){\usebox{\plotpoint}}
\put(1386.94,287.74){\usebox{\plotpoint}}
\multiput(1389,287)(19.838,-6.104){0}{\usebox{\plotpoint}}
\put(1406.78,281.64){\usebox{\plotpoint}}
\put(1426.67,275.72){\usebox{\plotpoint}}
\multiput(1429,275)(19.957,-5.702){0}{\usebox{\plotpoint}}
\put(1446.51,269.65){\usebox{\plotpoint}}
\put(1456,266){\usebox{\plotpoint}}
\end{picture}
\end	{center}
\vskip 0.15in
\caption{As in Fig.\ref{fig_Kbetasu3} but using the mean-field
improved coupling, $\beta_I$, in place of $\beta$.}
\label{fig_KbetaIsu3}
\end 	{figure}

The lattice corrections in eqn(\ref{APB3}) are precisely what
the use of a better coupling should improve - by reducing
their coefficients. How well does that work here? If we use
eqn(\ref{APB2}) to define $\beta_I$ we can plot $\beta_I a \surd \sigma$
against $1/\beta_I$ as in 
Fig.\ref{fig_KbetaIsu2} and Fig.\ref{fig_KbetaIsu3}.
It is immediately apparent from a comparison with 
Fig.\ref{fig_Kbetasu2} and Fig.\ref{fig_Kbetasu3}.
that in terms of the `improved' 
coupling the higher order lattice corrections are dramatically
reduced. More quantitatively, if we perform fits as before
but with $\beta_I$ replacing $\beta$, we obtain
the following results. In the case of $SU(2)$ we obtain
good fits with just the leading $O(1/\beta_I)$ correction
for the much larger range $\beta \geq 3.0$ while for $SU(3)$
excellent fits are possible for  $\beta \geq 8.175$
(and reasonable ones all the way down to $\beta=6.0$).
Moreover these fits
\begin{equation}
\beta_I a \surd \sigma = 1.341(7) - {0.421(50) \over \beta_I}
 \ \ \ \ \ \ \ : SU(2) 
\label{APB8}
\end{equation}
\begin{equation}
\beta_I a \surd \sigma = 3.318(12) - {2.43(22) \over \beta_I}
 \ \ \ \ \ \ \ : SU(3)
\label{APB9}
\end{equation}
display much smaller corrections to the leading asymptotic
terms than was the case in eqns(\ref{APB4}) and (\ref{APB5}). 
Since these fits are so good, there is no real reason
to include higher order corrections. However if we do so then
we obtain
\begin{equation}
\beta_I a \surd \sigma = 1.336(9) - {0.35(9) \over \beta_I}
- {0.18(15) \over {\beta_I}^2}
 \ \ \ \ \ \ \ : SU(2) 
\label{APB10}
\end{equation}
\begin{equation}
\beta_I a \surd \sigma = 3.323(28) - {2.57(80) \over \beta_I}
+ {0.7(3.7) \over {\beta_I}^2}
 \ \ \ \ \ \ \ : SU(3) 
\label{APB11}
\end{equation}
We observe that the coefficients of the higher order 
terms are small: so there is no reason to worry about the
next correction. A second and related observation is that the
asymptotic values are little changed with the inclusion of the
$O(1/{\beta_I}^2)$ correction - in contrast to what happened
when we used $\beta$ as our expansion parameter. Indeed even the
coefficients of the $1/\beta_I$ terms are insensitive to the
inclusion of a higher order term. All this represents a
substantial improvement in the perturbative control of the
continuum limit.

The fact that we can extrapolate $\beta_I a\surd\sigma$
with fits involving just two parameters, means that
we do not need to perform calculations at more than four 
values of $\beta$ in the case of SU(4) and SU(5). This
represents a substantial saving in computational effort.

From fits such as the above we can extract the continuum
mass ratios shown in eqn(\ref{C7}). We remark that it is both
because we are in 3 dimensions, where the bare coupling decreases
linearly with the scale $a$ rather than just logarithmically,
and because of the extent and accuracy of our lattice calculations, 
that it is possible to perform reasonably accurate extrapolations 
to the continuum limit even with the `bad' lattice bare coupling. 
This has enabled us to quantify, in a way that is
not yet possible in 4 dimensions, how much the
mean-field improved coupling actually improves the approach to
the continuum limit of the lattice spacing, $a$.

\vspace{1.0cm}

\noindent{\Large{\bf Appendix D : Calculations
with an asymmetric lattice action}} 

\vspace{0.25cm}

In this Appendix we present an SU(2) calculation of the mass
spectrum on lattices with timelike and spacelike lattice
spacings related by $a_t \simeq a_s/4$. As discussed in 
Section 5.1 the primary purpose of this study is to check
explicitly that our criteria for which effective masses 
adequately reflect the actual masses, are in fact accurate.
A second reason is that this, being a calculation with
a different action to the one we have used so far, will
provide us with some test of universality. We
shall first discuss some of the features that are
peculiar to such calculations. We then present our
results.

\vspace{0.25cm}

\noindent{\large{\bf Preliminaries}}

\vspace{0.25cm}

To allow different spatial and temporal lattice spacings
we use the action in eqn(\ref{B3}). What is the relation
between our choice of $\beta_s,\beta_t$ and the lattice
asymmetry? Suppose we are aiming for a particular ratio
\begin{equation}
r = {{a_t} \over {a_s}}.
\label{APD1}
\end{equation}
In the limit $a_s,a_t \to 0$ we have 
\begin{eqnarray}
\{1-{1\over N_c}ReTr U_{p_s}\} & \to & a_s^4 {1\over{2N_c}} Tr F^2 
\nonumber \\
\{1-{1\over N_c}ReTr U_{p_t}\} & \to & a_s^2 a_t^2 {1\over{2N_c}} Tr F^2
\nonumber \\ 
\beta S & \to & {1 \over {g^2}} \int d^2 x dt {1\over 2} Tr F^2
\label{APD2}
\end{eqnarray}
where $F^2$ is the continuum field strength squared.
Since the integration measure gives a factor $a_s^2 a_t$
when discretised, we see that the choice
\begin{equation}
\beta_s = r \beta \ \ ; \ \ \ 
\beta_t = {1 \over r} \beta \ \ ; \ \ \
\beta = {{2N_c} \over {a_s g^2}}
\label{APD3}
\end{equation}
is what is needed, at least at the $classical$ level, to achieve
the asymmetry, $r$, defined in eqn(\ref{APD1}).

In practice there will be quantum corrections to these classical
relations. Three related questions immediately arise.

{\noindent}(a) What do we need to know about $a_s$ and $a_t$?

{\noindent}(b) In a given simulation, how can we calculate $a_s$ and
$a_t$ directly?

{\noindent}(c) Can we easily `improve' upon the 
relations in eqn(\ref{APD3})?

Before considering each of these questions in turn, we
need to remark that the classical relations in 
eqn(\ref{APD3}) should remain a roughly reliable 
guide in the full theory. This is because our theory
is super-renormalisable and this is in contrast to the 
situation in 4 dimensions. Nonetheless we do expect
significant corrections, as we saw when considering
the $\beta$ dependence of $\beta a \surd \sigma$.
We saw there (see Appendix C) that the corrections
to the classical relation $\beta = 2N_c/ag^2$ are quite
large and can be drastically decreased by the use
of a mean-field improved coupling.   

In the rest of this Appendix we shall assume that
the asymmetric lattice has been chosen so that
$a_t \ll a_s$. This means that we shall systematically
ignore any $O(a_t^2)$ corrections as compared to
to ones that are  $O(a_s^2)$.

\vspace{0.10cm}

{\noindent}{\bf (a) How well do we need to know $a_s$ and $a_t$?}

\vspace{0.05cm}

The first thing we need to establish is whether
we actually need to know $a_s$ and $a_t$ any more
accurately than we already know them through using
eqn(\ref{APD3}).

If we just wanted to calculate some lattice mass ratios, 
$m_i/m_0$, then these could be obtained without knowing 
the lattice spacing at all: our usual procedure would give us
estimates of $a_tm_i$ and the lattice spacing then cancels
in the ratio. However if we want to extrapolate to the
continuum limit then the leading correction will
be $O(a_s^2)$, assuming $a_t \ll a_s$ as will be the
case here, and so we need to know $a_s m$ for some mass
$m$ in order to provide the correction term in the
analogue of eqn(\ref{D2}). This can, however, be finessed by
using eqn(\ref{D3}) instead; although 
$1/\beta^2 \simeq a^2 g^4/4N_c^2 + O(a^3)$, 
the correction has a small enough coefficient that
it should not significantly degrade the accuracy of 
our extrapolations. 

Of course, we also need to control finite volume effects 
in an efficient way -- that is to say, more efficient than
doing a detailed finite volume study at each coupling.
So we need to be able to compare the lattice size,
$L \equiv a_sL_s$, at different couplings. For this
we need to know $a_s$ to a reasonable approximation;
great accuracy is not needed because we usually 
include a margin of safety in our choice of the volume.

There is however at least one place where we do need 
accurate values for the lattice spacings. This is
in our calculation of the string tension.
Our usual procedure is to calculate the mass, 
$a_t m_P(L)$, of a flux loop of length $L=a_sL_s$ that 
winds around the spatial torus. This mass, $a_t m_P(L)$, 
can be written, using eqn(\ref{C1}), as
\begin{eqnarray}
a_t m_P(L) & = & a_t \times L \sigma(L) 
\nonumber \\
& = & a_t L \Bigl( \sigma(\infty) - 
{{\pi}\over{6L^2}} \Bigr)
\nonumber \\
& = & a_t a_s L_s \sigma - 
{{\pi}\over{6L_s}}{a_t \over a_s} 
\label{APD4}
\end{eqnarray}
Clearly we need to know $r=a_t/a_s$ very accurately
if we are to be able to calculate $a_t\surd\sigma$ with
the accuracy we are used to. A similar situation arises
if we calculate potentials using Wilson loops.

\vspace{0.10cm}

{\noindent}{\bf (b) How do we calculate $a_t/a_s$ directly?}

\vspace{0.05cm}

There are two obvious methods that we can use to
calculate $r=a_t/a_s$. The first involves calculating
the energies of states with non-zero momenta. Suppose
we have a particle of mass $m$. The allowed 
momenta are $a_s{\vec p} = (2\pi n_x/L_s,2\pi n_y/L_s)$
and the corresponding energies that we obtain from
our correlators may be written as $a_t E(p)$. For small
momenta we expect the continuum dispersion relation,
$E^2=p^2+m^2$, to be accurately satisfied. (We have 
explicitly seen that this is so on the symmetric lattices
that we have used in our main calculations in this paper.)
So we expect to have
\begin{eqnarray}
a^2_t E^2(p) & = & a_t^2 (p^2 + m^2)   
\nonumber \\
& = & \Bigl({a_t \over a_s}\Bigr)^2 \Bigl({{2\pi}\over{L_s}}\Bigr)^2 
(n_x^2 + n_y^2) + (a_t m)^2 
\label{APD5}
\end{eqnarray}
Therefore, from our calculated values of $a_t E(p)$ and 
$a_t m$ we can obtain, using eqn(\ref{APD5}), a value
for $a_t/a_s$. 

Our second method is even more direct. Normally we
calculate correlators in the $t$-direction. We could
instead calculate our correlators in, say, the $x$-direction.
In that case our space would be $(y,t)$ in place of $(x,y)$. As 
long as $a_tL_t$ is large, as it will always be, this new
spatial volume will also be large and we can assume
there are no finite volume effects. Thus we obtain
the same mass in both calculations, up to lattice spacing
corrections. That is to say, we obtain $a_t m$ from our
$t$-correlators and  $a_s m$ from our $x$-correlators.
Equating the two we obtain $a_t/a_s$.

We can combine the above two methods by calculating
glueball correlation functions in the $x$-direction
with non-zero momenta in either the $y$ or $t$ directions.
Comparing the energies of such states gives us another
direct estimate of $a_t/a_s$. If we include a range of
momenta one can attempt to tune this ratio so that
the dispersion relations in $p_t$ and $p_y$ coincide.
This allows us to fix the asymmetry without assuming
the continuum dispersion relation.

In practice, we shall not use this last method, and
we shall only consider the lowest two momenta in
applying our first method. While this reduces the
precision with which we can estimate the ratio
$a_t/a_s$, it suffices for our purposes.

Both of the above methods will suffer from lattice spacing
corrections. The continuum dispersion relation will 
only be valid up to corrections of order $(a_sp)^2$ and
the eigenstates of the transfer matrices defined on the
$x,y$ and $y,t$ spatial tori will differ by order $a_s^2$
corrections because the latter torus has a lattice spacing
$a_t$ rather than $a_s$ in one of the two directions.
However as long as we are consistent in the method used
to estimate $a_t/a_s$ we can absorb this correction into the
correction term used in taking the continuum limit.

Before turning to some explicit calculations of the above kind
it is worth pointing out that
although the second method described above seems more direct,
it is in practice more awkward to implement. The reason is that
we need to produce blocked link matrices in order to
have useful operators, and this has to be carefully
tailored in the case where one spatial lattice spacing
is very different from the other. And this is in addition
to the fact that using two different spatial planes means
producing two sets of blocked links. For these reasons
our calculations using the second method will be on
only a subset of our lattices.

\vspace{0.10cm}

{\noindent}{\bf (c) Can we `improve' upon the estimate of $r$?}

\vspace{0.05cm}

We have seen in our previous calculations that we get
much smaller corrections to the limit
$\beta a \surd\sigma \to 2N_c \surd\sigma/g^2$
as $a \to 0$, if we use a mean-field improved inverse 
coupling, $\beta_I$, in place of $\beta$. One might hope
that a similar approach with an asymmetric lattice action
would improve our control over the value of $r$.
The straightforward implementation of this idea in the
context of the action in eqn(\ref{B3}) would be to 
define `improved' values of $r$ and $\beta$ by
\begin{eqnarray}
r_I \beta_I & = & r \beta 
\langle{1\over N_c}Tr U_{p_s}\rangle 
\nonumber \\
{1 \over {r_I}} \beta_I & = & {1 \over r} 
\beta \langle{1\over N_c}Tr U_{p_t}\rangle
\label{APD6}
\end{eqnarray}
This is in the spirit of the approach suggested in
\cite{asymnew}
although there one effectively replaces 
${1\over N_c}\langle Tr U_{p_t}\rangle$
by unity. We shall calculate $r_I$ below and explicitly check 
how much of an improvement one really obtains.

\newpage

\vspace{0.25cm}

\noindent{\large{\bf The calculation}}

\vspace{0.25cm}

We perform calculations at $\beta=4.0,5.3$ and $8.0$. 
In all three cases we choose $r=0.25$ in the action, as
given in eqn(\ref{B3}). The lattice sizes are $12^2 60$,
$16^2 64$ and $24^2 96$. If the classical relations
in  eqn(\ref{APD3}) were valid, then the value of 
$a_s$ would be exactly what we obtained at the corresponding
values of $\beta$ on symmetric lattices. The reader will
note that our lattice sizes are somewhat larger than 
would be necessary if this were the case; this is to give
us some margin in case the quantum corrections to these
relations are significant.

In addition to these calculations we also perform
calculations on somewhat smaller lattices,
$8^2 60$, $12^2 64$ and $16^2 96$ respectively.
It is on these lattices that we calculate correlators
in both $t$ and $x$ directions. We shall only calculate
the mass of the periodic flux loop, $m_P(L)$, on these lattices.
Smaller lattices are preferable for this purpose because
$m_P(L)$ will be smaller and so we will have more
accurate values. (Obviously this is only important for
the correlators in the $x$-direction where $a_s m_P$
will not be small.)

\begin	{figure}[p]
\begin	{center}
\leavevmode
\input	{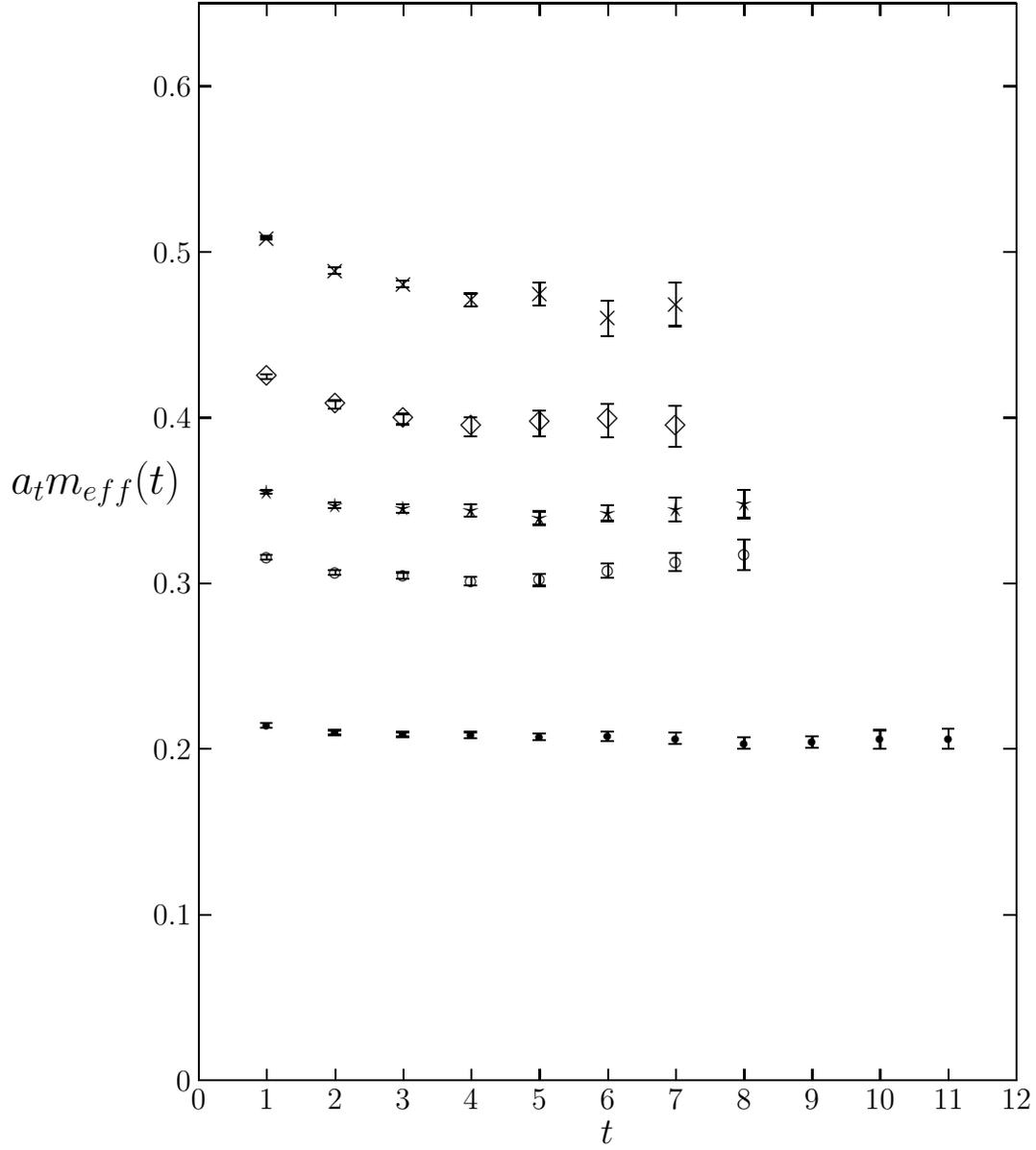}
\end	{center}
\vskip 0.15in
\caption{The effective masses obtained on a $24^2 96$
lattice at $\beta=8$ in SU(2), with a very small
temporal lattice spacing: $a_t \sim 0.25 a_s$.
States are  the $0^{++}$($\bullet$),
the $0^{++\ast}$($\circ$), the $2^{++}$($\star$), the 
$2^{++\ast}$($\diamond$) and the  $1^{++}$($\times$).}
\label{fig_meffasym}
\end 	{figure}

Our original motivation for performing such calculations
was to have a finer resolution on the effective mass
plot, so as to see whether the typical heavier mass could
really be extracted from its effective masses between
$t=a$ to $t=2a$. Since we have $a_t \sim a_s/4$,
this question becomes: have the effective masses in
the region $4a_t \leq t \leq 8a_t$ already reached 
their asymptotic plateau? So without further ado
we plot the effective masses from the $24^2 96$
lattice in Fig.\ref{fig_meffasym}. Note that if we wished to
obtain such a small lattice spacing in our usual
symmetric lattice calculation, we would have had to do it on
a $96^3$ lattice at $\beta \sim 30$: a daunting prospect!
We see from Fig.\ref{fig_meffasym} that there is reasonably
good evidence in all cases that an effective mass extracted
over the interval $4a_t \leq t \leq 8a_t$ provides an
unbiased estimate of the asymptotic mass. We infer from this
that extracting masses from the range $t=a$ to $t=2a$ on
symmetric lattices in the neighbourhood of $\beta=8$,
where the heavier states are in the noise for $t > 2a$, 
is in fact justified. Our $\beta=5.3$ calculations
also support this way of calculating masses (although 
with less precision) thus reassuring us that the
estimates we have used in this paper are indeed 
unbiased over the whole range of $\beta$ relevant to 
our continuum extrapolations.

Since these lattice actions are different from the
symmetric ones that we have used in the body of the
paper -- indeed, as we have already remarked, 
one may regard them as being close
to the Hamiltonian limit -- it is interesting to
extract a continuum mass spectrum from them, so
testing universality, to some extent. 

In Table~\ref{table_masym} we list the masses we
have extracted at our three value of $\beta$.
They are all in units of $a_t$. In addition to the
glueball masses we also list the mass of the flux
loop that winds around the spatial torus. As discussed
above, we need to know the value of $a_t/a_s$ in order
to extract $a_t \surd\sigma$ (or  $a_s \surd\sigma$).
And we need to know $a_t \surd\sigma$ if we are to
calculate $m_G/\surd\sigma$ for comparison with our
previous calculations. We therefore turn to this next.

In order not to confuse different quantities, we shall
continue to use $r$ for the parameter in the action,
and we shall choose $r=0.25$ here. Classically, but only
classically, we know that $r=a_t/a_s$. The `true' value of 
$a_t/a_s$ is the one that we explicitly calculate using
the methods described earlier in the Appendix: this we
shall either label $r_{meas}$ or simply refer to as
$a_t/a_s$. Finally there is $r_I$ as defined by eqn(\ref{APD6}).

We calculate $a_t/a_s$ by comparing the flux loop energy
as calculated from two momenta, $p_1$ and $p_2$, and then 
using eqn(\ref{APD5}). We do so using the lowest three
momenta which we shall refer to as $p_i = 0,1,2$ for 
convenience. This we shall do on both our larger and
smaller lattices at each value of $\beta$. Note that the
smaller $p$ the more reliable will be the continuum
energy-momentum dispersion relation. Since the lowest
momenta decrease as the lattice size increases, the values 
obtained on the larger lattices will have smaller systematic 
errors although larger statistical errors. The values thus
obtained are listed in Table~\ref{table_rasym}. We also show
there the values we get for $a_t m_P/a_s m_P$; these have
only been calculated for the smaller lattices. From these
results we infer the following values of the asymmetry:
\begin{equation}
{ {a_t} \over {a_s} } =
\left\{ \begin{array}{ll}
0.245(5)   & \ \ \ \ \ \beta=8.0 \\
0.237(6)   & \ \ \ \ \ \beta=5.3 \\
0.230(10)  & \ \ \ \ \ \beta=4.0
\end{array}
\right.
\label{APD7}
\end{equation}
The deviation from the classical value, $a_t/a_s=r=0.25$, is
not large. We also note that although the `improved' value, $r_I$,
is shifted in the right direction, it overshoots so that it
is no closer to the `true' value than is $r$. This is even
more so if one replaces the time-like plaquettes by unity
in eqn(\ref{APD6}). Thus it seems that the most naive
mean-field `improvement' is not an improvement here. 

Using in  eqn(\ref{APD4}) the values
of $a_t/a_s$ in eqn(\ref{APD7}) and the values of $a_t m_P$
in Table~\ref{table_masym}, we obtain
\begin{equation}
 a_t \surd\sigma =
\left\{ \begin{array}{ll}
0.04408(56)    & \ \ \ \ \ \beta=8.0 \\
0.06806(102)   & \ \ \ \ \ \beta=5.3 \\
0.0919(21)     & \ \ \ \ \ \beta=4.0
\end{array}
\right.
\label{APD8}
\end{equation}
The uncertainty in $a_t/a_s$ has roughly doubled the
error on $a_t \surd\sigma$; thus it is no longer the
most accurately calculated quantity (as it was in case
of the symmetric action) and if we were to calculate
mass ratios from scratch we might prefer to use
the scalar glueball mass as our basic scale. 
We now extrapolate these values (multiplied by $\beta$) 
to the continuum limit, using a $1/\beta$ correction 
just as in eqn(\ref{C4}). The fit has a very good 
confidence level and gives us the continuum value
of $4a_t \surd\sigma/a_s g^2$. In the continuum
limit $a_t/a_s = r =0.25$ and so we finally obtain
\begin{equation}
 {{\surd\sigma} \over {g^2}} = 0.3375(130).
\label{APD9}
\end{equation}
This is certainly consistent with the value of $0.3353(18)$
in eqn(\ref{C7}) which was obtained with $r=1$, but
the error is much larger. In large part this is just
because the present calculation is a much smaller one.
But in some part it is due to our uncertainty in
the value of $a_t/a_s$. Without this uncertainty
our error in eqn(\ref{APD9}) would have been smaller
by about a factor of 1.5. We have performed other
continuum extrapolations as well. If we use the classical
value $a_t/a_s=r=0.25$ in our calculations, then we
again obtain a good fit, this time with a continuum 
value of 
\begin{equation}
 {{\surd\sigma} \over {g^2}} = 0.3288(50).
\label{APD10}
\end{equation}
This is consistent with our previous values, as it should be
because the corrections to $r$ can be absorbed into the
$1/\beta$ correction. Indeed we find that the coefficient 
of the $1/\beta$ term is larger in the latter case: 
$\simeq 0.22(3)$ versus $\simeq 0.12(8)$ when we use 
the values in eqn(\ref{APD7}). We have performed other
extrapolations as well: the errors vary but the values are
consistent with each other. It is worth remarking that the 
fit leading to eqn(\ref{APD9}) has a much smaller correction
than one finds in the symmetric case when using $\beta$;
indeed it is about the same size as one obtains using the
improved coupling, $\beta_I$. This suggests that lattice
corrections are smaller on very asymmetric lattices, and
perhaps explains why we did not gain anything from using
the mean-field recipe.

We can now take the values of $a_t m_G$ in Table~\ref{table_masym},
the values of  $a_t \surd\sigma$ in eqn(\ref{APD8}), form 
ratios, and extrapolate to the continuum limit using
eqn(\ref{D3}). We obtain the continuum mass ratios shown
in Table~\ref{table_compasym}. We also show there the
symmetric lattice values that have been obtained elsewhere
in this paper. We observe that they are consistent within errors.
We note that if we extrapolate with an $O(a^2\sigma)$ correction 
as in eqn(\ref{D2}), we obtain almost identical results.
This is also the case if we use string tensions calculated using 
$a_t/a_s=0.25$, except that the fits tend to be significantly 
worse -- as one would expect if this involved an error that
was really $O(1/\beta)$ rather than  $O(1/\beta^2)$.

We observe in Table~\ref{table_compasym} that the errors
on the lighter masses are larger in the asymmetric case.
This is no surprise since our symmetric calculations
are $very$ much larger. What is striking is that for the
heaviest masses, such as the $1^{\pm +}$, the asymmetric 
errors are actually smaller. This displays the power of such 
calculations for determining the masses of heavier states.

Two final asides on the spectrum. For reasons we do not
entirely understand, we seem to have no problem in obtaining
a set of excited $2^{\pm +}$ masses that continue well
to $a=0$ and, indeed, are consistent with what one finds
for higher groups -- see Table~\ref{table_mcont}. This
is in contrast to the case of a symmetric lattice action.
We also note in Table~\ref{table_masym} that the mass of
the $1^{-+}$ is larger than that of the  $1^{-+\ast}$.
Our ordering of the states is determined by the effective
mass at $t=a$, which in this case seems to be unreliable.
This is undoubtedly related to the peculiarities in the 
$J=1$ sector that we have previously noted during our
symmetric action calculations. Obviously it is hard to
argue with using the lower mass for the ground state 
$1^{-+}$, and if we do so we obtain the value in square
brackets in  Table~\ref{table_compasym}. It is amusing
that this value fits better with parity doubling and with 
the values obtained for the higher groups.

\vfill
\eject

\noindent{\Large{\bf Appendix E :
The U(1) mass spectrum}}

\vspace{0.25cm}

In this Appendix we calculate the mass spectrum in the
D=2+1 U(1) theory. U(1) is as far as one can get from
U($\infty$), so this should provide us with a useful
contrast to the SU($N_c$) mass spectra which we
have calculated in this paper.

Calculating the U(1) spectrum might seem pointless;
the continuum limit should be a theory of free,
non-interacting photons. While this would certainly
provide a contrast to our SU($N_c$) spectra, it would 
hardly be very illuminating. 

Although the continuum limit is indeed trivial, in the sense 
that there will be no bound state whose mass is finite 
in units of the mass scale $g^2$, there is nonetheless
interesting dynamics at finite values of the lattice
spacing, $a$. This arises from the presence of magnetic
monopoles in the theory. (To be more precise, in D=3+1
these would be magnetic monopoles. Here they are pointlike
instantons whose fields are identical to the spatial
fields of a static Dirac monopole. Hence we shall
follow the usual custom and refer to them as magnetic
monopoles, even in D=2+1.) These monopoles clearly have
an action that is $\simeq c_M/ag^2$ where $c_M$ is a constant 
that depends on the particular lattice action being used.
Thus a monopole has a weighting $\sim \exp\{-c_M/ag^2\}$
and hence the average distance between monopoles will
be $d_M \sim a\exp\{c_M/3ag^2\}$. (Up to weakly varying
factors that come from integrating small fluctuations
around the monopoles.) This provides a
scale for the theory that is different from $g^2$.
This scale is interesting because the monopoles change
the physics in a qualitative fashion. As is well known
\cite{confU1}
they produce a linear confining force between external
static charges. One could also expect them to produce
a nontrivial mass spectrum. At the very least, there
will be a massive `photon'. Of course, in the continuum
limit $d_M \to \infty$ in units of $1/g^2$ and so on
the latter scale the monopoles disappear to $r=\infty$
as we approach the continuum limit. 

Actually, the above description represents an 
over-simplification. There is not just one 
new scale introduced by the monopoles. There is also 
a scale associated with the string tension, 
$l_{\sigma} \equiv 1/\surd\sigma$, and a scale associated
with the screening mass, $l_s \equiv 1/m_s$. These
scales mutually diverge in the continuum limit:
$ l_s \sim l_{\sigma}^2 \sim d_M^{4/3}$ up to constants
and powers of $\beta$. The origins of this peculiar
situation lie in the fact that the monopoles are
singular objects.

For these reasons we will not try to calculate a
`continuum' mass spectrum. Rather we shall calculate 
the spectrum for lattice spacings that are small
compared to the obvious dynamical length scale,
$a \ll 1/\surd\sigma$. 

The results of our calculation are presented in 
Table~\ref{table_massU1}. How do they compare to the
SU($N_c$) spectra listed in Tables~\ref{table_mcont}
and ~\ref{table_masskN}? An immediate difference with 
all of the SU($N_c$) spectra is that the $0^{++}$ is no 
longer the lightest particle; the $0^{--}$ is about half 
the mass. We also note that while the ratio
$m_{0^{++}}/\surd\sigma$ was increasing as $N_c$
decreased, the U(1) value is about as much below
the SU($\infty$) value as the SU(2) ratio is above.
Apart from these striking differences, the rest of
the spectrum seems quite similar (albeit within
the large errors). This is particularly so if we
compare to the $0^{++}$ mass rather than to $\surd\sigma$.
For example, the $2^{++}$ to $0^{++}$ mass ratio is
close to its SU($N_c$) value. We note also that
we have approximate parity doubling for $J\not=0$,
thus confirming that $a$ is indeed small enough for
continuum rotational symmetry to have been restored on 
hadronic length scales.

\vspace{1.0cm}

\newpage

\begin{table}
\begin{center}
\begin{tabular}{|c|c|c|c|c|}\hline
cools & $n_G=4$ &  $n_G=5$ & $n_G=6$ & $n_G=8$  \\ \hline
 0  &  0.2010 & 0.2085 & 0.2072 & 0.2088 \\ 
 1  &  0.1272 & 0.1199 & 0.1067 & 0.0874 \\ 
 2  &  0.0813 & 0.0662 & 0.0494 & 0.0266 \\ 
 5  &  0.0596 & 0.0417 & 0.0254 & 0.0073 \\
10  &  0.0516 & 0.0328 & 0.0168 & 0.0028 \\
15  &  0.0483 & 0.0291 & 0.0132 & 0.0015 \\
20  &  0.0463 & 0.0269 & 0.0111 & 0.0010 \\ \hline 
\end{tabular}
\caption{\label{table_cools} 
Average action per plaquette when a thermalised SU(5)
field is cooled using $n_G$ SU(2) subgroups.}
\end{center}
\end{table}

\begin{table}
\begin{center}
\begin{tabular}{|c|c|c|}\hline
 state & $R_o=0$ &  $R_o=5$   \\ \hline
 flux loop & 0.0054 & 0.0050 \\
           & 0.024  & 0.030  \\ \hline
  $0^{++}$ & 0.0098 & 0.0097 \\
           & 0.056  & 0.051  \\ \hline
  $0^{--}$ & 0.019  & 0.024  \\
           & 0.22   & 0.20   \\ \hline
  $2^{++}$ & 0.036  & 0.031  \\
           & 0.29   & 0.32   \\ \hline
\end{tabular}
\caption{\label{table_or11su3} 
Errors on SU(3) masses on a $12^2 16$ lattice at $\beta=11$;
$R_o$ is the number of over-relaxed sweeps for every 
heatbath sweep.}
\end{center}
\end{table}

\begin{table}
\begin{center}
\begin{tabular}{|c|c|c|c|c|c|}\hline
 state & $R_o=0$ & $R_o=2$ & $R_o=5$ & $R_o=10$ & $R_o=20$ \\ \hline
 flux loop & 0.0036 & 0.0053 & 0.0033 & 0.0044 & 0.0043 \\
           & 0.0085 & 0.0123 & 0.0078 & 0.0074 & 0.0094 \\ \hline
  $0^{++}$ & 0.0086 & 0.0132 & 0.0075 & 0.0139 & 0.0085 \\
           & 0.034  & 0.040  & 0.037  & 0.038  &  0.033 \\ \hline
  $0^{--}$ & 0.017  & 0.026  & 0.020  & 0.018  &  0.022 \\
           & 0.081  & 0.113  & 0.086  & 0.082  & 0.125  \\ \hline
  $2^{++}$ & 0.025  & 0.015  & 0.015  & 0.024  & 0.021 \\ 
           & 0.090  & 0.092  & 0.092  & 0.079  & 0.070 \\ \hline
\end{tabular}
\caption{\label{table_or15su3} 
Errors on SU(3) masses on a $12^2 16$ lattice at $\beta=15$.
$R_o$ is the number of over-relaxed sweeps for every 
heatbath sweep.}
\end{center}
\end{table}

\begin{table}
\begin{center}
\begin{tabular}{|c|c|c|c|c|c|}\hline
 state & $R_o=0$ & $R_o=3$ & $R_o=5$ & $R_o=7$ & $R_o=10$ \\ \hline
 flux loop & 0.0053 & 0.0044 & 0.0040 & 0.0038 & 0.0030 \\
          & 0.0089 & 0.0075 & 0.0070 & 0.0057 & 0.0069 \\ \hline
 $0^{++}$ & 0.0085 & 0.0072 & 0.0058 & 0.0056 & 0.0095 \\
          & 0.0170 & 0.0172 & 0.0138 & 0.0132 & 0.0142 \\ \hline
 $0^{--}$ & 0.0123 & 0.0072 & 0.0090 & 0.0150 & 0.0103 \\
          & 0.030  & 0.035  & 0.035  & 0.030  & 0.034 \\ \hline
 $2^{++}$ & 0.0148 & 0.0106 & 0.0103 & 0.0119 & 0.0120 \\
          & 0.044  & 0.029  & 0.032  & 0.041  & 0.045 \\ \hline
 $2^{-+}$ & 0.0094 & 0.0132 & 0.0103 & 0.0102 & 0.0094 \\
          & 0.043  & 0.044  & 0.026  & 0.044  & 0.054 \\ \hline
\end{tabular}
\caption{\label{table_or21su3} 
Errors on SU(3) masses on a $24^3$ lattice at $\beta=21$.
$R_o$ is the number of over-relaxed sweeps for every 
heatbath sweep.}
\end{center}
\end{table}

\begin{table}
\begin{center}
\begin{tabular}{|c|c|c|c|c|}\hline
 state & $R_o=0$ & $R_o=5$ & $R_o=9$ & $R_o=49$ \\ \hline
 flux loop  & 0.004 & 0.003 & 0.002 & 0.002 \\
         & 0.006 & 0.004 & 0.003 & 0.004 \\ \hline
 $0^+$   & 0.007 & 0.004 & 0.006 & 0.007 \\
         & 0.015 & 0.014 & 0.009 & 0.009 \\ \hline
 $2^+$   & 0.011 & 0.007 & 0.010 & 0.010 \\
         & 0.019 & 0.019 & 0.017 & 0.021 \\ \hline

\end{tabular}
\caption{\label{table_or9su2} 
Errors on SU(2) masses on a $12^2 24$ lattice at $\beta=9$.
$R_o$ is the number of over-relaxed sweeps for every 
heatbath sweep.}
\end{center}
\end{table}

\begin{table}
\begin{center}
\begin{tabular}{|c|c|c|c|c|}\hline
 $B_l$ & $SU(3);\beta=21$ & $SU(3);\beta=15$ & $SU(3);\beta=11$  
 &  $SU(2);\beta=9$ \\ \hline
 1 &  1.28(15) & 1.00(13) & 1.29 &  1.01(14) \\ \hline
 2 &  0.92(5) & 0.63(6) & 0.88 &  1.02(5) \\ \hline
 3 &  0.64(5) & 0.63(4) & 0.81 &  0.78(3) \\ \hline
 4 &  0.75(4) & 0.69(4) & 0.79 &  0.71(6) \\ \hline
 5 &  1.01(10) &        &      &          \\ \hline
\end{tabular}
\caption{\label{table_orplaq}
Ratio of errors with and without over-relaxation; for
`plaquettes' at blocking level  $B_l$.}
\end{center}
\end{table}

\begin{table}
\begin{center}
\begin{tabular}{|c|c|c|c|c|c|}\hline
\multicolumn{6}{|c|}{$\min_{\phi} \ \{{\rm am_{eff}(t=a)}\}$} \\ \hline
 state & $\gamma_d=0.25$ & $\gamma_d=0.50$ & $\gamma_d=1.0$ 
& $\gamma_d=2.0$ & $\gamma_d=4.0$ \\ \hline
 flux loop  & 0.622(11) & 0.612(10) & 0.603(10) & 0.601(9) & 0.630(8) \\ 
  $0^{++}$  & 0.979(15) & 0.967(13) & 0.963(12) & 0.969(11) & 1.008(11)  \\ 
  $2^{++}$  & 1.53(4) & 1.55(4) & 1.56(3) & 1.56(5) & 1.63(3) \\ 
  $2^{-+}$  & 1.71(4) & 1.61(3) & 1.58(3) & 1.58(3) & 1.63(3) \\ 
  $0^{-+}$  & 2.51(6) & 2.13(4) & 1.99(3) & 2.01(4) & 2.12(5) \\ 
  $1^{++}$  & 2.80(6) & 2.32(7) & 2.19(5) & 2.20(4) & 2.34(4) \\ 
  $1^{-+}$  & 3.01(19) & 2.43(8)  & 2.27(8) & 2.30(6) & 2.48(6) \\  \hline
\end{tabular}
\caption{\label{table_blockT}
Effective masses at $t=a$ from the `best' operators in different
blocking schemes, as described in Appendix B.} 
\end{center}
\end{table}

\begin{table}
\begin{center}
\begin{tabular}{|c|c|c|}\hline
\multicolumn{3}{|c|}{${\rm am_{eff}(t=a)}$} \\ \hline
 state & with $\phi_Y$ & without $\phi_Y$ \\ \hline
  $0^{++}$          & 1.08(2) & 1.08(2) \\
  $0^{++\ast}$      & 1.67(5) & 1.67(5) \\ 
  $0^{++\ast\ast}$  & 2.03(4) & 2.05(3) \\ 
  $0^{++\ast\ast\ast}$  & 2.26(6) &  2.26(5) \\  \hline
  $2^{++}$          & 1.78(2) & 1.78(2) \\ 
  $2^{++\ast}$      & 2.15(6) & 2.15(6) \\ 
  $2^{++\ast\ast}$  & 2.33(7) & 2.34(7) \\
  $2^{++\ast\ast\ast}$  & 2.56(7) & 2.61(7) \\  \hline
  $1^{--}$          & 2.60(6) &  2.60(6) \\ 
  $1^{--\ast}$      & 2.62(5) &  2.63(5) \\ 
  $1^{--\ast\ast}$  & 2.74(8) &  2.78(9) \\
  $1^{--\ast\ast\ast}$  & 2.92(10) &  3.02(7) \\  \hline
\end{tabular}
\caption{\label{table_Y}
Effective masses at $t=a$ obtained from bases with and
without the operator, $\phi_Y$, which contains two
`baryonic, vertices. From a $16^3$ lattice at $\beta=15$
in SU(3).}
\end{center}
\end{table}

\clearpage

\begin{table}
\begin{center}
\begin{tabular}{|c|c|c|}\hline
$\beta$ & $L,L^{\prime}$ & $c_{eff}$  \\ \hline
 12   &   32,48    &  0.66(20)  \\ \hline
  9   &   24,32    &  0.79(26)  \\ \hline
      &   16,24    &  0.63(9)   \\ \hline
      &   12,16    &  0.47(7)   \\ \hline
      &    8,12    &  0.10(2)   \\ \hline
      &    6,8     &  0.00(2)   \\ \hline
  6   &   24,32    & -1.7(2.9)  \\ \hline
      &   16,24    &  1.26(70)  \\ \hline
      &   12,16    &  0.87(27)  \\ \hline
      &    8,12    &  0.41(8)   \\ \hline
      &    6,8     &  0.28(5)   \\ \hline
\end{tabular}
\caption{\label{table_univcorr} 
Coefficient of effective leading $1/L$ correction 
in the flux loop mass: as extracted from loops
of length $L$ and $L^{\prime}$ using eqn(\ref{C2}).}
\end{center}
\end{table}

\begin{table}
\begin{center}
\begin{tabular}{|c|c|c|c|c|c|c|c|}\hline
$\beta$ & $L$ & $n_{mom}$ & $am(a)$ & $am(2a)$ &  $am(3a)$ &
$am(4a)$ & $am(5a)$ \\ \hline\hline
14.5 & 40 & 0 & 0.373(1) & 0.364(2) & 0.364(2) & 0.364(3) & 0.363(5) \\
 &    & 1 & 0.375(1) & 0.365(2) & 0.363(2)  & 0.361(3) & 0.358(4) \\
\hline\hline
12.0 & 48 & 0 & 0.685(2) & 0.668(5) & 0.656(9) & 0.628(16) & 0.632(36) \\
 &    & 1 & 0.690(2) & 0.670(4) & 0.671(7)  & 0.677(13) & 0.677(25) \\ \hline
 & 32 & 0 & 0.443(2) & 0.434(2) & 0.429(3)  & 0.426(4)  & 0.426(7) \\
 &    & 1 & 0.447(1) & 0.434(2) & 0.430(3)  & 0.425(4) & 0.425(6) \\
\hline\hline
9.0 & 32 & 0 & 0.846(3) & 0.826(5) & 0.832(10) & 0.837(21) & \\
 &    & 1 & 0.846(2) & 0.823(4) & 0.821(7)  & 0.832(17) &  \\ \hline
 & 24 & 0 & 0.617(2) & 0.605(4) & 0.601(6)  & 0.590(9)  & 0.593(17) \\
 &    & 1 & 0.622(1) & 0.609(2) & 0.600(5)  & 0.599(10) & 0.597(17) \\ \hline
 & 16 & 0 & 0.389(2) & 0.381(2) & 0.378(4)  & 0.378(5)  & 0.379(7)  \\
 &    & 1 & 0.398(2) & 0.383(2) & 0.378(4)  & 0.370(5)  & 0.364(9)  \\ \hline
 & 12 & 0 & 0.274(2) & 0.269(2) & 0.268(3)  & 0.271(3)  & 0.272(4)  \\
 &    & 1 & 0.291(3) & 0.274(6) & 0.261(10) & 0.268(14) &  \\ \hline
 &  8 & 0 & 0.174(1) & 0.172(1) & 0.172(1)  & 0.172(2)  & 0.173(2)  \\
 &    & 1 & 0.215(6) & 0.182(10) & 0.163(25) & &  \\ \hline
 &  6 & 0 & 0.130(1) & 0.129(1) & 0.129(1)  & 0.128(2)  & 0.128(2) \\
 &    & 1 & 0.093(16) & & & &  \\ \hline
\end{tabular}
\caption{\label{table_massloop} 
Effective loop masses for various loop lengths $L$ and from the
lowest 2 momenta, $p=2\pi n_{mom}/L$, in SU(2) at the values of
$\beta$ shown.}
\end{center}
\end{table}

\begin{table}
\begin{center}
\begin{tabular}{|c|c|c|c|c|}\hline
$N_c$ & $\beta$ & $L$ & $am_P$ & $c_{eff}$  \\ \hline
  5  & 44 & 8  & 0.4490(33)  &          \\
     &    & 12 & 0.7314(71)  & 0.55(8)  \\
     &    & 16 & 1.0110(56)  & 0.75(23) \\
     & 33 & 8  & 0.997(8)    &          \\
     &    & 12 & 1.551(25)   & 0.53(26) \\ \hline
  4  & 28 & 8  & 0.4257(45)  &          \\
     &    & 12 & 0.7139(66)  & 0.72(9)  \\
     &    & 16 & 0.9857(66)  & 0.70(23) \\ \hline
  3  & 15 & 8  & 0.4425(27)  &          \\
     &    & 12 & 0.7391(46)  & 0.72(6)  \\
     &    & 16 & 1.0103(92)  & 0.51(23) \\ 
    &    & 24 & 1.5636(180) & 0.92(44) \\ \hline
\end{tabular}
\caption{\label{table_linearsun}
Flux loop masses as a function of the loop length, $L$,
for SU(3), SU(4) and SU(5). Also shown is $c_{eff}$, the 
coefficient of the $1/L$ correction in eqns(\ref{C1},\ref{C2}).}
\end{center}
\end{table}

\begin{table}
\begin{center}
\begin{tabular}{|c|c|c||c|c|c|}\hline
$\beta$ & $L$ & $a\surd\sigma$ & $\beta$ & $L$ & $a\surd\sigma$  \\ \hline
14.5 & 40 & 0.09713(20) & 6.0   & 16 & 0.2538(10) \\
12.0 & 48 & 0.1179(8)   & 5.0   & 16 & 0.3129(20) \\ 
12.0 & 32 & 0.1179(5)   & 4.5   & 12 & 0.3527(30) \\ 
 9.0 & 32 & 0.1622(4)   & 3.75  &  8 & 0.4487(33) \\ 
 9.0 & 24 & 0.1616(6)   & 3.47  &  8 & 0.4889(56) \\ 
6.56 & 24 & 0.2297(10)  & 3.0   &  6 & 0.584(16)  \\ 
 6.0 & 32 & 0.2529(33)  & 2.5   &  4 & 0.709(11)  \\ 
 6.0 & 24 & 0.2562(26)  & 2.08$\dot{3}$ &  4 & 0.852(50)  \\ \hline
\end{tabular}
\caption{\label{table_Ksu2}
SU(2) string tensions as extracted from flux loop
masses of length $L$ using eqn(\ref{C1}).}
\end{center}
\end{table}

\begin{table}
\begin{center}
\begin{tabular}{|c|c|c||c|c|c|}\hline
$\beta$ & $L$ & $a\surd\sigma$ & $\beta$ & $L$ & $a\surd\sigma$  \\ \hline
34.0 & 40 & 0.10379(26) & 8.175 & 6 & 0.5598(31) \\
28.0 & 32 & 0.12753(20) & 7.5   & 6 & 0.591(23)  \\ 
21.0 & 24 & 0.17479(38) & 7.5   & 4 & 0.633(3)   \\ 
15.0 & 24 & 0.2570(15)  & 7.0   & 4 & 0.698(5)   \\ 
15.0 & 16 & 0.2553(12)  & 6.5   & 4 & 0.782(26)  \\ 
11.0 & 12 & 0.3748(23)  & 6.0   & 4 & 0.835(41)  \\ \hline 
\end{tabular}
\caption{\label{table_Ksu3}
SU(3) string tensions as extracted from flux loop
masses of length $L$ using eqn(\ref{C1}).}
\end{center}
\end{table}

\begin{table}
\begin{center}
\begin{tabular}{|c|c|c||c|c|c|}\hline
\multicolumn{3}{|c||}{SU(4)} & \multicolumn{3}{c|}{SU(5)} \\ \hline
$\beta$ & $L$ & $a\surd\sigma$ & $\beta$ & $L$ & $a\surd\sigma$  \\ \hline
51.0 & 32 & 0.12859(23) & 82.0 & 32 & 0.12715(27) \\
40.0 & 24 & 0.16804(30) & 64.0 & 24 & 0.1664(4)   \\ 
28.0 & 16 & 0.2523(8)   & 44.0 & 16 & 0.2554(7)   \\ 
21.0 & 12 & 0.3597(24)  & 33.0 & 12 & 0.3645(29)  \\ \hline
\end{tabular}
\caption{\label{table_Ksu45}
SU(4) and SU(5) string tensions as extracted from flux loop
masses of length $L$ using eqn(\ref{C1}).}
\end{center}
\end{table}

\begin{table}
\begin{center}
\begin{tabular}{|c|c|c||c|c|c|}\hline
$\beta$ & $L$ & plaq & $\beta$ & $L$ & plaq  \\ \hline
14.5 & 40 & 0.929803(3)  & 6.0   & 16 & 0.824744(33) \\
12.0 & 48 & 0.914824(3)  & 5.0   & 16 & 0.786850(20) \\ 
12.0 & 32 & 0.914823(3)  & 4.5   & 12 & 0.760841(45) \\ 
 9.0 & 32 & 0.885445(5)  & 3.75  &  8 & 0.706986(48) \\ 
 9.0 & 24 & 0.885438(7)  & 3.47  &  8 & 0.680058(59) \\ 
6.56 & 24 & 0.840548(22) & 3.0   &  6 & 0.624023(62) \\ 
 6.0 & 32 & 0.824772(10) & 2.5   &  4 & 0.54737(20)  \\ 
 6.0 & 24 & 0.824782(16) & 2.08$\dot{3}$ &  4 & 0.47100(13)  \\ \hline
\end{tabular}
\caption{\label{table_plaqsu2}
Average SU(2) plaquette values.}
\end{center}
\end{table}

\begin{table}
\begin{center}
\begin{tabular}{|c|c|c||c|c|c|}\hline
$\beta$ & $L$ & plaq & $\beta$ & $L$ & plaq  \\ \hline
34.0 & 40 & 0.919680(2)  & 8.175 & 6 & 0.620730(45) \\
28.0 & 32 & 0.901903(2)  & 7.5   & 6 & 0.57810(19)  \\ 
21.0 & 24 & 0.867671(5)  & 7.5   & 4 & 0.57801(8)   \\ 
15.0 & 24 & 0.810773(9)  & 7.0   & 4 & 0.54118(11)  \\ 
15.0 & 16 & 0.810767(11) & 6.5   & 4 & 0.50060(13)  \\ 
11.0 & 12 & 0.733401(18) & 6.0   & 4 & 0.45736(19)  \\ \hline 
\end{tabular}
\caption{\label{table_plaqsu3}
Average SU(3) plaquette values.}
\end{center}
\end{table}

\begin{table}
\begin{center}
\begin{tabular}{|c|c|c||c|c|c|}\hline
\multicolumn{3}{|c||}{SU(4)} & \multicolumn{3}{c|}{SU(5)} \\ \hline
$\beta$ & $L$ & plaq & $\beta$ & $L$ & plaq  \\ \hline
51.0 & 32 & 0.898791(1)  & 82.0 & 32 & 0.899245(1)   \\
40.0 & 24 & 0.869608(3)  & 64.0 & 24 & 0.869510(3)   \\ 
28.0 & 16 & 0.809339(9)  & 44.0 & 16 & 0.805322(6)   \\ 
21.0 & 12 & 0.737628(18) & 33.0 & 12 & 0.731640(17)  \\ \hline
\end{tabular}
\caption{\label{table_plaqsu45}
Average SU(4) and SU(5) plaquette values.}
\end{center}
\end{table}

\begin{table}
\begin{center}
\begin{tabular}{|c|c|c|c|c|}\hline
group & $c_0$ & $c_1$ & CL($\%$) & $\beta\geq$   \\ \hline
SU(2) & 1.341(7)  & -0.421(51)  & 60 & 3.0   \\
SU(3) & 3.318(12) & -2.43(22)   & 90 & 8.175 \\
SU(4) & 6.065(32) & -7.74(1.10) & 90 & 21.0  \\
SU(5) & 9.657(54) & -21.4(2.7)  & 70 & 33.0  \\ \hline
\end{tabular}
\caption{\label{table_Kcont}
Continuum extrapolations of 
$\beta_I a\surd\sigma \to 2N_c \surd\sigma/g^2$ 
as in eqn(\ref{C6}), with confidence 
level of best fit, and range of $\beta$ fitted.}
\end{center}
\end{table}

\begin{table}
\begin{center}
\begin{tabular}{|c|ccccc|}\hline
state  & $am(a)$ & $am(2a)$ &  $am(3a)$ & $am(4a)$ 
& $am(5a)$ \\ \hline\hline
$0^{++}$  & 0.541(3) & 0.533(5) & 0.539(7) & 0.523(11) 
& 0.533(14) \\ \hline
$0^{++\ast}$ & 0.821(3) & 0.798(7) & 0.796(14) & 0.799(29) 
& 0.795(65) \\ \hline
$0^{++\ast\ast}$ & 1.033(4) & 1.009(10) & 1.028(28) 
& 1.013(66) & 1.18(22) \\ \hline 
$0^{--}$ & 0.793(3) & 0.779(7) & 0.765(10) & 0.750(30) 
& 0.71(5) \\ \hline
$0^{--\ast}$ & 1.016(4) & 1.002(8) & 0.969(26) & 1.054(62)
& 1.09(17) \\ \hline
$0^{--\ast\ast}$ & 1.202(5) & 1.193(14) & 1.14(4) 
& 1.20(18) & \\ \hline
$0^{-+}$ & 1.208(5) & 1.155(11) & 1.146(43) & 1.02(11) & \\ \hline
$0^{+-}$ & 1.326(6) & 1.256(14) & 1.256(59) & 1.59(26) & \\ \hline
$2^{++}$ & 0.913(3) & 0.891(7) & 0.905(19) & 0.911(40) & \\ \hline
$2^{++\ast}$ & 1.103(4) & 1.072(9) & 1.074(31) & 1.00(10) & \\ \hline
$2^{-+}$ & 0.906(3) & 0.879(7) & 0.853(14) & 0.847(43) & \\ \hline
$2^{-+\ast}$ & 1.094(4) & 1.076(9) & 1.069(32) & 1.04(10) & \\ \hline
$2^{--}$ & 1.081(4) & 1.054(10) & 1.051(28) & 1.10(10) & \\ \hline
$2^{--\ast}$ & 1.271(5) & 1.225(16) & 1.218(46) & 1.17(11) & \\ \hline
$2^{+-}$ & 1.075(4) & 1.043(9) & 1.049(30) & 1.10(11) & \\ \hline
$2^{+-\ast}$ & 1.276(4) & 1.230(17) & 1.218(49) & 1.36(17) & \\ \hline
$1^{++}$ & 1.344(4) & 1.294(14) & 1.284(48) & 1.14(18) & \\ \hline
$1^{-+}$ & 1.355(5) & 1.300(10) & 1.327(46) & 1.33(16) & \\ \hline
$1^{--}$ & 1.255(3) & 1.212(9)  & 1.176(34) & 1.18(10) & \\ \hline
$1^{+-}$ & 1.286(5) & 1.255(14) & 1.231(44) & 1.15(19) & \\ \hline
\end{tabular}
\caption{\label{table_meff5} 
Effective masses for the states shown, on a $32^3$ lattice
at $\beta=82$ in SU(5).}
\end{center}
\end{table}

\clearpage

\begin{table}
\begin{center}
\begin{tabular}{|c|c|c|c|c|c|c|}\hline
\multicolumn{7}{|c|}{SU(2) ; $\beta=6$} \\ \hline
state & $L=6$ & $L=8$ & $L=12$ & $L=16$ & $L=24$ & $L=32$ \\ \hline
$0^{++}$ & 0.99(2) & 1.12(2) & 1.21(2) & 1.19(2) & 1.20(2) 
& 1.18(2) \\ \hline
$2^{++}$ & 0.92(2) & 1.17(2) & 1.71(5) & 1.80(8) & 1.99(12) 
& 1.87(10) \\ \hline
$2^{-+}$ & 2.20(13) & 2.17(15) & 2.15(21) & 1.81(14) & 2.01(16)
& 1.91(11) \\ \hline
$2 \times l_P$  & 0.62(1) & 0.90(1) & 1.43(1) & 2.00(2) & 3.12(6)
& 4.11(15) \\ \hline
\end{tabular}
\caption{\label{table_Vsu2}
How the lightest SU(2) glueball masses depend on the spatial volume.
Twice the mass of the periodic flux loop is also shown.}
\end{center}
\end{table}

\begin{table}
\begin{center}
\begin{tabular}{|c|c|c|c|c|}\hline
\multicolumn{5}{|c|}{SU(3) ; $\beta=15$} \\ \hline
state         & $L=8$    & $L=12$  & $L=16$   & $L=24$   \\ \hline
$0^{++}$      & 0.99(4)  & 1.09(2) & 1.15(2)  & 1.10(2)  \\
$0_{ll}^{++}$ & 0.98(2)  & 1.49(4) & 1.92(9)  & 2.0(4)   \\ \hline
$0^{++\ast}$  & 1.24(4)  & 1.59(4) & 1.66(6)  & 1.65(5)  \\ \hline
$0^{--}$      & 1.72(7)  & 1.65(5) & 1.57(6)  & 1.57(4)  \\ \hline
$2^{++}$      & 0.95(5)  & 1.69(6) & 1.81(7)  & 1.86(11) \\
$2_{ll}^{++}$ & 0.95(2)  & 1.58(4) & 2.08(15) &          \\ \hline
$2^{-+}$      & 1.87(10) & 1.83(7) & 1.79(8)  & 1.76(6)  \\ \hline
\end{tabular}
\caption{\label{table_Vsu3}
How the lightest SU(3) masses depend on the spatial volume.}
\end{center}
\end{table}

\begin{table}
\begin{center}
\begin{tabular}{|c||c|c||c|c|}\hline
\multicolumn{1}{|c||}{} &
\multicolumn{2}{c||}{SU(4) ; $\beta=28$} & 
\multicolumn{2}{c|}{SU(5) ; $\beta=44$} \\ \hline
state         & $L=12$    & $L=16$  & $L=12$   & $L=16$   \\ \hline
$0^{++}$      & 1.08(2)  & 1.08(2) & 1.09(3)  & 1.05(2) \\ \hline 
$0^{++\ast}$  & 1.53(6)  & 1.62(4) & 1.64(6)  & 1.65(3) \\ \hline
$0^{--}$      & 1.60(8)  & 1.60(4) & 1.52(7)  & 1.58(5) \\ \hline
$2^{++}$      & 1.80(10) & 1.76(4) & 1.74(11) & 1.69(7) \\ \hline
$2^{-+}$      & 1.80(12) & 1.79(5) & 1.78(9)  & 1.60(6) \\ \hline
\end{tabular}
\caption{\label{table_Vsu45}
How the lightest SU(4) and SU(5) masses depend on the spatial volume.}
\end{center}
\end{table}

\begin{table}
\begin{center}
\begin{tabular}{|l|l|l|l|l|l|l|}\hline
state & $\beta=3.75$ & $\beta=4.5$ & $\beta=5$ & $\beta=6$ 
& $\beta=6$ & $\beta=6$ \\
      & L=8  & L=12  & L=16  & L=16  & L=24 & L=32  \\ \hline
$0^{++}$         & 2.07(9) & 1.642(43) & 1.478(24) & 1.193(18) 
& 1.191(18) & 1.170(23) \\
$0^{++\ast}$     & 2.7(3)  & 2.12(19) & 2.11(11) & 
& 1.67(6) & 1.60(6) \\ 
$0^{++\ast\ast}$ & & & 2.6(4) & & 2.07(12) & 2.10(13) \\
$0^{-+}$         & & & & 2.10(33) & 2.41(35) & 2.59(29) \\ 
$2^{++}$         & & & 2.26(17) & 1.80(8) & 1.94(11) & 1.87(11) \\ 
$2^{++\ast}$     & & & & & 2.00(14) & 1.94(12) \\ 
$2^{-+}$         & & & 2.20(15) & 1.81(14) & 1.77(11) & 1.91(9) \\ 
$2^{-+\ast}$     & & & & & 2.28(27) & 2.26(16) \\ 
$1^{++}$         & & & & 2.43(31) & 2.35(38) & 2.64(27) \\ 
$1^{-+}$         & & & & 2.9(7) & 2.4(5) & 3.0(7) \\ \hline
\end{tabular}
\caption{\label{table_msu2a}
The lightest SU(2) masses at lower values of $\beta$.}
\end{center}
\end{table}

\begin{table}
\begin{center}
\begin{tabular}{|l|l|l|l|l|l|}\hline
state & $\beta=9$ & $\beta=9$ & $\beta=12$ & $\beta=12$ & $\beta=14.5$ \\
      & L=24  & L=32  & L=32  & L=48  & L=40 \\ \hline
$0^{++}$         & 0.7643(60) & 0.7552(67) & 0.5572(36) & 0.5628(46)
& 0.4562(23) \\
$0^{++\ast}$     & 1.082(16) & 1.087(14) & 0.8072(53) & 0.8047(74)
& 0.6532(33) \\ 
$0^{++\ast\ast}$ & 1.271(21) & 1.340(22) & 0.949(8) & 0.982(14)
& 0.790(4) \\
$0^{-+}$         & 1.629(44) & 1.60(4) & 1.187(18) & 1.188(18)
& 0.965(8) \\ 
$0^{-+\ast}$     & 1.95(9)   & 1.75(7) & 1.302(20) & 1.350(28) 
& 1.178(13) \\ 
$2^{++}$         & 1.259(13) & 1.249(15) & 0.913(7) & 0.920(13) 
& 0.7557(35) \\ 
$2^{++\ast}$     & 1.396(22) & 1.484(27) & 0.971(8) & 1.055(11) 
& 0.846(7) \\
$2^{-+}$         & 1.276(24) & 1.286(23) & 0.928(9) & 0.910(10) 
& 0.7626(51) \\ 
$2^{-+\ast}$     & 1.532(32) & 1.480(31) & 1.089(13) & 1.094(18) 
& 0.958(10) \\ 
$1^{++}$         & 1.814(54) & 1.82(6) & 1.258(17) & 1.295(20) 
& 1.042(9) \\ 
$1^{++\ast}$     & 2.04(10)  & 1.94(8) & 1.493(24) & 1.491(31) 
& 1.240(15) \\ 
$1^{-+}$         & 1.892(54) & 1.81(8) & 1.356(18) & 1.317(20) 
& 1.096(10) \\ 
$1^{-+\ast}$     & 1.83(6) & 1.73(6) & 1.331(17) & 1.320(19) 
& 1.092(12) \\ \hline
\end{tabular}
\caption{\label{table_msu2b}
The lightest SU(2) masses at higher values of $\beta$.}
\end{center}
\end{table}

\begin{table}
\begin{center}
\begin{tabular}{|l|l|l|l|l|l|l|}\hline
state   & $\beta=11$ & $\beta=15$  & $\beta=15$ 
& $\beta=21$ & $\beta=28$  & $\beta=34$  \\
  & L=12 & L=16 & L=24 & L=24 & L=32 & L=40    \\ \hline
$0^{++}$         & 1.626(36) & 1.123(16) & 1.095(14) & 0.7561(62) 
& 0.5517(38) & 0.4482(36) \\
$0^{++\ast}$     & 2.19(14)  & 1.66(6)   & 1.652(44) & 1.124(15)  
& 0.823(6)   & 0.6737(45) \\ 
$0^{++\ast\ast}$ &           & 2.06(9)   & 2.04(11)  & 1.411(23)  
& 1.034(9)   & 0.8512(67) \\
$0^{--}$         & 2.30(15) & 1.568(53) & 1.569(38) & 1.101(14) 
& 0.8133(57) & 0.6682(48) \\ 
$0^{--\ast}$     &          & 2.05(16) & 2.00(10) & 1.385(21) 
& 1.025(10) & 0.8386(49) \\ 
$0^{--\ast\ast}$ &          & 2.44(41) & 2.40(30) & 1.596(39) 
& 1.191(16) & 0.9969(85) \\
$0^{-+}$         & & & 2.32(24) & 1.627(41) & 1.206(15) & 0.9634(81) \\ 
$0^{-+\ast}$     & & & & 1.835(73) & 1.322(17) & 1.194(14) \\ 
$0^{+-}$         & & & 2.08(23) & 1.826(59) & 1.330(15) & 1.088(10) \\ 
$0^{+-\ast}$     & & & & 1.98(11)  & 1.582(31) & 1.315(16) \\ 
$2^{++}$         & 2.31(21) & 1.81(7) & 1.86(11) & 1.218(16) 
& 0.9123(70) & 0.7354(43) \\ 
$2^{++\ast}$     & & 2.15(14) & 2.18(15) & 1.520(27) 
& 1.057(10) & 0.9134(79) \\
$2^{-+}$         & 2.64(29) & 1.786(82) & 1.758(58) & 1.257(18) 
& 0.937(8) & 0.7526(48) \\ 
$2^{-+\ast}$     & & 2.10(16) & 2.19(15) & 1.618(43) 
& 1.109(10) & 0.9142(65) \\ 
$2^{--}$         & & 2.05(15) & 1.95(14) & 1.475(23) 
& 1.0928(86) & 0.8913(63) \\ 
$2^{--\ast}$     & & 2.21(19) & 2.35(26) & 1.705(42) 
& 1.254(15) & 1.0452(85) \\ 
$2^{+-}$         & & 2.04(16) & 1.89(11) & 1.539(27) 
& 1.114(11) & 0.8867(59) \\
$2^{+-\ast}$     & & & & 1.813(54) & 1.325(15) & 1.085(11) \\
$1^{++}$         & & & & 1.738(40) & 1.298(11) & 1.0513(70) \\ 
$1^{++\ast}$     & & & & 1.933(61) & 1.481(19) & 1.212(10) \\ 
$1^{-+}$         & & & & 1.881(51) & 1.350(12) & 1.082(8) \\ 
$1^{-+\ast}$     & & & & 1.902(48) & 1.352(13) & 1.096(9) \\ 
$1^{--}$         & & & & 1.780(34) & 1.269(11) & 1.036(7) \\ 
$1^{--\ast}$     & & & & 1.877(51) & 1.371(15) & 1.075(9) \\ 
$1^{+-}$         & & & & 1.788(37) & 1.297(11) & 1.074(9) \\ 
$1^{+-\ast}$     & & & & 1.996(56) & 1.404(16) & 1.103(7) \\ \hline
\end{tabular}
\caption{\label{table_msu3}
The lightest SU(3) masses.}
\end{center}
\end{table}

\clearpage

\begin{table}
\begin{center}
\begin{tabular}{|l|l|l|l|l|}\hline
state & $\beta=21$ & $\beta=28$ & $\beta=40$ & $\beta=51$   \\
      & L=12 & L=16 & L=24  & L=32   \\ \hline
$0^{++}$         & 1.525(36) & 1.083(14) & 0.7109(52) & 0.5466(40) \\
$0^{++\ast}$     & 2.31(22)  & 1.616(39) & 1.080(10)  & 0.821(6) \\ 
$0^{++\ast\ast}$ &           & 1.99(8)   & 1.364(21)  & 1.032(9) \\
$0^{--}$         & 2.10(13)  & 1.599(36) & 1.039(13)  & 0.8040(46) \\ 
$0^{--\ast}$     &           & 2.00(7)   & 1.301(18)  & 1.010(8) \\ 
$0^{--\ast\ast}$ &           & 2.38(17)  & 1.544(24)  & 1.186(10) \\
$0^{-+}$         &           & 2.35(16)  & 1.575(27)  & 1.200(10) \\ 
$0^{-+\ast}$     &           &           & 1.80(6)    & 1.325(14) \\ 
$0^{+-}$         &           & 2.66(32)  & 1.76(6)    & 1.340(14) \\ 
$0^{+-\ast}$     &           &           & 1.85(6)    & 1.564(22) \\ 
$2^{++}$         & 2.08(18)  & 1.76(4)   & 1.168(14)  & 0.9122(56) \\ 
$2^{++\ast}$     &           & 2.09(10)  & 1.408(17)  & 1.085(10) \\
$2^{-+}$         & 2.15(19)  & 1.79(5)   & 1.205(18)  & 0.8936(67) \\ 
$2^{-+\ast}$     &           & 1.99(10)  & 1.429(23)  & 1.095(12) \\ 
$2^{--}$         &           & 2.08(12)  & 1.430(23)  & 1.067(12) \\ 
$2^{--\ast}$     &           & 2.38(18)  & 1.57(4)    & 1.244(13) \\ 
$2^{+-}$         &           & 2.17(10)  & 1.394(24)  & 1.077(9) \\
$2^{+-\ast}$     &           & 2.53(29)  & 1.59(4)    & 1.296(13) \\
$1^{++}$         &           & 2.66(28)  & 1.720(31)  & 1.300(11) \\ 
$1^{++\ast}$     &           &           & 2.05(4)    & 1.454(15) \\ 
$1^{-+}$         &           & 2.68(26)  & 1.67(3)    & 1.340(13) \\ 
$1^{-+\ast}$     &           &           & 1.76(3)    & 1.344(10) \\ 
$1^{--}$         &           & 2.48(13)  & 1.70(3)    & 1.251(9) \\ 
$1^{--\ast}$     &           &           & 1.74(5)    & 1.320(13) \\ 
$1^{+-}$         &           & 2.42(18)  & 1.73(5)    & 1.264(10) \\ 
$1^{+-\ast}$     &           &           & 1.79(4)    & 1.313(16) \\ \hline
\end{tabular}
\caption{\label{table_msu4}
The lightest SU(4) masses.}
\end{center}
\end{table}

\begin{table}
\begin{center}
\begin{tabular}{|l|l|l|l|l|}\hline
state & $\beta=33$ & $\beta=44$ & $\beta=64$ & $\beta=82$   \\
      & L=12 & L=16 & L=24  & L=32   \\ \hline
$0^{++}$         & 1.550(47) & 1.054(15) & 0.695(5)  & 0.5325(48) \\
$0^{++\ast}$     & 2.4(3)    & 1.654(30) & 1.053(11) & 0.798(7) \\ 
$0^{++\ast\ast}$ &           & 2.06(11)  & 1.342(27) & 1.009(10) \\
$0^{--}$         & 2.29(24)  & 1.581(44) & 1.041(11) & 0.765(10) \\
$0^{--\ast}$     &           & 2.05(11)  & 1.278(20) & 0.997(8)  \\
$0^{--\ast\ast}$ &           & 2.27(19)  & 1.555(34) & 1.19(2)  \\
$0^{-+}$         &           & 2.32(14)  & 1.478(23) & 1.155(11) \\
$0^{-+\ast}$     &           &           & 1.80(4)   & 1.347(18) \\
$0^{+-}$         &           & 2.54(32)  & 1.78(7)   & 1.256(14) \\
$0^{+-\ast}$     &           &           & 1.81(5)   & 1.580(33)  \\
$2^{++}$         &           & 1.69(7)   & 1.08(5)   & 0.8914(69)  \\
$2^{++\ast}$     &           & 2.05(9)   & 1.39(3)   & 1.072(9)  \\ 
$2^{-+}$         &           & 1.60(6)   & 1.07(5)   & 0.8785(69)  \\
$2^{-+\ast}$     &           & 2.06(14)  & 1.37(3)   & 1.075(9)  \\
$2^{--}$         &           & 2.14(11)  & 1.390(25) & 1.054(10)  \\
$2^{--\ast}$     &           & 2.57(28)  & 1.559(31) & 1.225(16)  \\
$2^{+-}$         &           & 2.08(12)  & 1.430(21) & 1.043(9  \\ 
$2^{+-\ast}$     &           & 2.51(36)  & 1.557(34) & 1.230(17)  \\
$1^{++}$         &           & 2.38(22)  & 1.718(30) & 1.294(14)  \\
$1^{++\ast}$     &           &           & 1.93(8)   & 1.421(14)  \\
$1^{-+}$         &           & 2.57(23)  & 1.697(31) & 1.300(10)  \\
$1^{-+\ast}$     &           &           & 1.78(5)   & 1.331(15)  \\
$1^{--}$         &           & 2.29(10)  & 1.654(32) & 1.212(9)  \\ 
$1^{--\ast}$     &           &           & 1.66(4)   & 1.27(2)  \\ 
$1^{+-}$         &           & 2.41(16)  & 1.71(5)   & 1.255(14)  \\
$1^{+-\ast}$     &           &           & 1.73(4)   & 1.29(2) \\ \hline
\end{tabular}
\caption{\label{table_msu5}
The lightest SU(5) masses.}
\end{center}
\end{table}

\begin{table}
\begin{center}
\begin{tabular}{|l|l|l|l|l|}\hline
\multicolumn{5}{|c|}{$m_G/\surd\sigma$} \\ \hline
state & SU(2) & SU(3) & SU(4) & SU(5) \\ \hline
$0^{++}$         & 4.718(43) & 4.329(41) & 4.236(50) & 4.184(55) \\
$0^{++\ast}$     & 6.83(10)  & 6.52(9)  & 6.38(13) & 6.20(13) \\
$0^{++\ast\ast}$ & 8.15(15)  & 8.23(17) & 8.05(22) & 7.85(22) \\ 
$0^{--}$         &           & 6.48(9)  & 6.271(95) & 6.03(18) \\ 
$0^{--\ast}$     &           & 8.15(16) & 7.86(20) & 7.87(25) \\ 
$0^{--\ast\ast}$ &           & 9.81(26) & 9.21(30) & 9.51(41) \\
$0^{-+}$         & 9.95(32)  & 9.30(25) & 9.31(28) & 9.19(29) \\
$0^{+-}$         &           & 10.52(28) & 10.35(50) & 9.43(75) \\
$2^{++}$         & 7.82(14)  & 7.13(12) & 7.15(13) & 7.19(20) \\
$2^{++\ast}$     &           &          & 8.51(20) & 8.59(18) \\
$2^{-+}$         & 7.86(14)  & 7.36(11) & 6.86(18) & 7.18(16) \\
$2^{-+\ast}$     &           & 8.80(20) & 8.75(28) & 8.67(24) \\
$2^{--}$         &           & 8.75(17) & 8.22(32) & 8.24(21) \\
$2^{--\ast}$     &           & 10.31(27) & 9.91(41) & 9.79(45) \\
$2^{+-}$         &           & 8.38(21)  & 8.33(25) & 8.02(40) \\
$2^{+-\ast}$     &           & 10.51(30) & 10.64(60) & 9.97(55) \\
$1^{++}$         & 10.42(34) & 10.22(24) & 9.91(36) & 10.26(50) \\
$1^{-+}$         & 11.13(42) & 10.19(27) & 10.85(55)& 10.28(34) \\
$1^{--}$         &           & 9.86(23)  & 9.50(35) & 9.65(40) \\
$1^{+-}$         &           & 10.41(36) & 9.70(45) & 9.93(44) \\ \hline
\end{tabular}
\caption{\label{table_mcont}
Glueball masses in units of the string tension: in the continuum limit.}
\end{center}
\end{table}

\begin{table}
\begin{center}
\begin{tabular}{|l|l|l|l|l|}\hline
\multicolumn{5}{|c|}{best fit confidence level (\%)} \\ \hline
state & SU(2) & SU(3) & SU(4) & SU(5) \\ \hline
$0^{++}$         & 85(25) & 70(20) & 70(20) & 60(20) \\
$0^{++\ast}$     & 25(5)  & 80(25) & 90(25) & 100(25) \\
$0^{++\ast\ast}$ & 15(3)  & 90(25) & 40(10) & 60(20) \\ 
$0^{--}$         &  & 95(25) & 35(10) & 15(3) \\ 
$0^{--\ast}$     &  & 90(25) & 35(10) & 23(5) \\
$0^{--\ast\ast}$ &  & 65(20) & 70(20) & 70(20) \\
$0^{-+}$         & 90(25) & 45(15) & 85(25) & 35(10) \\
$0^{+-}$         &  & 85(25) & 95(25) & 13(2) \\
$2^{++}$         & 95(25) & 40(10) & 25(6)  & 17(3) \\
$2^{++\ast}$     &  &        & 90(25) & 90(25) \\
$2^{-+}$         & 60(20) & 50(15) & 15(3)  & 30(8) \\ 
$2^{-+\ast}$     &  & 12(3)  & 14(10) & 65(20) \\
$2^{--}$         &  & 80(25) & 25(6)  & 85(25) \\
$2^{--\ast}$     &  & 70(20) & 40(10) & 30(10) \\
$2^{+-}$         &  & 25(5)  & 45(10) & 3(0.3) \\
$2^{+-\ast}$     &  & 80(25) & 12(3)  & 40(10) \\
$1^{++}$         & 60(20) & 40(10) & 100(25) & 22(4) \\
$1^{-+}$         & 60(20) & 60(15) & 10(2)  & 100(25) \\
$1^{--}$         &  & 45(10) & 10(2)  & 2(0.1) \\
$1^{+-}$         &  & 25(5)  & 12(3)  & 15(3) \\ \hline
\end{tabular}
\caption{\label{table_clcont}
Confidence levels of the best fits in Table~\ref{table_mcont};
in brackets those of the fits that provide the errors.}
\end{center}
\end{table}

\begin{table}
\begin{center}
\begin{tabular}{|c|l|l|l|}\hline
\multicolumn{4}{|c|}{ $a_t m$ with  $a_t \simeq 0.25 a_s$} \\ \hline
state & $\beta=4.0$ & $\beta=5.3$ & $\beta=8.0$  \\
      & $12^2 60$   & $16^2 64$   & $24^2 96$    \\ \hline
  flux loop      & 0.431(7)  & 0.305(5)  & 0.185(2) \\
$0^{++}$         & 0.443(5)  & 0.321(4)  & 0.207(2) \\
$0^{++\ast}$     & 0.63(1)   & 0.448(12) & 0.302(3) \\
$0^{++\ast\ast}$ & 0.77(2)   & 0.582(7)  & 0.368(7) \\
$0^{-+}$         & 1.01(3)   & 0.68(3)   & 0.427(9) \\
$0^{-+\ast}$     & 1.14(3)   & 0.73(3)   & 0.527(5) \\
$2^{++}$         & 0.717(10) & 0.532(8)  & 0.344(4) \\
$2^{++\ast}$     & 0.86(2)   & 0.636(9)  & 0.395(6) \\
$2^{-+}$         & 0.750(13) & 0.538(5)  & 0.339(4) \\
$2^{-+\ast}$     & 0.93(3)   & 0.647(8)  & 0.407(5) \\
$1^{++}$         & 1.03(3)   & 0.75(5)   & 0.471(4) \\
$1^{++\ast}$     & 1.3(1)    & 0.76(6)   & 0.551(7) \\
$1^{-+}$         & 1.12(4)   & 0.80(6)   & 0.499(6) \\
$1^{-+\ast}$     & 1.22(5)   & 0.80(5)   & 0.477(12) \\ \hline\hline
$\langle Tr U_{p_s}/N_c \rangle$ & 0.64060(7) & 0.73195(3) & 0.82418(1) \\
$\langle Tr U_{p_t}/N_c \rangle$ & 0.91436(2) & 0.93597(1) &
0.95795(1) 
\\ \hline
\end{tabular}
\caption{\label{table_masym}
Masses with asymmetric SU(2) action; $r=0.25$ in eqn(\ref{B3}).
Also shown are the average timelike and spacelike palquettes.}
\end{center}
\end{table}

\begin{table}
\begin{center}
\begin{tabular}{|c|c|l|l|l|l|l|}\hline
$\beta$ & $L_s$ & $r$ & $r_I$ &  $p=0,1$ & $p=0,2$ & $a_tm_l/a_sm_l$ \\ \hline  
8.0 & 24 & 0.25 & 0.232 & 0.249(11) & 0.252(5) &  \\
    & 16 & 0.25 & 0.232 & 0.244(9)  & 0.241(4) & 0.235(8) \\
5.3 & 16 & 0.25 & 0.221 & 0.239(25) & 0.235(7) & \\
    & 12 & 0.25 & 0.221 & 0.235(10) & 0.238(4) & 0.241(6) \\
4.0 & 12 & 0.25 & 0.209 & 0.236(28) & 0.241(8) & \\
    &  8 & 0.25 & 0.209 & 0.223(7)  & 0.219(5) & 0.241(12) \\ \hline
\end{tabular}
\caption{\label{table_rasym}
Various estimates of $a_t/a_s$ as described in Appendix D. Also
shown is $r_I$, the mean-field improved value of $r$.}
\end{center}
\end{table}

\begin{table}
\begin{center}
\begin{tabular}{|l|l|l|}\hline
state            & SU(2) ; r=1.0 & SU(2) ; r=0.25 \\ \hline
$0^{++}$         & 4.718(43) & 4.65(10) \\
$0^{++\ast}$     & 6.83(10)  & 6.83(20) \\
$0^{++\ast\ast}$ & 8.15(15)  & 8.39(33) \\
$0^{-+}$         & 9.95(32)  & 9.23(38) \\
$2^{++}$         & 7.82(14)  & 7.81(20) \\
$2^{++\ast}$     &           & 8.86(30) \\
$2^{-+}$         & 7.86(14)  & 7.54(20) \\
$2^{-+\ast}$     &           & 8.94(27) \\
$1^{++}$         & 10.42(34) & 10.51(24) \\
$1^{-+}$         & 11.13(42) & 11.03(30) \\
                 &           & [10.38(44)] \\ \hline
$\surd\sigma/g^2$      & 0.3353(18) & 0.3375(130) \\ \hline
\end{tabular}
\caption{\label{table_compasym}
Comparison between the continuum mass ratios, $m_G/\surd\sigma$, 
obtained with the asymmetric r=0.25 SU(2) action, and our previous
$r=1$ SU(2) results. Also shown is $\surd\sigma/g^2$ for both cases.} 
\end{center}
\end{table}

\begin{table}
\begin{center}
\begin{tabular}{|l||c|l|c|l|}\hline
state & $\lim_{N_c\to\infty} m/g^2N_c$ & slope & $N_c\geq$ 
& CL\% \\ \hline \hline
$0^{++}$         & 0.808(11) & -0.070(79)  & 2 & 90(25) \\ 
$0^{++\ast}$     & 1.227(25)  & -0.31(18) & 2 & 65(20) \\
$0^{++\ast\ast}$ & 1.581(42)  & -0.84(28) & 2 & 50(15) \\ 
$0^{-+}$         & 1.787(60)  & -0.50(51) & 2 & 85(25) \\
$2^{++}$         & 1.365(33)  & -0.25(28) & 2 & 35(10) \\
$2^{-+}$         & 1.369(36)  & -0.20(27) & 2 & 15(4)  \\
$2^{-+\ast}$     & 1.704(70)  & -0.74(88) & 3 & 95(25) \\
$1^{++}$         & 1.98(8)   & -0.90(57) & 2 & 80(25) \\
$1^{-+}$         & 1.99(8)   & -0.61(70) & 2 & 35(10)  \\ \hline
$0^{--}$         & 1.167(42)  &  0.26(50) & 3 & 65(20) \\
$0^{--\ast}$     & 1.508(72)  & -0.07(87) & 3 & 65(20) \\
$0^{--\ast\ast}$ & 1.77(13)  &  0.24(161) & 3 & 30(8) \\
$0^{+-}$         & 1.87(23)  &  0.63(245) & 3 & 45(10) \\
$2^{--}$         & 1.57(8)   &  0.40(93)  & 3 & 55(15) \\
$2^{--\ast}$     & 1.87(12)  &  0.23(143) & 3 & 90(25) \\
$2^{+-}$         & 1.59(10)  & -0.37(117) & 3 & 65(20) \\
$2^{+-\ast}$     & 1.97(17)  & -0.28(188) & 3 & 50(15) \\
$1^{--}$         & 1.85(13)  & -0.33(149) & 3 & 55(15) \\
$1^{+-}$         & 1.87(16)  &  0.37(200) & 3 & 45(10) \\ \hline
\end{tabular}
\caption{\label{table_massgN}
The large $N_c$ limit of the mass spectrum in units of 
$g^2N_c$; with the slope of the linear fit 
when plotted against $1/N_c^2$. Also the 
range of colours fitted and the confidence level of the fits.}
\end{center}
\end{table}

\begin{table}
\begin{center}
\begin{tabular}{|l||c|l|c|l|}\hline
state & $\lim_{N_c\to\infty} m/\surd\sigma$ & slope & $N_c\geq$ 
& CL\% \\ \hline \hline
$0^{++}$         & 4.065(55) & 2.58(42)  & 2 & 80(25) \\ 
$0^{++\ast}$     & 6.18(13)  & 2.68(100) & 2 & 70(20) \\
$0^{++\ast\ast}$ & 7.99(22)  & 0.79(160) & 2 & 50(15) \\ 
$0^{-+}$         & 9.02(30)  & 3.52(275) & 2 & 85(25) \\
$2^{++}$         & 6.88(16)  & 3.50(134) & 2 & 30(10) \\
$2^{-+}$         & 6.89(21)  & 3.13(162) & 2 & 20(5)  \\
$2^{-+\ast}$     & 8.62(38)  & 1.69(165) & 3 & 90(25) \\
$1^{++}$         & 9.98(25)  & 1.78(203) & 2 & 80(25) \\
$1^{-+}$         & 10.06(40) & 3.58(365) & 2 & 30(8)  \\ \hline
$0^{--}$         & 5.91(25)  & 5.24(300) & 3 & 55(15) \\
$0^{--\ast}$     & 7.63(37)  & 4.61(460) & 3 & 70(20) \\
$0^{--\ast\ast}$ & 8.96(65)  & 7.2(80)   & 3 & 35(10) \\
$0^{+-}$         & 9.47(116) & 9.7(12.4) & 3 & 40(10) \\
$2^{--}$         & 7.89(35)  & 7.6(44)   & 3 & 60(20) \\
$2^{--\ast}$     & 9.46(66)  & 7.6(77)   & 3 & 95(25) \\
$2^{+-}$         & 8.04(50)  & 3.2(60)   & 3 & 60(20) \\
$2^{+-\ast}$     & 9.97(91)  & 5.1(10.0) & 3 & 50(15) \\
$1^{--}$         & 9.36(60)  & 4.4(70)   & 3 & 60(20) \\
$1^{+-}$         & 9.43(75)  & 8.4(98)   & 3 & 50(15) \\ \hline
\end{tabular}
\caption{\label{table_masskN}
The large $N_c$ limit of the mass spectrum in units of the
string tension; with the slope of the linear fit when plotted 
against $1/N_c^2$. Also the 
range of colours fitted and the confidence level of the fits.}
\end{center}
\end{table}

\begin{table}
\begin{center}
\begin{tabular}{|l||c|c|c|}\hline
\multicolumn{4}{|c|}{$m_G/\surd\sigma$ : U(1)} \\ \hline
state & $\beta=2.0$ & $\beta=2.2$ & $\beta=2.3$ \\ \hline 
$0^{++}$ & 3.54(9) & 3.29(23)  & 3.36(17) \\
$0^{--}$ & 1.97(7) & 1.52(5)   & 1.50(5) \\ 
$0^{-+}$ & 7.1(9)  & 8.1(4) & 8.8(3) \\ 
$0^{+-}$ &         & 8.5(4) & 10.4(5) \\ 
$2^{++}$ & 5.2(7) & 4.64(30) & 5.12(27)  \\ 
$2^{-+}$ & 4.8(11) & 4.9(9)  & 5.45(40) \\ 
$2^{--}$ & 6.1(4) & 7.0(3) & 6.3(4) \\ 
$2^{+-}$ & 7.0(5) & 6.7(3) & 6.7(6) \\ 
$1^{++}$ &  & 9.6(6) & 11.0(6) \\ 
$1^{-+}$ &  & 9.7(6) & 11.6(6) \\ 
$1^{--}$ & 7.8(8)  & 7.9(3) &  8.7(3) \\ 
$1^{+-}$ & 8.2(5) & 8.0(3) & 8.6(3) \\ \hline
$a\surd\sigma$ & 0.2251(18) & 0.1734(16) & 0.1505(15) \\ \hline 
\end{tabular}
\caption{\label{table_massU1}
The U(1) mass spectrum in units of the string tension,
at several values of $\beta$. In the last row is the
string tension in lattice units.}
\end{center}
\end{table}

\clearpage

\begin	{figure}[p]
\begin	{center}
\leavevmode
\input	{plot_KbetaIsu2}
\end	{center}
\vskip 0.15in
\caption{As in Fig.\ref{fig_Kbetasu2} but using the mean-field
improved coupling, $\beta_I$, in place of $\beta$.}
\label{fig_KbetaIsu2}
\end 	{figure}

\begin	{figure}[p]
\begin	{center}
\leavevmode
\input	{plot_mcpsuN}
\end	{center}
\vskip 0.15in
\caption{Some of the $C=+$ glueball masses for 2,3,4 and 5
colours, in units of $g^2N_c$ and plotted against $1/N_c^2$:
$0^{++}$($\bullet$), $0^{++*}$($\times$), $2^{++}$($\star$), 
$0^{-+}$($\diamond$), $1^{++}$($\circ$). The best linear
extrapolations to the $N_c = \infty$ limit are also shown.}
\label{fig_mcpsuN}
\end 	{figure}

\begin	{figure}[p]
\begin	{center}
\leavevmode
\input	{plot_mcnsuN}
\end	{center}
\vskip 0.15in
\caption{Some of the $C=-$ glueball masses for 3,4 and 5
colours, in units of $g^2N_c$ and plotted against $1/N_c^2$:
$0^{--}$($\bullet$), $0^{--*}$($\times$), $2^{--}$($\star$), 
$1^{--}$($\circ$). The best linear
extrapolations to the $N_c = \infty$ limit are also shown.}
\label{fig_mcnsuN}
\end 	{figure}

\end{document}